\documentclass[12pt,preprint]{aastex}
\shorttitle{2175-{\AA} Feature in Mg~II Absorbers}
\shortauthors{Jiang et al.}
\begin{document}
\title{Towards Detecting The 2175-{\AA} Dust Feature Associated With Strong High Redshift Mg~II Absorption Lines}
\author{Peng Jiang\altaffilmark{1,2}, Jian Ge\altaffilmark{2}, Hongyan Zhou\altaffilmark{1},
Junxian Wang\altaffilmark{1}, and Tinggui Wang\altaffilmark{1}}
\altaffiltext{1}{Key Laboratory for Research in Galaxies and Cosmology, The University of Science
and Technology of China, Chinese Academy of Sciences, Hefei, Anhui, 230026, China}
\altaffiltext{2}{Astronomy Department, University of Florida, 211 Bryant Space
Science Center, P. O. Box 112055, Gainesville, FL 32611}
\email{jpaty@mail.ustc.edu.cn}

\begin{abstract}
We report detections of 39 2175-{\AA} dust extinction bump candidates associated with strong
Mg~II absorption lines at z$\sim$ 1--1.8 on quasar spectra in Sloan Digital Sky Survey (SDSS) DR3.
These strong Mg~II absorption line systems are detected among 2,951 strong Mg~II absorbers 
with the rest equivalent width $W_r\lambda2796 >$ 1.0{\AA} at $1.0 < z < 1.86$, which is part of a 
full sample of 7,421 strong Mg~II absorbers compiled by Prochter et al. (2006).
The redshift range of the absorbers is chosen to allow the 2175-{\AA} extinction features to be completely
covered within the SDSS spectrograph operation wavelength range.
An upper limit of the background quasar emission redshift at z$=$2.1 is set to prevent the Ly$\alpha$ forest
lines from contaminating the sensitive spectral region for the 2175-{\AA} bump measurements. 
The FM90 (Fitzpatrick \& Massa 1990) parameterization is applied to model the Optical/UV
extinction curve in the rest frame of Mg~II absorbers of the 2175-{\AA} bump candidates.
The simulation technique developed by Jiang et al. (2010a, b) is used to derive
the statistical significance of the candidate 2175-{\AA} bumps.
A total of 12 absorbers are detected with 2175-{\AA} bumps at a 5$\sigma$ level of
statistical significance, 10 are detected at a 4$\sigma$ level and 17 are detected at 
a 3$\sigma$ level. Most of the candidate bumps in this work are similar to the relatively
weak 2175-{\AA} bumps observed in the Large Magellanic Clouds (LMC) LMC2 supershell
rather than the strong ones observed in the Milky Way (MW).
This sample has greatly increased the total number of 2175-{\AA} extinction bumps
measured on SDSS quasar spectra. Follow-up observations may rule out some of possible false detections
and reveal the physical and chemical natures of 2175-{\AA} quasar absorbers.
\end{abstract}


\section{Introduction}
Quasar absorption lines (QALs) provide us a unique tool to study abundances,
physical conditions and kinematics of gas in a wide variety of astrophysical environments.
Many of the early detected QALs are associated systems with redshifts close to the quasar emission 
redshifts (e.g., Burbidge et al. 1966; Schmidt 1966). The  discovery of intervening QALs
 was first reported by Bahcall et al. (1966). Bahcall (1968) developed the  first systematic method 
 for identifying absorption systems on quasar spectra. 
To date, thousands of QALs have been detected (e.g., Nestor et al. 2005; Prochter et al. 2006) 
thanks to the huge  quasar database from SDSS (York et al. 2000).

The existence of dust content in quasar Damped Ly$\alpha$ Absorbers (DLAs) was first detected through
reddening measurements of quasar continuum spectra (Fall et al. 1989; Pei et al. 1991). 
It was later confirmed  by measurements of depletion of different metal elements in the gas phase onto
dust grains (e.g., Pettini et al. 1994, 1997, 1999; Lu et al. 1996).
Murphy \& Liske (2004) further investigated dust reddening of DLAs on background
quasars by comparing the spectral index distribution of a sample of 70 DLAs at $z \sim 3$
to that of a large control sample, but they found no evidence for dust reddening.
Vladilo et al. (2008) revisited dust reddening of DLAs by deriving the color excess
of background quasars, and they found significant color excess caused by the foreground
absorbers in their sample of 248 DLAs at $2.2 < z < 3.5$. By studying
the average extinction curves on quasar spectra with Mg~II absorption lines, York et al. (2006)
found clear evidence for presence of dust in intervening absorption systems
with $1 < z < 2$. Recently, M\'{e}nard et al. (2008) confirmed dust reddening in
a much larger absorber sample with nearly 7000 strong Mg~II absorption systems at $0.4 < z < 2.2$.
All these dust reddening studies show that, on average, the dust grain size distribution
of quasar absorption systems is similar to that of the Small Magellanic Cloud (SMC; Pei 1992).
The extinction curve does not show presence of the broad bump around 2175-{\AA}
found in the Milky Way (MW) interstellar clouds (e.g., Stecher 1965; Savage \& Mathis 1979; Fitzpatrick 1989).
The 2175-{\AA} dust feature has been found in a handful of local galaxies, where 
the reddening law of their inter-stellar medium (ISM) can be studied spatially
(Keel et al. 2001). The 2175-{\AA} bumps in the MW extinction curves appear to be the strongest among all
of the known galaxies.  The strength of the bump in the extinction
curves from the Large Magellanic Clouds 2 (LMC2) Supershell (near the 30 Dor star formation region; Gordon et al. 2003)
is quite moderate while the bumps are nearly absent in the SMC extinction curves.
It is very likely that the MW-like dust is a minor population among ISM dust in the universe.
Therefore, the 2175-{\AA} dust extinction bumps cannot be recovered by using any average extinction curves method.

The detections of the 2175-{\AA} extinction bump in individual quasar absorption systems
were well reviewed by Wang et al. (2004). Recently, several new detections have been reported
(e.g., Srianand et al. 2008; Noterdaeme et al. 2009; Zhou et al. 2010). All the recent ones
were detected on SDSS spectra by comparing the reddened quasar spectrum with the SDSS
composite quasar spectrum (Vanden Berk et al. 2001). In nearly parallel efforts,
the analysis of Gamma-Ray Bursts (GRB) afterglow spectra has also revealed several positive detections
of 2175-{\AA} dust feature from intervening absorbers and from gas in the GRB host galaxies
(e.g., Ellison et al. 2006; El{\'i}asd{\'o}ttir et al. 2009; Prochaska et al. 2009).
In this paper, we report detections of 39 2175-{\AA} dust extinction bump candidates associated
with strong Mg~II absorption line systems on quasar spectra in SDSS DR3 based on the FM90 parameterization
modeling of Optical/UV extinction curves.

This paper is organized as follows:
in \S2 we describe the parameterization of Optical/UV extinction curves;
in \S3 we introduce the procedures of searching for 2175-{\AA} absorbers in a 
strong Mg~II absorber sample from SDSS DR3 and report 39 candidates;
in \S4 further investigations and follow-up observations in future are discussed;
more relevant discussions and a short summary are given in \S5.

\section{Parameterization of Extinction Curves}
Previous studies by Cardelli et al. (1989; CCM) show that most of the variation in the 
MW extinction curves can be described as an empirical relationship based on a single parameter,
$R_V=A_V/E(\bv)$. Since $R_V$ is a rough measurement of an average dust grain size,
the $R_V$ dependent CCM relationship gives a physical basis for the variation in the
MW extinction curves. However, the width and the peak position of the Galactic 2175-{\AA}
extinction bump are uncorrelated with $R_V$ (see Figure 10 in Cardelli et al. 1989 and Figure 20
in Fitzpatrick \& Massa 2007). This indicates that the carriers of bumps are likely
independent of the dust grains responsible for $R_V$ variations. Thus, it is necessary to use
an independent parameterization model to properly describe the 2175-{\AA} dust extinction bump.

We use a parameterized extinction curve (FM90 parameterization) constituted by
a linear component and a Drude component to describe the Optical/UV extinction
curve in the rest frame of an absorber on SDSS spectra for quasar absorption line systems.
The linear component is used to model the underlying extinction\footnote{The linear
component also accounts for the variation of the intrinsic quasar spectral slopes.},
while the Drude component is used to model
the possible 2175-{\AA} extinction bump. The parameterized extinction curve is
written as
\begin{equation}
A(\lambda)=c_1+c_2x+c_3D(x,x_0,\gamma)
\end{equation}
where $x=\lambda^{-1}$.
And $D(x,x_0,\gamma)$ is a Drude profile, which is expressed as
\begin{equation}
D(x,x_0,\gamma)={\frac{x^2}{(x^2-x_0^2)^2+x^2\gamma^2}}
\end{equation}
where $x_0$ and $\gamma$ is the peak position and FWHM
of the Drude profile, respectively. Dust reddened SDSS quasar spectra are modeled by reddening
the SDSS composite quasar spectrum with a parameterized
extinction curve in the rest frame of the absorber of interest.
Our aim is to unveil the 2175-{\AA}
absorption feature associated with absorption line systems on quasar spectra.
Here we have no intention to derive the absolute extinction curve. Instead, our derived curve
is a relative extinction curve without being normalized by E(\bv). Therefore, we cannot precisely
extract the conventionally defined extinction parameters, such as $A_V$, E(\bv) and $R_V$, from the 
derived extinction curve. However, all the features of 2175-{\AA} absorption bump are preserved
in our derived curve. The strength of bump can be measured by the area of the bump
$A_{bump}=\pi c_3/(2\gamma)$, which can be interpreted as rescaling the integrated apparent optical
depth of the bump absorption ($A_{\lambda}={\frac{2.5}{ln 10}}\tau_{\lambda}$).

\section{Searching for 2175-{\AA} Extinction Bumps}
In this section, we present the procedures for searching for 2175-{\AA} extinction bumps
associated with strong Mg~II absorption lines on SDSS DR3 quasar spectra.

\subsection{A Redshift Selected Sample of Strong Mg~II Absorbers in SDSS DR3}
The full sample of strong Mg~II absorption line systems in SDSS DR3 was compiled by Prochter et al. (2006).
It contains 7,421 confirmed Mg~II absorption systems with the rest equivalent width
\footnote{Unless otherwise stated, $W_r$ refers to the rest equivalent width of the Mg~II $\lambda$2796
transition} $W_r >$ 1.0 \AA, spanning a redshift range of 0.35$ < z < $2.3.
We applied two redshift criteria to select spectra for searching for the 2175-{\AA} extinction
bumps on them. We restricted the redshifts of Mg~II absorption lines to within 1.0$\le z_{abs} \le1.86$
to allow the potential 2175-{\AA} extinction feature to lie completely within the SDSS spectrograph
operation wavelength region (3,800{\AA}--9,200{\AA}; York et al. 2001). We also set an upper limit of z=2.1
for the background quasar redshifts to avoid the Ly$\alpha$ forest line contamination in the 
potential 2175-{\AA} bump features. The final sample which meets both criteria has a total of 2,951 strong
Mg~II absorbers.

The spectroscopic and photometric data in this work
are extracted from the SDSS DR7 database (Abazajian et al. 2009) to take advantage
of the new photometric and spectrophotometric calibrations.
These calibrations are critical for determining the overall shape of spectra.
We adopted the redshifts and $W_r$s of Mg~II absorption lines measured
by Prochter et al. (2006) based on SDSS DR3 spectra.
Although the wavelength calibration has been updated in SDSS DR7, the measurement
of absorption redshift in Prochter et al. (2006) is still adequate for this work
since the width of expected 2175-{\AA} extinction bumps are so broad that our result is not
sensitive to the accuracy of absorption redshifts. The error of $W_r$s is dominated by systematic
uncertainty due to continuum fitting and contamination by absorption lines from 
other redshift absorbers and the Earth's atmosphere. The expected typical error on $W_r$s is
0.2--0.3 {\AA} (Prochter et al. 2006). We measured the $W_r$s for a few Mg~II absorbers by
using their DR7 spectra to examine whether the new calibrations affect the measurement of $W_r$.
Our results are consistent with the values reported in Prochter et al. (2006) within the expected
error (0.2--0.3 {\AA}). Therefore, we take the $W_r$s measured by Prochter et al. (2006) and put
conservative errorbars (0.3 {\AA}) on the measurements uniformly in this paper.

We assumed that the quasar spectra are free of contamination of starlight from the host galaxy
because the quasars in this work have redshifts $z>1$. All the photometric and spectroscopic SDSS
data are corrected for Galactic reddening by using the dust map of Schlegel et al. (1998).

\subsection{SDSS DR7 Quasar Composite Spectrum}
The SDSS quasar composite spectrum plays an important role in our method as mentioned in \S2.
The median quasar composite spectrum combined by Vanden Berk et al. (2001) was employed
in our previous works (Jiang et al. 2010a, b). That composite spectrum was created by
combining 2,200 quasar spectra in the SDSS EDR database (Stoughton et al. 2002). In this paper,
we updated the SDSS quasar composite spectrum by combining 105,783 quasar spectra in the
SDSS DR7 database (Schneider et al. 2010). There are two main considerations to update it:
(1) the change of spectral calibration method between EDR and DR7; (2) the change of quasar
selection criteria. The first one will improve the precision of individual spectrum and
the later one will make the distribution of quasar redshifts more uniform. We follow the
method in Vanden Berk et al. (2001) to create the composite SDSS DR7 quasar spectrum by
using median combining. The combined spectrum is shown in Figure 1 and Table 1.
The analysis of the new composite spectrum and comparison with the old one are beyond
the scope of this paper.  

\subsection{Searching Procedures}
In this subsection, we describe the procedures for searching for 2175-{\AA} extinction
features on quasar spectra in the redshift selected Mg~II absorber sample.
The basic idea is to fit the quasar spectra by reddening
the SDSS composite quasar spectrum with a parameterized extinction curve in the rest
frame of a Mg~II absorber and then determine the significance of positive bump candidates with
the simulation technique developed by Jiang et al. (2010a, b).

First, we fitted every spectrum of the selected Mg~II absorbers with a reddened composite quasar spectrum.
The only continuum of a quasar spectrum was fitted while the regions with strong emission lines
(see Figure 1a) and known strong absorption lines are masked without fitting.
We selected the preliminary 2175-{\AA} extinction bump candidates with three criteria:
(1). the peak position is in the range of 4.4 $\mu m^{-1}< x_0<$ 4.8 $\mu m^{-1}$; (2).
the bump width is in the range of 0.5 $\mu m^{-1}$ $<\gamma<$ 2.7 $\mu m^{-1}$; (3). the
height of the bump $c3$ is positive. The constraints on the peak position and width of a bump are
determined according to the distribution of the parameters measured on 328 Galactic 2175-{\AA}
extinction bumps by Fitzpatrick \& Massa (2007) and 9 bumps in the LMC2 supershell region by
Gordon et al. (2003). We obtained a total of 259 preliminary bump candidates from our initial
search.

Pitman et al. (2000) pointed out that the variation of broad Fe II emission
multiplets can mimic the feature of an extinction bump on a quasar spectrum. Many of our
preliminary candidates may be false positives caused by the broad iron emission lines. Therefore, 
it is important to determine the statistical significance of the bump candidates
with considerations of possible broad iron emission feature contamination in the spectra.
The simulation technique developed by Jiang et al. (2010a, b) was applied to determine the
detection significance. The simulation begins with the selection of a control sample of SDSS
quasar spectra at the similar redshift to the quasar of interest. In this work,
we selected the control quasar spectra with an I band signal-to-noise ratio of {\tt SNR$\ge$}5 
and emission redshifts in the range of $z_{em}-0.05 < z < z_{em}+0.05$, where $z_{em}$ is the
emission redshift of the studied quasar, in the SDSS DR7 database (Schneider et al. 2010).
The average size of the control samples is about 5,000.
We fitted all the spectra in each control sample by reddening the composite
quasar spectrum with a parameterized extinction curve at the redshift of the Mg~II absorber
of interest. The parameters $x_0$ and $\gamma$ in the parameterized extinction
curve are fixed to the best values fitting the studied 2175-{\AA} extinction bump candidate.
We then collected all the best fitted bump strengths of the spectra in the control sample.
The distribution of these strengths describes the random fluctuation of the quasar
continuum and the variation of broad Fe II emission multiplets by assuming no spectrum in
the control sample possesses a real extinction bump feature.
This strength distribution can be fitted by a Gaussian profile. 
If the bump strength of the studied candidate is far away from the Gaussian strength
distribution of its control sample, then the bump candidate has statistical
significance. We consider preliminary bump candidates with a significant level of
$>$ 3$\sigma$\footnote{$\sigma$ is the standard deviation of the Gaussian profile.}
as candidates while rejecting those with a confidence level less than 3$\sigma$.
We then performed a visual examination on the selected candidates and
removed one broad absorption line (BAL) quasar and one quasar which spectrum were greatly 
trimmed in the SDSS data. All the rejected candidates are listed in Table 2. This final round
study results in 39 2175-{\AA} extinction bump candidates.

\subsection{The 39 2175-{\AA} extinction bump candidates}
We split the 39 candidates into three groups: the 12 high confidence candidates
with a statistical significance level of greater than 5$\sigma$; the 10 median confidence
candidates with a level of 4$\sigma$; the 17 low confidence candidates with a level of
3$\sigma$. The best fitted parameters of the extinction curves of the high, median and
low confidence candidates are listed in Table 3, 4 and 5, respectively.
Figure 2 is a series of plots of the high confidence 2175-{\AA} absorber candidates.
The best fitted models are overplotted with their SDSS data in panel (a).
To emphasize the requirement of an absorption bump on the extinction curve, we also overplot
the reddened composite quasar spectrum with only the linear component of the best fitting model
(green solid line). The derived extinction curve is plotted in panel (b). 
The extracted distribution of bump strengths in the simulations is presented in panel (c).
The 3, 4, 5$\sigma$ boundaries of the distributions are indicated with vertical dashed lines.
Figure 3 shows a series of plots of the median confidence candidates and Figure 4 shows the
low confidence ones.

From literatures, we found 6 2175-{\AA} bumps had been detected on SDSS quasar spectra
previously (Srianand et al. 2008; Noterdaeme et al. 2009; Zhou et al. 2010; Jiang et al. 2010b). 
Among the 6 2175-{\AA} absorbers, J085042.24+515911.6 (Srianand et al. 2008) is the only one which
is included in our redshift selected SDSS DR3 Mg~II absorber subsample. The others are excluded
because they were observed later than SDSS DR3 quasars or their redshifts are out of our selected
range. J0850+5159 is identified as a high confidence candidate in this work. We refer the readers to
Jiang et al. (2010a) for the details of quantitative analysis for the other 5 previously detected
bumps.

Figure 5 illustrates the distribution of widths and strengths of 257 preliminary 2175-{\AA}
bump candidates\footnote{There are 259 preliminary 2175-{\AA} bump candidates as stated above.
Here, the BAL quasar and the quasar with a largely trimmed spectrum are excluded.}.
For comparison, Figure 5 also includes the bumps observed in MW
(Fitzpatrick \& Massa 2007; empty circles) and LMC2 (Gordon et al. 2003; filled black
circles).\footnote{Note that the
area of bump defined in Fitzpatrick \& Massa (2007) and Gordon et al. (2003) is
different from that in this work. Since their extinction curves have been normalized
by E(\bv), $A_{bump}$=E(\bv)$\times A_{bump}^{*}$, where $A_{bump}^{*}$ is the area
defined in Fitzpatrick \& Massa (2007) and Gordon et al. (2003).}
The bump candidates are labeled with filled color circles. Their bump strengths are much
weaker than those of the MW bumps on average, but similar to the LMC2 supershell bumps.
Among the 6 previous bump detections on SDSS quasar spectra, there are two strong
MW-like bumps: the 2175-{\AA} absorber toward the quasar J100713.68+285348.4 at z $\sim$ 0.9
(Zhou et al. 2010) and the other one toward the quasar J145907.19+002401.2 at z $\sim$ 1.4
(Jiang et al. 2010b).

Figure 5 also shows that the strengths of low confidence candidates
(filled green circles) increase with the increasing bump widths.
This trend can be explained naturally with the simulation technique.
For a spectrum without a real extinction feature in the control sample, the fitted
bump strength is actually an integration of random fluctuations over
the wavelength range covered by it. In principle, a broader integration
makes the resultant fitted bump strengths to appear stronger. 
Therefore, a broad bump must be stronger than a narrow one in order to be
significantly distinguished from random fluctuations of quasar spectra.
In other word, the detection threshold is higher for broader bumps.
We also notice that some rejected bump candidates are even stronger than
some of the confidence candidates with the same widths. This is because that the 
distribution of fitted bump strengths in the simulations is also dependent on
x$_0^{qso}$, which is the center of the bump in the rest frame of quasar emissions.
Since we mask the region with strong emission lines when fitting quasar spectra,
the bumps on/near broad emission lines cannot be well constrained. The
deviation of the resultant strength distribution becomes larger when the bump center 
is close to a strong broad emission line. Therefore, the detection threshold
depends on x$_0^{qso}$ as well as the width $\gamma$. In summary, every bump
candidate needs to be analyzed individually to derive its statistical significance.

In order to investigate how x$_0^{qso}$ and $\gamma$ affect the detection
threshold quantitatively, we designed a set of simulations by using all the quasar
spectra with redshifts in the range of $1.0 < z < 2.0$ in SDSS DR7.
We conducted simulations for a fairly wide range of x$_0^{qso}$ and $\gamma$
listed in Table 6. First, we collected all the spectra which could fully contain
a 2175-{\AA} bump with the specified pair of x$_0^{qso}$ and $\gamma$ as a sample.
We then derived the distribution of fitted bump strengths by fitting every quasar
spectrum with a composite quasar spectrum reddened by a FM09 parameterized
Optical/UV extinction curve. The resultant distribution of strengths is fitted with a Gaussian
profile. Basically the procedures are almost the same as those of significance
simulations described above, except for the selection of the quasar sample.
In Table 6, we list the 3$\sigma$ and 5$\sigma$ detection thresholds for different
pairs of x$_0^{qso}$ and $\gamma$. Figure 6 illustrates the simulation results.
The dotted spectra are SDSS quasar composite spectra redshifted to $z=1.95$.
Then we redden the composite spectra by an extinction bump
of a x$_0^{qso}$ and $\gamma$ pair with the corresponding 5 $\sigma$ threshold
strength. We do not apply any underlying linear extinction on composite spectra.
The spectra for different x$_0^{qso}$s are organized in separated panels; the
spectra in the same panel are for increasing $\gamma$ from top to bottom.

In Figure 7, we study the correlationship between bump strength of the candidates
and $W_r$ of Mg~II$\lambda$2796 and the correlationship between strength and relative
color $\Delta (g-i)$ of the background quasar. The color of quasars is redshift dependent,
since the broad emission features on underlying continuum move in/out the photometric
passbands at different redshift (Richards et al. 2001). Richards et al. (2003) introduced
a relative color $\Delta (g-i)$ to determine the underlying continuum color of quasars
by subtracting the median colors of quasars at the redshift of each quasar from the
measured colors of each quasar. The distribution of relative colors should be a Gaussian,
assuming a Gaussian distribution of power-law spectral indices of quasars.
However, $\Delta (g-i)$ shows a significant asymmetric tail to the red end. The objects
in this tail are reddened by dust (Richards et al. 2003). Figure 7a illuminates the
distribution of the 2,951 redshift selected Mg~II absorbers and the 39 2175-{\AA}
absorber candidates in the space of $W_r$ and $\Delta (g-i)$. The reddened tail
of the distribution of $\Delta (g-i)$ for the Mg~II absorber sample is prominent. The average
$\Delta (g-i)$ of detected 2175-{\AA} absorbers is 0.39 and all of them have relative colors
$\Delta (g-i)$s of $>$ 0.
The 2175-{\AA} absorbers should be dusty since the 2175-{\AA} dust extinction feature
is already detected on their spectra. However, about one third of our confidence candidates
are not located in the dust reddening tail in figure 7a. It could be explained if the
background quasars of these relative blue 2175-{\AA} absorbers have very flat spectral index
intrinsically. Another possible explanation is that the underlying linear extinction curves
of these relative blue candidates are fairly flat in UV bands. In this work,
the linear component of FM90 extinction curve accounts for the variation of the
intrinsic quasar spectral slopes as well as the dust reddening. This is different
from the fitting method used by Srianand et al (2008) and
Noterdaeme et al. (2009). Both of them used the average MW and LMC2 extinction
curves when reddening the SDSS composite quasar spectrum. On these average curves, the
strength of the 2175-{\AA} bump and the slope of the linear component are fixed together.
A quasar spectrum possessing a strong bump feature has to have a very red $\Delta (g-i)$
in their cases. Otherwise, the quasar spectrum cannot be fitted with their method. In
Figure 7b, we plot the bump strength of our candidates versus $\Delta (g-i)$ of their
background quasars. We indeed obtain several fairly strong bumps in the region of small
positive $\Delta (g-i)$. No correlation appears in this plot.
Another difference between our fitting method and the average MW extinction curve method
is that the width of 2175-{\AA} bump is free.
The width of the bump is fixed in Srianand et al (2008) and Noterdaeme et al. (2009) to
the width of average MW bump (0.922$\mu$m$^{-1}$) or the width of average LMC2 bump
(0.945$\mu$m$^{-1}$). However, the widths of known MW and LMC2 bumps distribute widely
from 0.5$\mu$m$^{-1}$ to 2.7$\mu$m$^{-1}$. It is reasonable to expect some broad bump
features appearing on quasar spectra. These spectra cannot be well fitted by their
method (e.g., the 2175-{\AA} absorber toward J145907.19+002401.2 in Jiang et al. 2010b).
The bump strengths of our candidates are plotted versus $W_r$ is Figure 7c.
We found no correlationship between them.

\section{Future Work on Candidates}
In this paper, we use the SDSS composite quasar spectrum to model the "dereddened"
spectrum of background quasar. This approach has been applied by most of the researchers 
who study the 2175-{\AA} bumps on SDSS quasar spectra (e.g., Srianand et al. 2008;
Noterdaeme et al. 2009; Zhou et al. 2010). Composite quasar spectrum serves as a good
approximation for the individual quasar spectrum in these works. However, the quasar
spectra varies a lot in the spectral slope of nonthermal continuum, the width of broad
emission lines, the intensity of broad emission lines. The most important problem is the
pseudo-continuum due to Fe~II broad emission lines while investigating the 2175-{\AA}
absorption bumps. The fluctuation of pseudo-continuum can mimic a broad absorption bump
feature on a quasar spectrum in some cases (e.g., Pitman 2000; Noterdaeme et al. 2009).
Assuming the fluctuation is purely random, researchers tried to rule out the fake bumps
in statistical ways (e.g., Noterdaeme et al. 2009; Jiang et al. 2010b). We use the same
statistical method with Jiang et al. 2010b to gauge the significance of the 2175-{\AA}
bump candidates. In a conservative manner, we prefer to name the
selected 39 2175-{\AA} bumps in this work as candidates rather than detections.
We think further investigation is necessary to rule out some of possible false detections.
 
The spectrum fitting of continuum and emission lines of quasar has been widely studied
by researchers interested in quasar emissions (e.g., Boroson \& Green 1992;
Dong et al. 2005, 2008; Wang et al. 2009). There are three main components in the model.
The AGN continuum is usually modeled with a power law;  the broad Fe~II multiplets
(pseudo-continuum) is modeled with the UV Fe~II template, which was generated
by Tsuzuki et al. (2006) based on their measurements of I Zw1; the other emission lines
are modeled with Gaussians. For a reddened quasar, an extinction curve needs to be considered
in the model. In practice, the fitting with this full model is fairly complicated. In
most cases, the fitting cannot be solved. Although it is difficult, the full model fitting
provides us a possible tool to distinguish the real 2175-{\AA} absorption bump with the
pseudo-continuum due to broad Fe~II multiplets on a quasar spectrum. The high resolution
and high S/N follow-up observations could enable us to attempt the full model fitting on
the 2175-{\AA} bump candidates. If succeed, the existence of bumps can be
determined more firmly and their shape can be measured more precisely.

Spectroscopic follow-up observation may also reveal the chemical and physical natures of
2175-{\AA} bump absorbers, such as metallicity, dust depletion, temperature, ionization
state and density as well as velocity profile. 
Srianand et al. (2008) detected 21-cm absorption lines associated with two 2175-{\AA}
quasar absorbers. The 21-cm absorption is usually an indicator of cold dense gas clouds.
Jiang et al. (2010b) measured the relative abundance of two strong 2175-{\AA} absorbers and
found the high dust depletion [Fe/Zn]=-1.59 (i.e. a characteristic of cold dense clouds in the MW)
in both of them. In addition, Noterdaeme et al. (2009) measured a similar high dust depletion
level ([Fe/Zn]=-1.47) of a 2175-{\AA} absorber at $z$=1.64 toward the
quasar SDSS J160457.50+220300.5 and detected C~I and CO absorption lines associated with
this absorber. These are important clues to the cold ISM origin of the 2175-{\AA} dust
absorption feature. However, the carrier of the 2175-{\AA} bump is still not very clear 
(e.g. Draine 2003 and reference therein). The large polycyclic aromatic hydrocarbon (PAH)
molecules, which have strong $\pi \rightarrow \pi^*$ absorption in the 2000--2500 {\AA} region,
are proposed to be a promising candidate for the carrier of 2175-{\AA} absorption
(Li \& Draine 2001). In Figure 5, we see a wide distribution of
strengths and widths for the bumps observed in local galaxies. It
could be caused by the varying size of PAH molecules from one sightline to another. Draine
(2003) interpreted the observed variations in  bump widths (and small variations in peak positions)
of the Galactic 2175-\AA\ extinction bump profile can result from differences in the PAH mix.
This difference can also affect the oscillator strength per molecule and lead to the
deviation of bump strengths.

\section{Discussion and Summary}
The SDSS DR7 database contains 105,783 quasar spectra (Schneider et al. 2010). We expect
to find more than 50 significant 2175-{\AA} bumps associated with Mg~II absorption line systems
by implementing the same searching strategy in this paper. Another possible approach of
searching for the 2175-{\AA} extinction bump on quasar spectra is to compare the reddened
quasar with a blue quasar template (J. Wang et al., 2011, in preparation). The blue quasar
template is defined as the spectrum of an observed SDSS blue quasar with similar width and
strength of broad emission lines (including the broad FeII emission lines) with the reddened
quasar of interest. The only difference between the reddened quasar and its blue template is
considered to be dust reddening. This idea is very similar to the method used to extract the
extinction curves by comparing the spectrum of a reddened hot early-type star with a unreddened
star having the same stellar type in the MW (e.g. Trunpler 1930; Welty \& Fowler 1992).
This method may measure the 2175-{\AA} bump on quasar spectrum more precisely when compared
with the composite spectrum method since the emission lines are matched.
However, the selection of blue quasar template needs to be studied carefully.

Dust extinction and reddening effects can change the magnitude and color
of background quasars. It is very likely that some dust reddened quasars cannot be
recovered by the SDSS quasar target selection algorithm (Richards et al. 2002), which is
mainly based on the SDSS imaging magnitudes and colors.
In addition to searching for 2175-{\AA} absorbers in the current
SDSS quasar spectra database, we also plan to select the 2175-{\AA} quasar absorber
candidates in the SDSS image database for follow-up spectroscopic observations in future.

The wavelength coverage of SDSS spectrum is from 3800 {\AA} to 9200 {\AA}.
It allows us to detect Galactic 2175-{\AA} extinction feature up to redshift $z \sim 3$.
The highest redshift for identified 2175-{\AA} bump to date is $z = 3.03$, which was
detected by Prochaska et al. (2009) using the afterglow Optical/IR photometry of gamma-ray
burst GRB 080607. Another high redshift 2175-{\AA} bump was detected by
El{\'i}asd{\'o}ttir et al. (2009) on the afterglow optical spectrum of GRB 070802 at $z = 2.45$.
We are also studying the feasibility of using our method to search for high redshift 2175-{\AA}
bumps. The main difficulty to implement our method to high redshift bumps is that
the random Ly$\alpha$ limit absorptions are present blueward of the Ly$\alpha$
emission line on the spectrum of quasar beyond $z=2.1$. They make the spectrum fitting
procedure more complicated.
However, a further developed detection method for high redshift bumps would
be very helpful for us to search for 2175-{\AA} dust extinction bumps on the spectra
of $\sim$160,000 quasars at redshifts $2.2 < z < 3$ from the SDSS-III's Baryon Oscillation
Spectroscopic Survey (BOSS; Eisenstein et al. 2011)\footnote{http://www.sdss3.org/cosmology.php} in future.

In this paper, we searched for 2175-{\AA} dust extinction bumps associated with strong Mg~II
absorption lines on quasar spectra in SDSS DR3. The search results in 12 high confidence
candidates with a 5$\sigma$ level of statistical significance, 10 median confidence
candidates with a 4$\sigma$ level and 17 low confidence candidates with a 3$\sigma$ level.  
The total number of 2175-{\AA} bump detections have been largely increased by this work.
Follow-up observations may rule out some of possible false detections
and reveal the physical and chemical natures of 2175-{\AA} quasar absorbers.

\acknowledgements
The authors appreciate the enlightening suggestions from the anonymous referee, which help to
improve the quality of this paper largely. This work was partially supported by NSF with
grant NSF AST-0451407, AST-0451408 \& AST-0705139,
and a China NSF grant (NSF-10973012) and the University of Florida.
PJ acknowledges support from China Scholarship Council.
This research has also been partially supported by the
CAS/SAFEA International Partnership Program for Creative Research Teams.

Funding for the SDSS and SDSS-II has been provided by the Alfred P. Sloan
Foundation, the Participating Institutions, the National Science Foundation,
the U.S. Department of Energy, the National Aeronautics and Space Administration,
the Japanese Monbukagakusho, the Max Planck Society, and the Higher Education
Funding Council for England. The SDSS Web Site is http://www.sdss.org/.

The SDSS is managed by the Astrophysical Research Consortium for the
Participating Institutions. The Participating Institutions are the American
Museum of Natural History, Astrophysical Institute Potsdam, University of Basel,
University of Cambridge, Case Western Reserve University, University of Chicago,
Drexel University, Fermilab, the Institute for Advanced Study, the Japan
Participation Group, Johns Hopkins University, the Joint Institute for Nuclear
Astrophysics, the Kavli Institute for Particle Astrophysics and Cosmology,
the Korean Scientist Group, the Chinese Academy of Sciences (LAMOST),
Los Alamos National Laboratory, the Max-Planck-Institute for Astronomy (MPIA),
the Max-Planck-Institute for Astrophysics (MPA), New Mexico State University,
Ohio State University, University of Pittsburgh, University of Portsmouth,
Princeton University, the United States Naval Observatory, and the University of Washington.

\clearpage
\begin{figure}
\epsscale{1.0}
\plotone{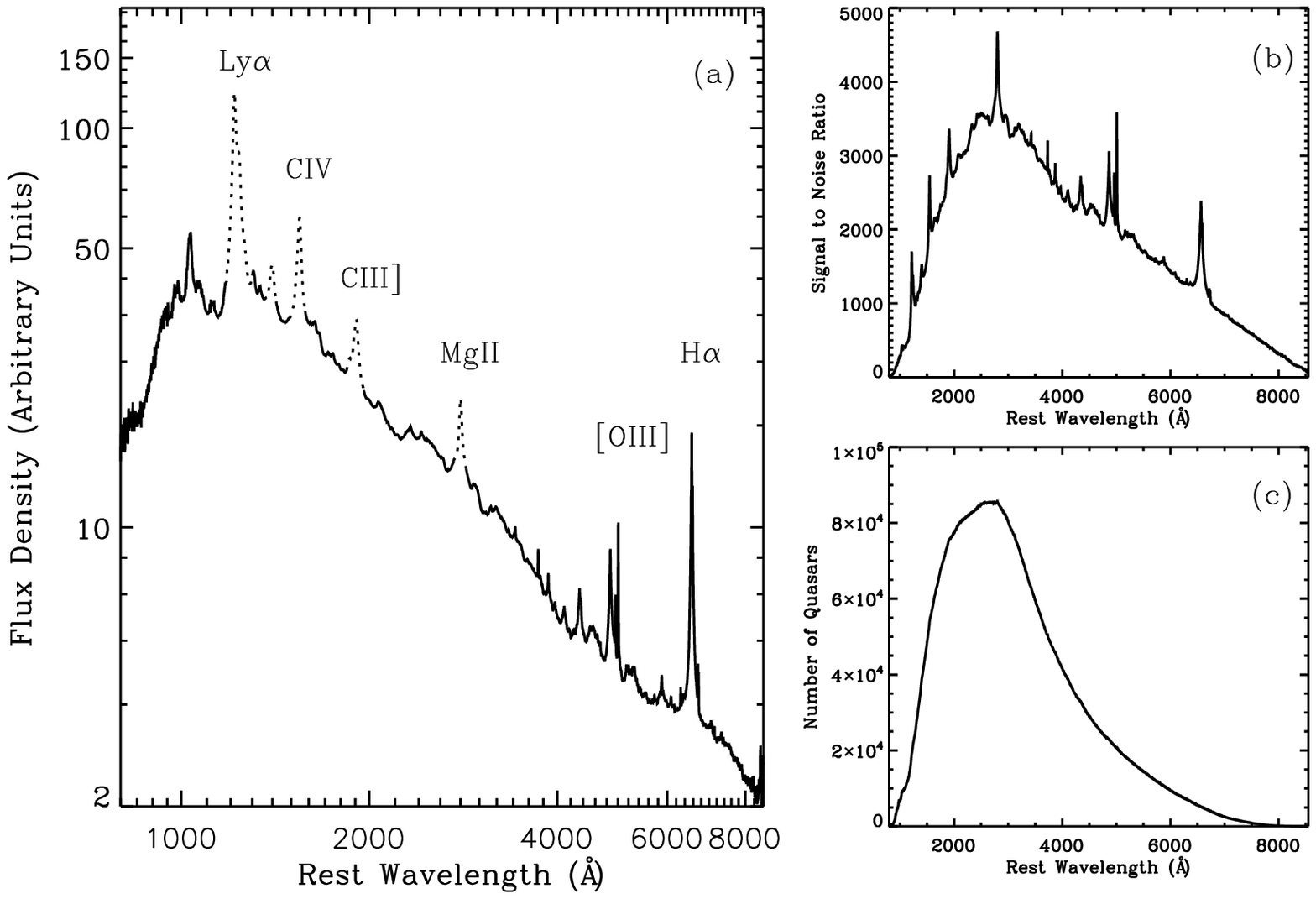}
\caption{Composite DR7 quasar spectrum by using median combining. (a). The regions with strong emission lines are masked
during spectrum fitting, which are marked with dotted lines. (b). Signal to noise ratio per 1 {\AA} bin for the median
composite quasar spectrum. (c). Number of quasar spectra combined in each 1 {\AA} bin of the composite spectrum.\label{fig1}}
\end{figure}

\clearpage
\begin{figure}
\epsscale{1.0}
\plotone{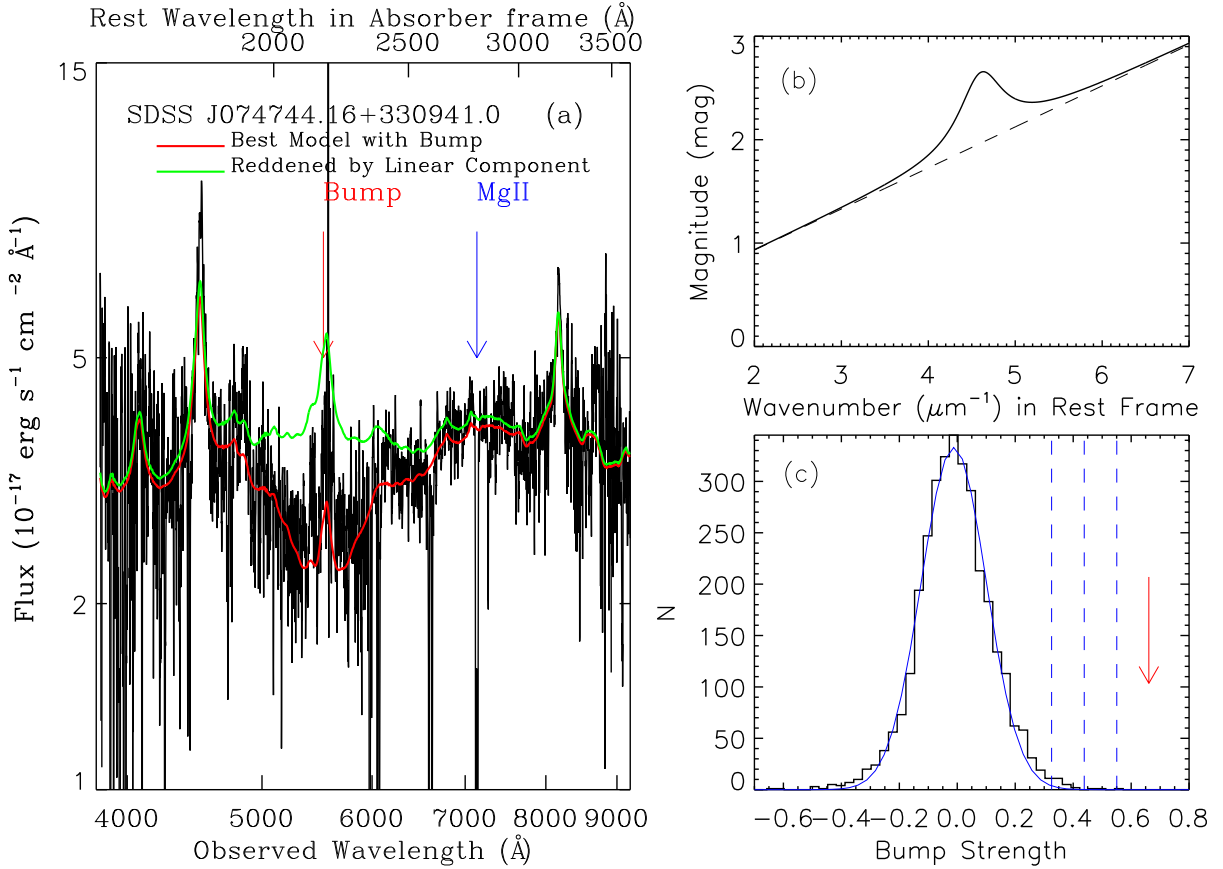}
\caption{The best fitted extinction model for J0747+3309. (a). Red solid line is the best fitted model.
Red arrow indicates the center of 2175-{\AA} extinction bump
and blue arrow indicates the Mg~II absorption lines.
Green solid line is reddened composite quasar spectrum
by using the linear component of best model only to emphasize the requirement of extinction bump. 
(b). The best fitted
extinction curve (see the parameters of extinction curves in Table 3). (c). Histogram of fitted bump
strength of the control sample for J0747+3309. The blue line is the best fitted Gaussian profile. Red
arrow indicates the strength of bump derived from spectrum of J0747+3309.
The three vertical blue dashed lines indicate the 3, 4 and 5$\sigma$ boundaries of the Gaussian.
[{\it{See the electronic edition of the Journal for all plots in}} Figure 2.]
\label{fig2}}
\end{figure}
\clearpage

\begin{figure}
\epsscale{1.0}
\plotone{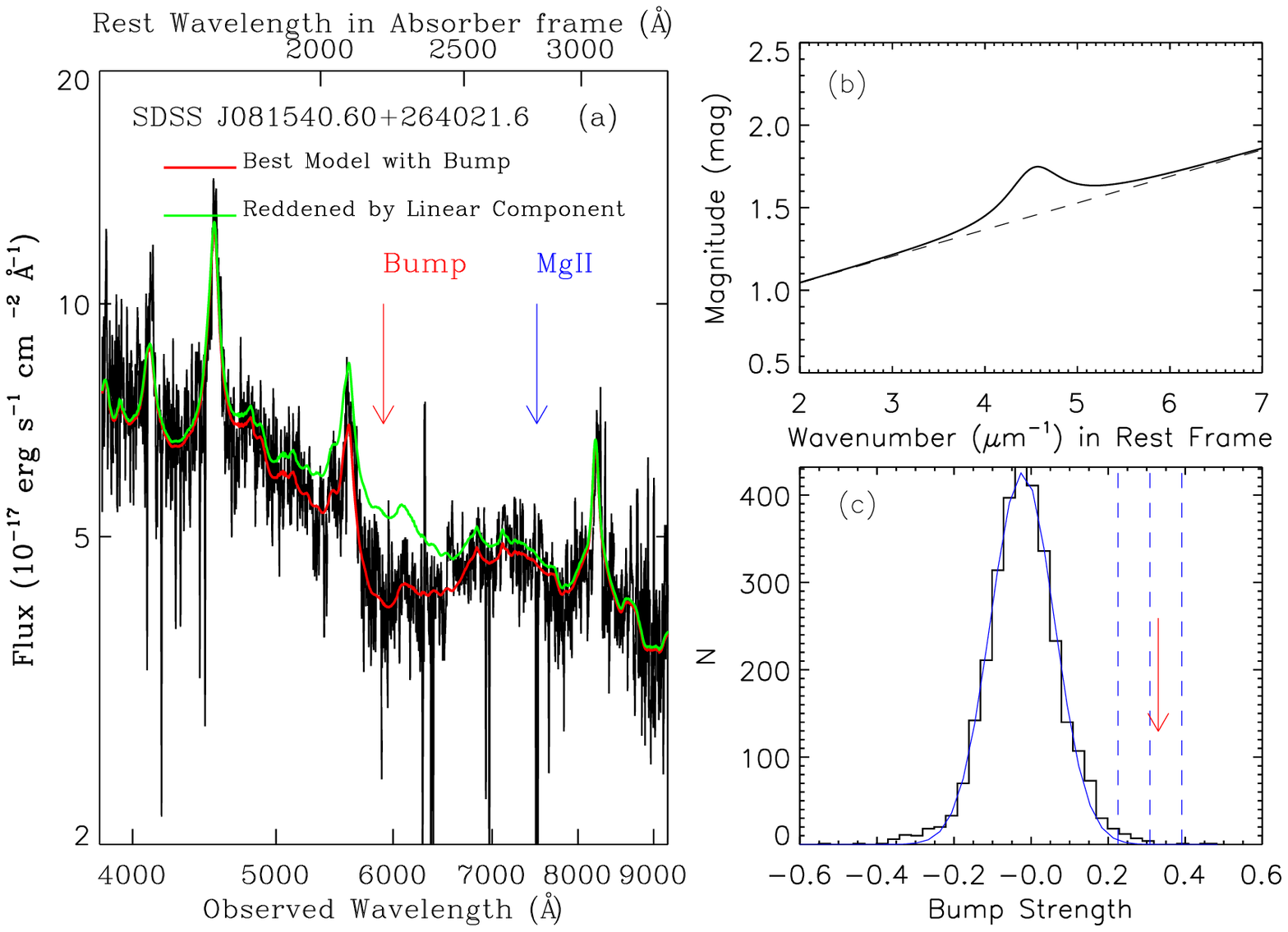}
\caption{The best fitted extinction model for J0815+2640. (a). Red solid line is the best fitted model.
Green solid line is reddened composite quasar spectrum
by using the linear component of best model only. (b). The best fitted
extinction curve. (c). Histogram of fitted bump strength of the control sample for J0815+2640.
[{\it{See the electronic edition of the Journal for all plots in}} Figure 3.]
\label{fig3}}
\end{figure}
\clearpage

\begin{figure}
\epsscale{1.0}
\plotone{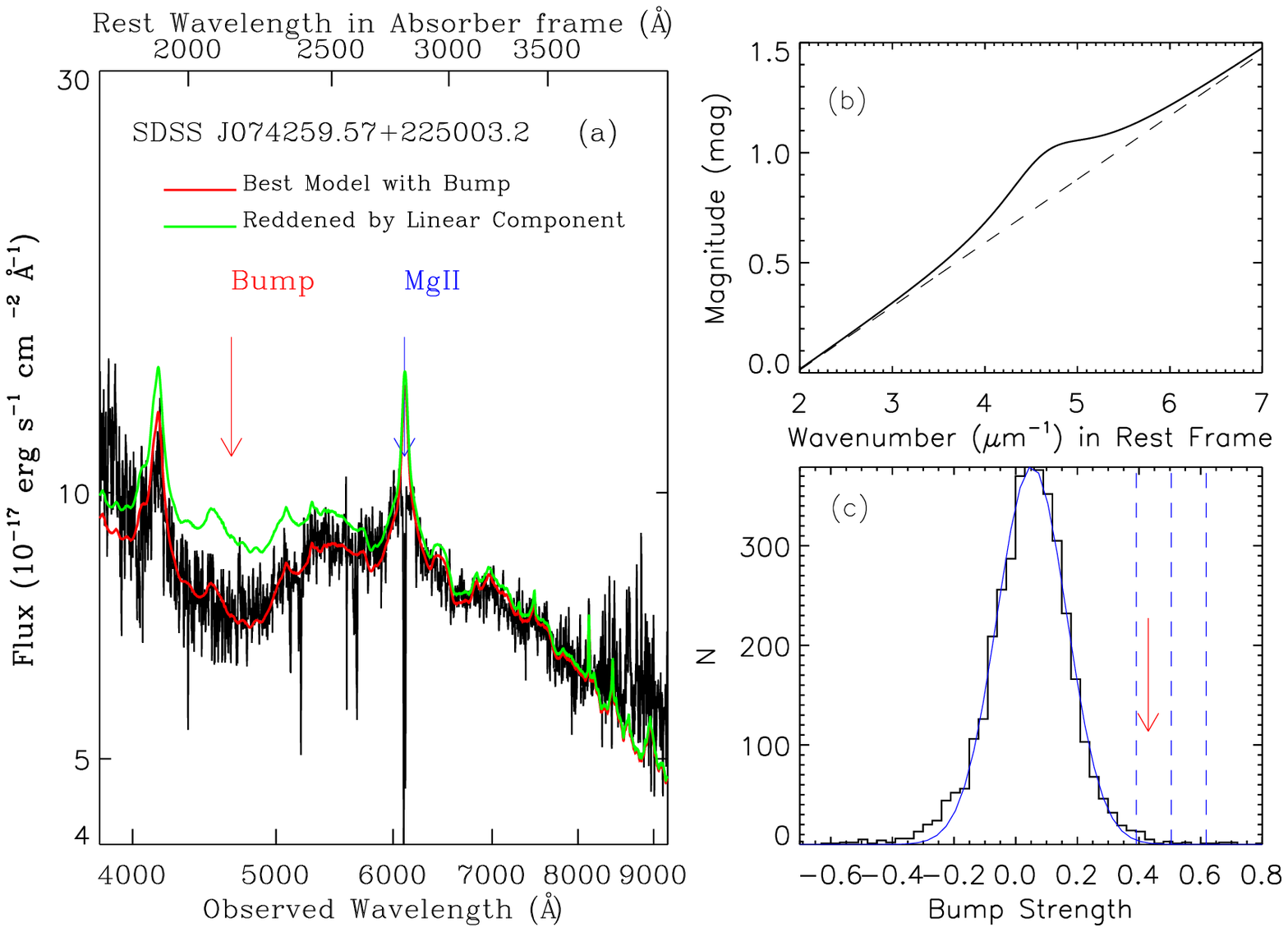}
\caption{The best fitted extinction model for J0742+2250. (a). Red solid line is the best fitted model.
Green solid line is reddened composite quasar spectrum
by using the linear component of best model only. (b). The best fitted
extinction curve. (c). Histogram of fitted bump strength of the control sample for J0742+2250.
[{\it{See the electronic edition of the Journal for all plots in}} Figure 4.]
\label{fig4}}
\end{figure}
\clearpage

\begin{figure}
\epsscale{1.0}
\plotone{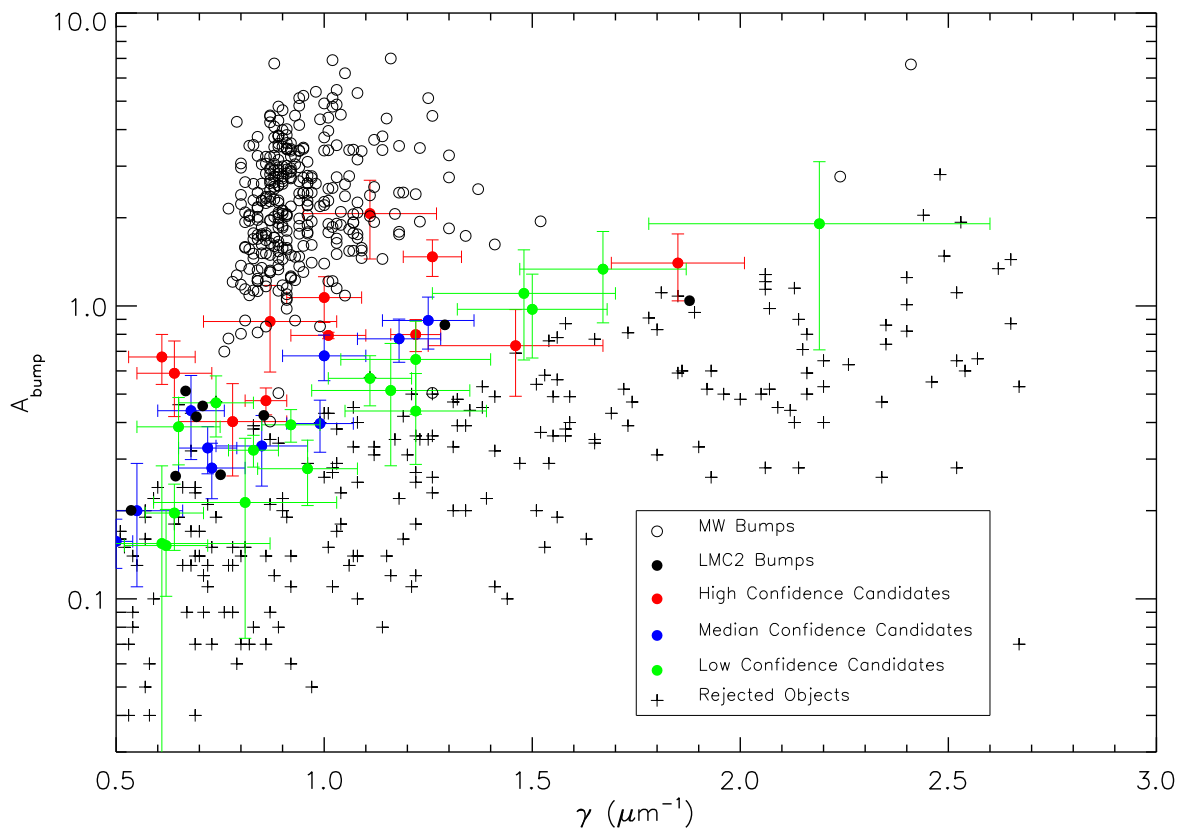}
\caption{Comparison of bump strength with 2175-{\AA} extinction bumps in MW (Fitzpatrick \& Massa 2007)
and LMC2 (Gordon et al. 2003) supershell. High, median and low confidence candidates in this work are
labeled with filled red, blue and green circles respectively. Bumps observed in MW are labeled with
empty circles and bumps observed in LMC2 supershell are labeled with filled black circles. The rejected
candidates are labeled with plus signs. For visual clarity, the errorbars on MW and LMC2 bumps and
rejected candidates are not plotted.
\label{fig5}}
\end{figure}
\clearpage

\begin{figure}
\epsscale{1.0}
\plotone{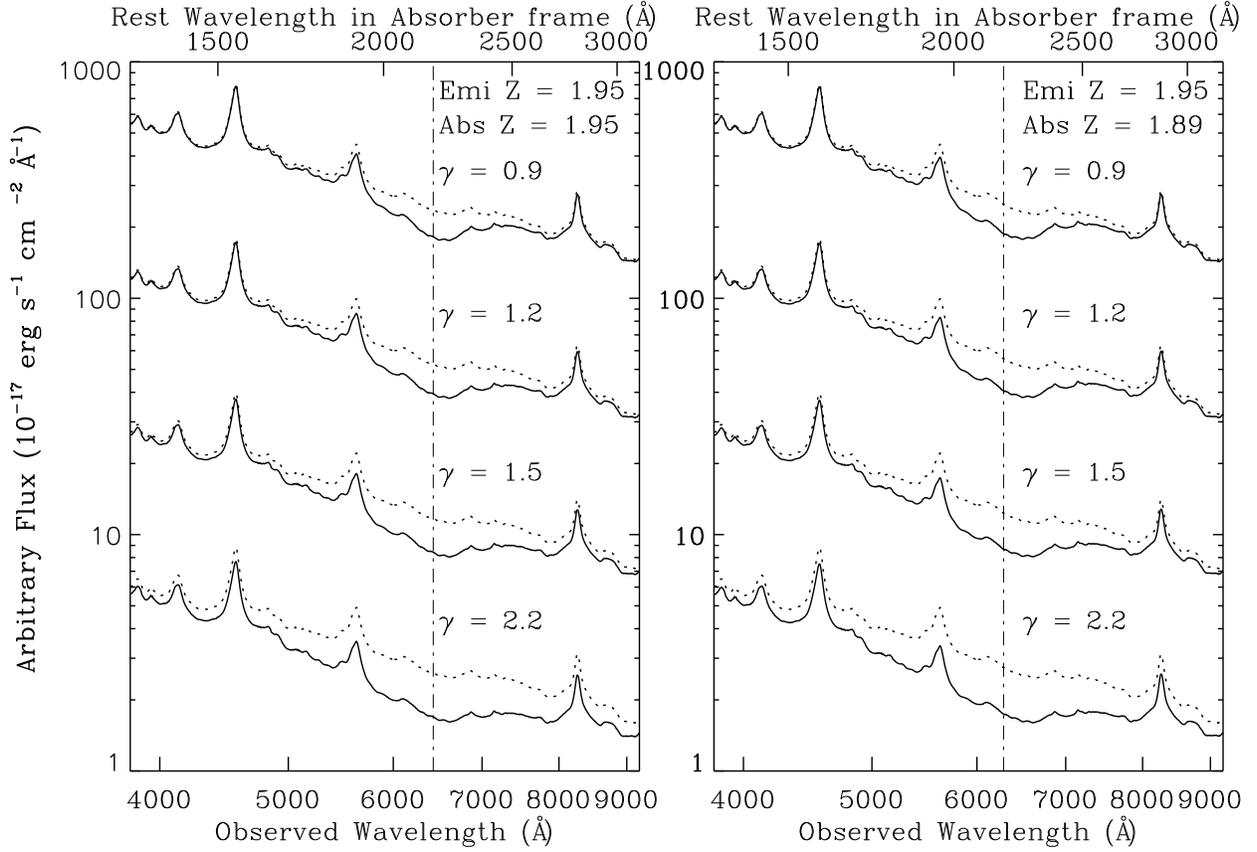}
\caption{The dotted spectra are SDSS
quasar composite spectra. The solid ones are the spectra reddened by bumps.
The spectra for different x$_0^{qso}$s are organized in separated panels; the
spectra in the same panel are for increasing $\gamma$ from top to bottom.
The dot-dashed lines indicate the center positions of the bump.
[{\it{See the electronic edition of the Journal for all plots in}} Figure 7.]
\label{fig6}}
\end{figure}
\clearpage

\begin{figure}
\epsscale{1.0}
\plotone{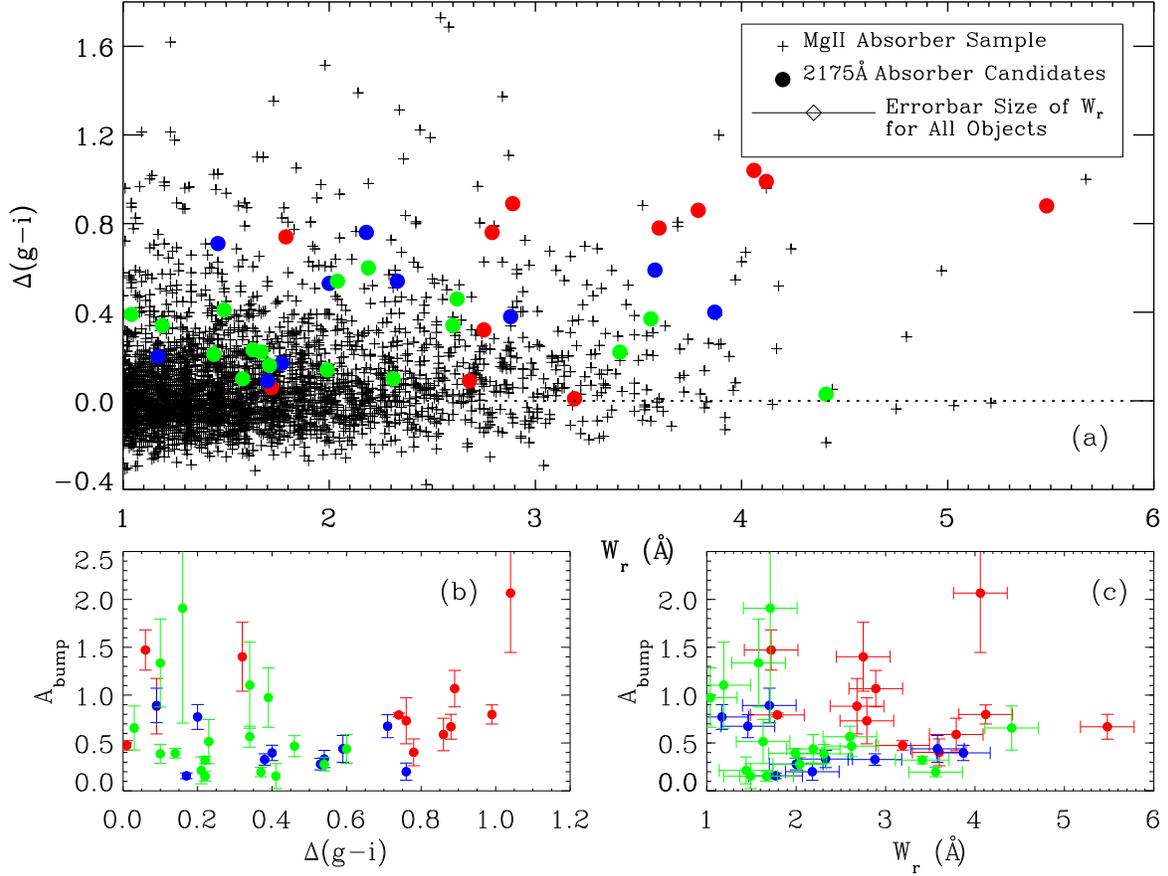}
\caption{(a). The distribution of the 2,951 redshift selected Mg~II absorbers and 2175-{\AA} absorber
candidates in the space of $W_r$ and $\Delta (g-i)$.
The horizontal dotted line indicates $\Delta (g-i)$=0. High, median and low confidence candidates are
labeled with filled red, blue and green circles respectively. (b). Bump strength of candidates
versus the rest equivalent widths of its associated Mg~II$\lambda$2796 absorption line. (c). Bump
strength of candidates with the relative color of background quasars. The expected typical
error of $W_r$ is of 0.3 {\AA}. The errorbars of $\Delta (g-i)$
are smaller than the radius of the color circles in the plots.
\label{fig7}}
\end{figure}
\clearpage

\begin{deluxetable}{ccc}
\tabletypesize{\normalsize}
\tablecaption{Median Composite DR7 Quasar Spectrum\label{tbl-1}}
\tablewidth{0pt}
\tablehead{
\colhead{$\lambda$} & \colhead{$f_{\lambda}$} & \colhead{$f_{\lambda}$ Uncertainty} \\
\colhead{({\AA})} & \colhead{(arbitrary units)} & \colhead{(arbitrary units)}
}
\startdata
800.0 & 14.634 & 1.009 \\
801.0 & 17.037 & 1.156 \\
802.0 & 17.622 & 1.128 \\
803.0 & 15.667 & 0.989 \\
804.0 & 15.867 & 1.040 \\
805.0 & 15.461 & 1.005 \\
806.0 & 17.926 & 1.130 \\
807.0 & 17.598 & 1.128 \\
808.0 & 16.062 & 0.994 \\
809.0 & 16.326 & 0.993 \\
810.0 & 17.917 & 1.105 \\
811.0 & 18.795 & 1.008 \\
812.0 & 15.443 & 0.833 \\
813.0 & 15.608 & 0.857 \\
814.0 & 16.658 & 0.962 \\
815.0 & 16.303 & 0.887 \\
\enddata
\tablecomments{Table 1 is available in its entirety in the electronic edition of the Astrophysical Journal.}
\end{deluxetable}

\begin{deluxetable}{lcccrccc}
\tabletypesize{\scriptsize}
\tablecaption{The List of Rejected Candidates\label{tbl-2}}
\tablewidth{0pt}
\tablehead{
\colhead{} & \colhead{} & \colhead{} & \colhead{$W_r^*$} & \colhead{$\Delta(g-i)$} & \colhead{$\gamma$} & \colhead{$A_{bump}$} & \colhead{} \\
\colhead{SDSS} & \colhead{Z$_{emi}$} & \colhead{Z$_{abs}$} & \colhead{(\AA)} & \colhead{(mag)} & ($\mu$m$^{-1}$) & \colhead{} & \colhead{Note}
}
\startdata
J000221.80+151454.5 & 1.825 & 1.435 & 1.29 & $-$0.01$\pm$0.02 & 2.40$\pm$0.28 & 0.82$\pm$0.26 & a \\
J000748.29+151746.0 & 1.237 & 1.044 & 1.83 & 0.25$\pm$0.02 & 2.13$\pm$0.40 & 1.15$\pm$0.72 & a \\
J001410.98$-$084429.2 & 1.772 & 1.292 & 2.32 & 0.59$\pm$0.03 & 0.68$\pm$0.09 & 0.35$\pm$0.10 & b \\
J002355.15+141900.9 & 1.116 & 1.012 & 1.46 & 0.19$\pm$0.02 & 0.70$\pm$0.14 & 0.17$\pm$0.08 & a \\
J003545.13+011441.2 & 1.541 & 1.550 & 1.54 & 0.39$\pm$0.04 & 0.55$\pm$0.07 & 0.13$\pm$0.03 & a \\
J003550.66$-$004301.8 & 1.681 & 1.344 & 1.19 & $-$0.06$\pm$0.03 & 1.22$\pm$0.34 & 0.25$\pm$0.16 & a \\
J004324.47$-$103922.3 & 1.681 & 1.454 & 1.01 & $-$0.10$\pm$0.03 & 0.97$\pm$0.62 & $<$0.11 & a \\
J011705.18+152931.8 & 1.845 & 1.518 & 1.18 & $-$0.12$\pm$0.02 & 2.00$\pm$0.53 & 0.48$\pm$0.32 & a \\
\enddata
\tablenotetext{*}{the expected typical error of $W_r$ is of 0.3 {\AA}}
\tablenotetext{a}{at a statistical confidence level of $< 3\sigma$}
\tablenotetext{b}{broad absorption line quasar}
\tablenotetext{c}{the quasar spectrum was trimmed a lot}
\tablecomments{The emission redshifts of quasars are extracted from Schneider et al. (2010).
The absorption redshifts are extracted from Prochter et al. (2006).
Table 2 is available in its entirety in the electronic edition of the Astrophysical Journal.}
\end{deluxetable}

\begin{deluxetable}{lccccrrccccc}
\tabletypesize{\scriptsize}
\rotate
\tablecaption{12 High Confidence 2175-{\AA} Absorber Candidates\label{tbl-3}}
\tablewidth{0pt}
\tablehead{
\colhead{} & \colhead{} & \colhead{} & \colhead{$W_r^b$} & \colhead{$\Delta$(g-i)} & \colhead{$c_1$} & \colhead{$c_2$} &
\colhead{$c_3$} & \colhead{$x_0$} & \colhead{$\gamma$} & \colhead{} & \colhead{} \\
\colhead{SDSS} & \colhead{Z$_{emi}^a$} & \colhead{Z$_{abs}^b$} & \colhead{(\AA)} & \colhead{(mag)} & \colhead{(mag)} & \colhead{(mag)} &
\colhead{(mag)} & \colhead{($\mu$m$^{-1}$)} & \colhead{($\mu$m$^{-1}$)} & \colhead{$\chi_{\nu}^2$} & \colhead{Significance}
}
\startdata
J074744.16+330941.0 & 1.916 & 1.549 & 5.48 & 0.88$\pm$0.03 & 0.14$\pm$0.03 &  0.40$\pm$0.01 & 0.26$\pm$0.05 & 4.61$\pm$0.01 & 0.61$\pm$0.08 & 1.42 & 6.0$\sigma$ \\
J085042.24+515911.6 & 1.893 & 1.327 & 4.12 & 0.99$\pm$0.03 & $-$0.95$\pm$0.02 &  0.45$\pm$0.01 & 0.62$\pm$0.08 & 4.52$\pm$0.01 & 1.22$\pm$0.06 & 1.14 & 5.0$\sigma$ \\
J090016.67+021445.8 & 1.992 & 1.051 & 4.06 & 1.04$\pm$0.04 & 0.09$\pm$0.08 &  0.47$\pm$0.03 & 1.46$\pm$0.44 & 4.66$\pm$0.04 & 1.11$\pm$0.16 & 1.16 & 9.2$\sigma$ \\
J095631.05+404628.2 & 1.510 & 1.323 & 3.79 & 0.86$\pm$0.05 & 0.56$\pm$0.04 &  0.35$\pm$0.01 & 0.24$\pm$0.07 & 4.60$\pm$0.03 & 0.64$\pm$0.09 & 1.28 & 8.4$\sigma$ \\
J101751.15+474940.0 & 1.220 & 1.118 & 2.75 & 0.32$\pm$0.03 & 0.16$\pm$0.03 & $-$0.02$\pm$0.02 & 1.65$\pm$0.42 & 4.55$\pm$0.03 & 1.85$\pm$0.16 & 1.11 & 5.2$\sigma$ \\
J103718.77+014430.7 & 1.482 & 1.126 & 1.72 & 0.06$\pm$0.02 & 1.04$\pm$0.03 & $-$0.10$\pm$0.01 & 1.18$\pm$0.17 & 4.49$\pm$0.01 & 1.26$\pm$0.07 & 1.29 & 6.3$\sigma$ \\
J105748.63+610910.8 & 1.276 & 1.203 & 3.60 & 0.78$\pm$0.04 & 0.51$\pm$0.04 &  0.27$\pm$0.02 & 0.20$\pm$0.07 & 4.54$\pm$0.03 & 0.78$\pm$0.13 & 1.32 & 6.1$\sigma$ \\
J120935.80+671715.7 & 2.030 & 1.843 & 3.19 & 0.01$\pm$0.03 & 0.69$\pm$0.02 & 0.13$\pm$0.01 & 0.26$\pm$0.03 & 4.44$\pm$0.01 & 0.86$\pm$0.05 & 1.51 & 6.7$\sigma$ \\
J143108.98+000725.1 & 1.842 & 1.153 & 1.79 & 0.74$\pm$0.03 & $-$0.25$\pm$0.03 &  0.36$\pm$0.01 & 0.51$\pm$0.01 & 4.65$\pm$0.02 & 1.01$\pm$0.09 & 1.10 & 8.1$\sigma$ \\
J144612.97+035154.4 & 1.945 & 1.511 & 2.89 & 0.89$\pm$0.03 & $-$0.02$\pm$0.04 &  0.57$\pm$0.01 & 0.68$\pm$0.12 & 4.52$\pm$0.02 & 1.00$\pm$0.09 & 1.22 & 6.7$\sigma$ \\
J153020.05+592217.0 & 1.689 & 1.405 & 2.68 & 0.09$\pm$0.04 & 2.05$\pm$0.06 & 0.10$\pm$0.02 & 0.49$\pm$0.16 & 4.79$\pm$0.03 & 0.87$\pm$0.16 & 1.10 & 6.1$\sigma$ \\
J214811.57$-$085904.6 & 1.642 & 1.643 & 2.79 & 0.76$\pm$0.05 & $-$0.12$\pm$0.03 &  0.46$\pm$0.01 & 0.68$\pm$0.22 & 4.56$\pm$0.04 & 1.46$\pm$0.21 & 1.40 & 5.0$\sigma$ \\
\enddata
\tablenotetext{a}{We adopt the emission redshifts of quasars measured by Schneider et al. (2010).}
\tablenotetext{b}{We adopt the absorption redshifts and $W_r$ measured by Prochter et al. (2006). The expected typical error of $W_r$ is of 0.3 {\AA}.}
\tablecomments{The best-fitted parameters of optical/UV extinction curve in the rest frame of the high confidence 2175-{\AA} absorber candidates.}
\end{deluxetable}

\begin{deluxetable}{lccccrrccccc}
\tabletypesize{\scriptsize}
\rotate
\tablecaption{10 Median Confidence 2175-{\AA} Absorber Candidates\label{tbl-4}}
\tablewidth{0pt}
\tablehead{
\colhead{} & \colhead{} & \colhead{} & \colhead{$W_r^b$} & \colhead{$\Delta$(g-i)} & \colhead{$c_1$} & \colhead{$c_2$} &
\colhead{$c_3$} & \colhead{$x_0$} & \colhead{$\gamma$} & \colhead{} & \colhead{} \\
\colhead{SDSS} & \colhead{Z$_{emi}^a$} & \colhead{Z$_{abs}^b$} & \colhead{(\AA)} & \colhead{(mag)} & \colhead{(mag)} & \colhead{(mag)} &
\colhead{(mag)} & \colhead{($\mu$m$^{-1}$)} & \colhead{($\mu$m$^{-1}$)} & \colhead{$\chi_{\nu}^2$} & \colhead{Significance}
}
\startdata
J081540.60+264021.6 & 1.936 & 1.680 & 2.88 & 0.38$\pm$0.03 & 0.72$\pm$0.02 &  0.16$\pm$0.01 & 0.15$\pm$0.03 & 4.53$\pm$0.02 & 0.72$\pm$0.07 & 1.24 & 4.3$\sigma$ \\
J091927.61+014603.0 & 1.277 & 1.274 & 2.18 & 0.76$\pm$0.04 & 0.05$\pm$0.03 &  0.43$\pm$0.01 & 0.07$\pm$0.03 & 4.80$\pm$0.03 & 0.55$\pm$0.11 & 1.33 & 4.8$\sigma$ \\
J102832.58+042354.1 & 1.725 & 1.461 & 1.70 & 0.09$\pm$0.03 & 1.28$\pm$0.02 & $-$0.02$\pm$0.01 & 0.71$\pm$0.14 & 4.72$\pm$0.02 & 1.25$\pm$0.11 & 1.10 & 4.3$\sigma$ \\
J105049.73+071554.8 & 1.913 & 1.262 & 2.33 & 0.54$\pm$0.03 & 0.40$\pm$0.02 &  0.17$\pm$0.01 & 0.18$\pm$0.05 & 4.69$\pm$0.02 & 0.85$\pm$0.11 & 1.13 & 4.5$\sigma$ \\
J110747.04+631607.1 & 1.952 & 1.250 & 3.87 & 0.40$\pm$0.02 & 0.10$\pm$0.01 &  0.07$\pm$0.01 & 0.25$\pm$0.05 & 4.60$\pm$0.02 & 0.99$\pm$0.08 & 1.08 & 4.3$\sigma$ \\
J111857.03+484750.1 & 1.993 & 1.639 & 1.17 & 0.20$\pm$0.04 & 0.88$\pm$0.02 &  0.13$\pm$0.01 & 0.58$\pm$0.10 & 4.71$\pm$0.02 & 1.18$\pm$0.10 & 1.28 & 4.3$\sigma$ \\
J124715.26+520801.0 & 1.812 & 1.050 & 3.58 & 0.59$\pm$0.05 & 0.57$\pm$0.04 &  0.19$\pm$0.01 & 0.19$\pm$0.06 & 4.57$\pm$0.02 & 0.68$\pm$0.08 & 1.13 & 4.5$\sigma$ \\
J133125.93+004414.0 & 2.021 & 1.310 & 1.77 & 0.17$\pm$0.02 & 0.50$\pm$0.01 &  0.07$\pm$0.01 & 0.05$\pm$0.01 & 4.64$\pm$0.01 & 0.50$\pm$0.04 & 1.25 & 4.2$\sigma$ \\
J145953.23+012944.2 & 1.659 & 1.623 & 2.00 & 0.53$\pm$0.03 & 0.33$\pm$0.02 &  0.21$\pm$0.01 & 0.13$\pm$0.03 & 4.68$\pm$0.02 & 0.73$\pm$0.08 & 1.25 & 4.0$\sigma$ \\
J233131.90$-$001940.1 & 1.845 & 1.391 & 1.46 & 0.71$\pm$0.04 & 0.29$\pm$0.03 &  0.26$\pm$0.01 & 0.43$\pm$0.08 & 4.43$\pm$0.02 & 1.00$\pm$0.10 & 1.26 & 4.1$\sigma$ \\
\enddata
\tablenotetext{a}{We adopt the emission redshifts of quasars measured by Schneider et al. (2010).}
\tablenotetext{b}{We adopt the absorption redshifts and $W_r$ measured by Prochter et al. (2006). The expected typical error of $W_r$ is of 0.3 {\AA}.}
\tablecomments{The best-fitted parameters of optical/UV extinction curve in the rest frame of the median confidence 2175-{\AA} absorber candidates.}
\end{deluxetable}

\begin{deluxetable}{lccccrrccccc}
\tabletypesize{\scriptsize}
\rotate
\tablecaption{17 Low Confidence 2175-{\AA} Absorber Candidates\label{tbl-5}}
\tablewidth{0pt}
\tablehead{
\colhead{} & \colhead{} & \colhead{} & \colhead{$W_r^b$} & \colhead{$\Delta$(g-i)} & \colhead{$c_1$} & \colhead{$c_2$} &
\colhead{$c_3$} & \colhead{$x_0$} & \colhead{$\gamma$} & \colhead{} & \colhead{} \\
\colhead{SDSS} & \colhead{Z$_{emi}^a$} & \colhead{Z$_{abs}^b$} & \colhead{(\AA)} & \colhead{(mag)} & \colhead{(mag)} & \colhead{(mag)} &
\colhead{(mag)} & \colhead{($\mu$m$^{-1}$)} & \colhead{($\mu$m$^{-1}$)} & \colhead{$\chi_{\nu}^2$} & \colhead{Significance}
}
\startdata
J074259.57+225003.2 & 1.182 & 1.181 & 2.19 & 0.60$\pm$0.02 & $-$0.56$\pm$0.03 &  0.29$\pm$0.01 & 0.34$\pm$0.12 & 4.68$\pm$0.04 & 1.22$\pm$0.17 & 1.17 & 3.3$\sigma$ \\
J092201.86+494010.6 & 1.488 & 1.068 & 1.04 & 0.39$\pm$0.03 & 0.28$\pm$0.04 &  0.10$\pm$0.02 & 0.93$\pm$0.30 & 4.70$\pm$0.03 & 1.50$\pm$0.18 & 1.83 & 3.9$\sigma$ \\
J092923.67+443343.2 & 1.296 & 1.072 & 1.63 & 0.23$\pm$0.03 & 0.30$\pm$0.04 &  0.11$\pm$0.02 & 0.38$\pm$0.17 & 4.55$\pm$0.04 & 1.16$\pm$0.19 & 1.00 & 3.4$\sigma$ \\
J094933.34+020657.0 & 1.531 & 1.215 & 4.41 & 0.03$\pm$0.03 & 2.13$\pm$0.04 & $-$0.12$\pm$0.02 & 0.51$\pm$0.18 & 4.53$\pm$0.03 & 1.22$\pm$0.18 & 1.60 & 3.1$\sigma$ \\
J102725.91+060505.8 & 1.219 & 1.002 & 1.71 & 0.16$\pm$0.03 & 0.77$\pm$0.11 & $-$0.10$\pm$0.05 & 2.66$\pm$1.67 & 4.72$\pm$0.09 & 2.19$\pm$0.41 & 1.08 & 3.0$\sigma$ \\
J104934.08+022118.9 & 1.831 & 1.594 & 2.60 & 0.34$\pm$0.03 & 0.85$\pm$0.02 &  0.12$\pm$0.01 & 0.40$\pm$0.08 & 4.61$\pm$0.02 & 1.11$\pm$0.10 & 1.06 & 3.8$\sigma$ \\
J105122.23+032300.6 & 2.015 & 1.527 & 2.62 & 0.46$\pm$0.03 & 0.45$\pm$0.02 &  0.30$\pm$0.01 & 0.22$\pm$0.05 & 4.43$\pm$0.02 & 0.74$\pm$0.09 & 1.10 & 3.9$\sigma$ \\
J113448.13+593945.7 & 1.345 & 1.335 & 3.56 & 0.37$\pm$0.03 & 0.48$\pm$0.01 &  0.13$\pm$0.01 & 0.08$\pm$0.02 & 4.55$\pm$0.02 & 0.64$\pm$0.07 & 1.33 & 3.8$\sigma$ \\
J121312.08+542748.0 & 1.703 & 1.070 & 1.44 & 0.21$\pm$0.03 & 1.05$\pm$0.04 &  0.06$\pm$0.01 & 0.11$\pm$0.07 & 4.54$\pm$0.04 & 0.81$\pm$0.22 & 1.09 & 3.7$\sigma$ \\
J135731.06+022726.5 & 1.773 & 1.368 & 2.31 & 0.10$\pm$0.04 & 1.37$\pm$0.03 & 0.17$\pm$0.01 & 0.16$\pm$0.04 & 4.40$\pm$0.03 & 0.65$\pm$0.10 & 1.19 & 3.9$\sigma$ \\
J142820.59$-$005348.3 & 1.533 & 1.517 & 2.04 & 0.54$\pm$0.03 & $-$0.01$\pm$0.02 &  0.27$\pm$0.01 & 0.17$\pm$0.04 & 4.51$\pm$0.03 & 0.96$\pm$0.12 & 1.24 & 3.2$\sigma$ \\
J143512.94+042036.9 & 1.944 & 1.657 & 3.41 & 0.22$\pm$0.03 & 0.24$\pm$0.01 &  0.13$\pm$0.01 & 0.17$\pm$0.02 & 4.58$\pm$0.02 & 0.83$\pm$0.06 & 1.23 & 3.0$\sigma$ \\
J144046.93+012041.0 & 1.397 & 1.030 & 1.58 & 0.10$\pm$0.02 & 0.87$\pm$0.06 & $-$0.11$\pm$0.03 & 1.42$\pm$0.49 & 4.59$\pm$0.04 & 1.67$\pm$0.20 & 1.35 & 3.5$\sigma$ \\
J154734.12+033314.4 & 1.369 & 1.007 & 1.19 & 0.34$\pm$0.03 & 0.55$\pm$0.06 &  0.07$\pm$0.03 & 1.04$\pm$0.42 & 4.48$\pm$0.04 & 1.48$\pm$0.22 & 1.19 & 3.6$\sigma$ \\
J203926.15+005444.9 & 1.256 & 1.118 & 1.49 & 0.41$\pm$0.03 & 0.24$\pm$0.04 &  0.24$\pm$0.01 & 0.06$\pm$0.05 & 4.70$\pm$0.08 & 0.61$\pm$0.26 & 1.07 & 3.1$\sigma$ \\
J214324.36+003502.8 & 2.037 & 1.604 & 1.99 & 0.14$\pm$0.02 & 0.48$\pm$0.01 & 0.17$\pm$0.01 & 0.23$\pm$0.03 & 4.43$\pm$0.01 & 0.92$\pm$0.07 & 1.33 & 3.0$\sigma$ \\
J232713.03$-$100027.7 & 1.981 & 1.266 & 1.67 & 0.22$\pm$0.03 & 0.65$\pm$0.02 & 0.14$\pm$0.01 & 0.06$\pm$0.02 & 4.50$\pm$0.02 & 0.62$\pm$0.10 & 1.30 & 3.0$\sigma$ \\
\enddata
\tablenotetext{a}{We adopt the emission redshifts of quasars measured by Schneider et al. (2010).}
\tablenotetext{b}{We adopt the absorption redshifts and $W_r$ measured by Prochter et al. (2006). The expected typical error of $W_r$ is of 0.3 {\AA}.}
\tablecomments{The best-fitted parameters of optical/UV extinction curve in the rest frame of the low confidence 2175-{\AA} absorber candidates.}
\end{deluxetable}

\begin{deluxetable}{lcccccccc}
\tabletypesize{\scriptsize}
\rotate
\tablecaption{Sensitivity Simulations\label{tbl-6}}
\tablewidth{0pt}
\tablehead{
\colhead{x$_0^{qso}$} & \multicolumn{4}{c}{3 $\sigma$ threshold} & \multicolumn{4}{c}{5 $\sigma$ threshold} \\
\cline{2-5} \cline{6-9}\\
\colhead{({\AA})} & \colhead{$\gamma=$0.9$\mu$m$^{-1}$} & \colhead{$\gamma=$1.2$\mu$m$^{-1}$}
& \colhead{$\gamma=$1.5$\mu$m$^{-1}$} & \colhead{$\gamma=$2.2$\mu$m$^{-1}$}
& \colhead{$\gamma=$0.9$\mu$m$^{-1}$} & \colhead{$\gamma=$1.2$\mu$m$^{-1}$}
& \colhead{$\gamma=$1.5$\mu$m$^{-1}$} & \colhead{$\gamma=$2.2$\mu$m$^{-1}$}
}
\startdata
2180 & 0.25 & 0.37 & 0.53 & 1.01 & 0.41 & 0.60 & 0.85 & 1.62 \\
2130 & 0.27 & 0.40 & 0.56 & 1.06 & 0.44 & 0.64 & 0.90 & 1.70 \\
2080 & 0.29 & 0.43 & 0.60 & 1.13 & 0.47 & 0.70 & 0.97 & 1.81 \\
2030 & 0.31 & 0.47 & 0.66 & 1.22 & 0.51 & 0.77 & 1.10 & 1.96 \\
1980 & 0.35 & 0.53 & 0.73 & 1.35 & 0.58 & 0.86 & 1.20 & 2.17 \\
1930 & 0.41 & 0.60 & 0.83 & 1.49 & 0.69 & 0.98 & 1.35 & 2.41 \\
1880 & 0.43 & 0.64 & 0.89 & 1.61 & 0.72 & 1.06 & 1.47 & 2.62 \\
1830 & 0.36 & 0.59 & 0.87 & 1.68 & 0.61 & 0.99 & 1.45 & 2.75 \\
1780 & 0.29 & 0.50 & 0.78 & 1.68 & 0.49 & 0.83 & 1.29 & 2.76 \\
1730 & 0.25 & 0.41 & 0.66 & 1.59 & 0.41 & 0.68 & 1.08 & 2.61 \\
1680 & 0.22 & 0.35 & 0.54 & 1.41 & 0.37 & 0.57 & 0.89 & 2.30 \\
1630 & 0.28 & 0.39 & 0.55 & 1.26 & 0.47 & 0.65 & 0.89 & 2.04 \\
1580 & 0.40 & 0.55 & 0.71 & 1.28 & 0.67 & 0.91 & 1.16 & 2.06 \\
1530 & 0.55 & 0.73 & 0.91 & 1.51 & 0.91 & 1.20 & 1.48 & 2.43 \\
\enddata
\tablecomments{The results of sensitivity simulations are listed in the grid of $x_0^{qso}$ and $\gamma$.}
\end{deluxetable}
\clearpage
\begin{figure}\epsscale{1.0}
\plotone{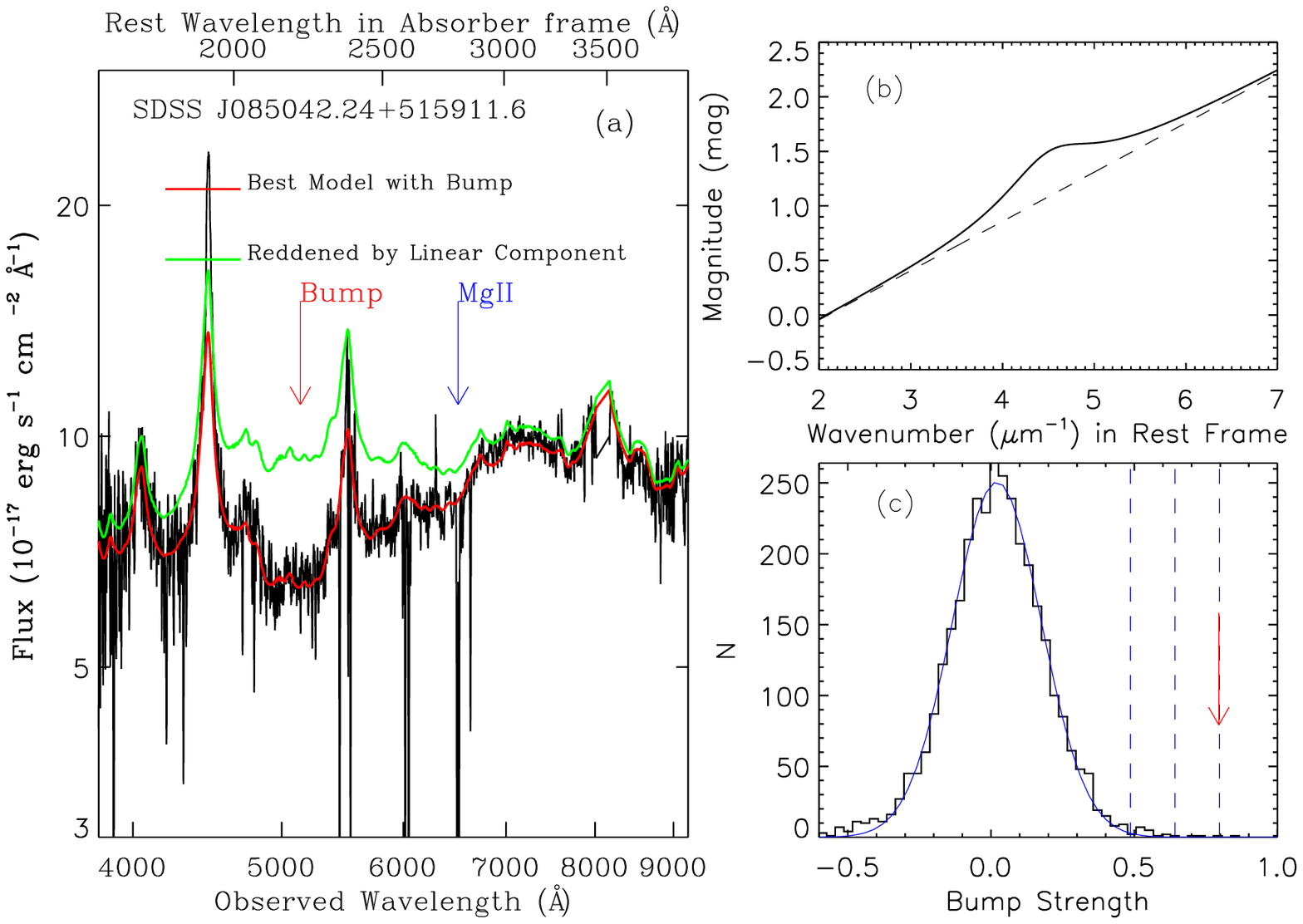}
\caption{(ONLINE ONLY)
The best fitted extinction model for J0850+5159. (a). Red solid line is the best fitted model.
Green solid line is reddened composite quasar spectrum
by using the linear component of best model only. (b). The best fitted
extinction curve. (c). Histogram of fitted bump strength of the control sample for J0850+5159.
\label{fig9}}
\end{figure}
\clearpage

\clearpage
\begin{figure}\epsscale{1.0}
\plotone{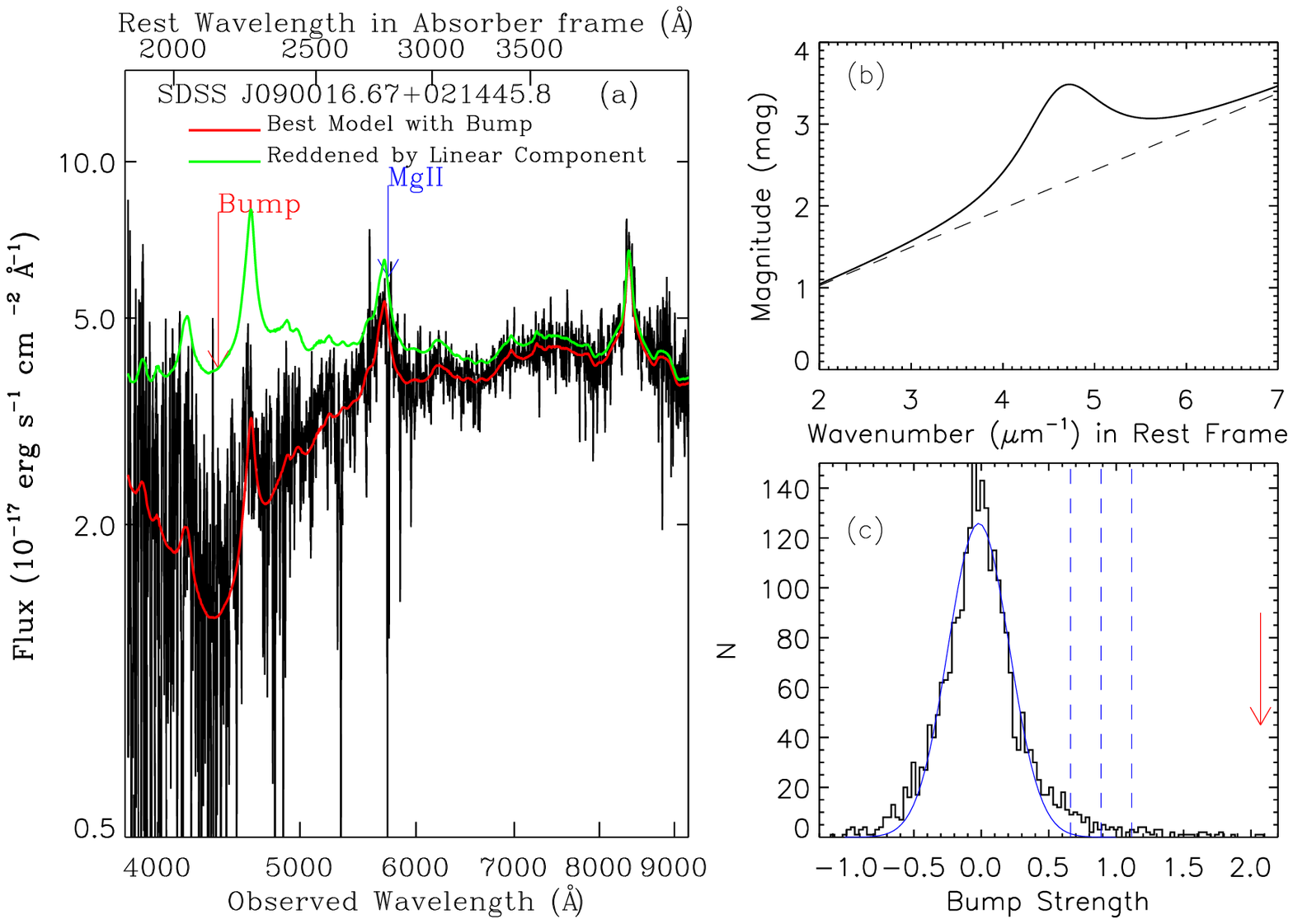}
\caption{(ONLINE ONLY)
The best fitted extinction model for J0900+0214. (a). Red solid line is the best fitted model.
Green solid line is reddened composite quasar spectrum
by using the linear component of best model only. (b). The best fitted
extinction curve. (c). Histogram of fitted bump strength of the control sample for J0900+0214.
\label{fig10}}
\end{figure}
\clearpage

\clearpage
\begin{figure}\epsscale{1.0}
\plotone{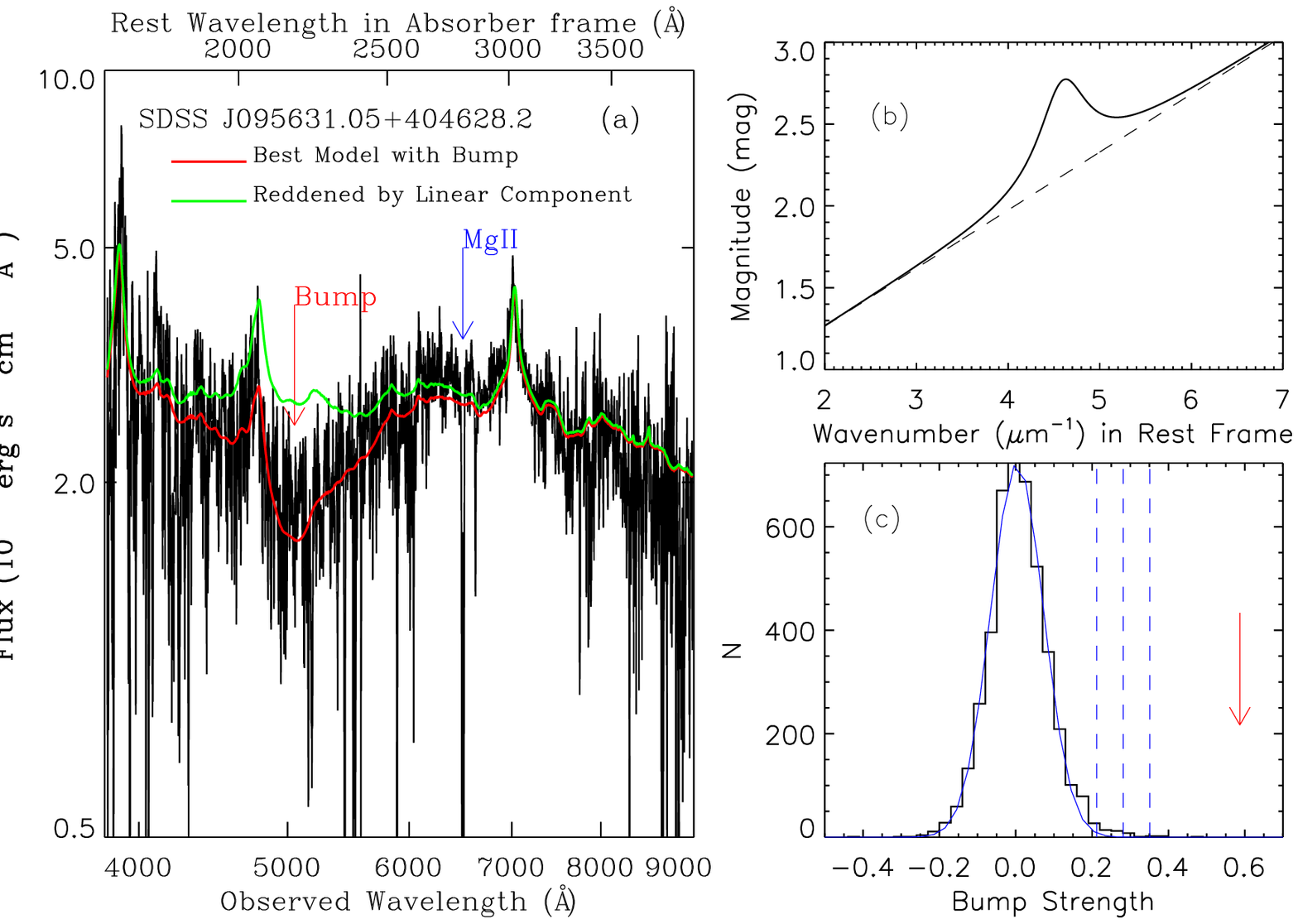}
\caption{(ONLINE ONLY)
The best fitted extinction model for J0956+4046. (a). Red solid line is the best fitted model.
Green solid line is reddened composite quasar spectrum
by using the linear component of best model only. (b). The best fitted
extinction curve. (c). Histogram of fitted bump strength of the control sample for J0956+4046.
\label{fig11}}
\end{figure}
\clearpage

\clearpage
\begin{figure}\epsscale{1.0}
\plotone{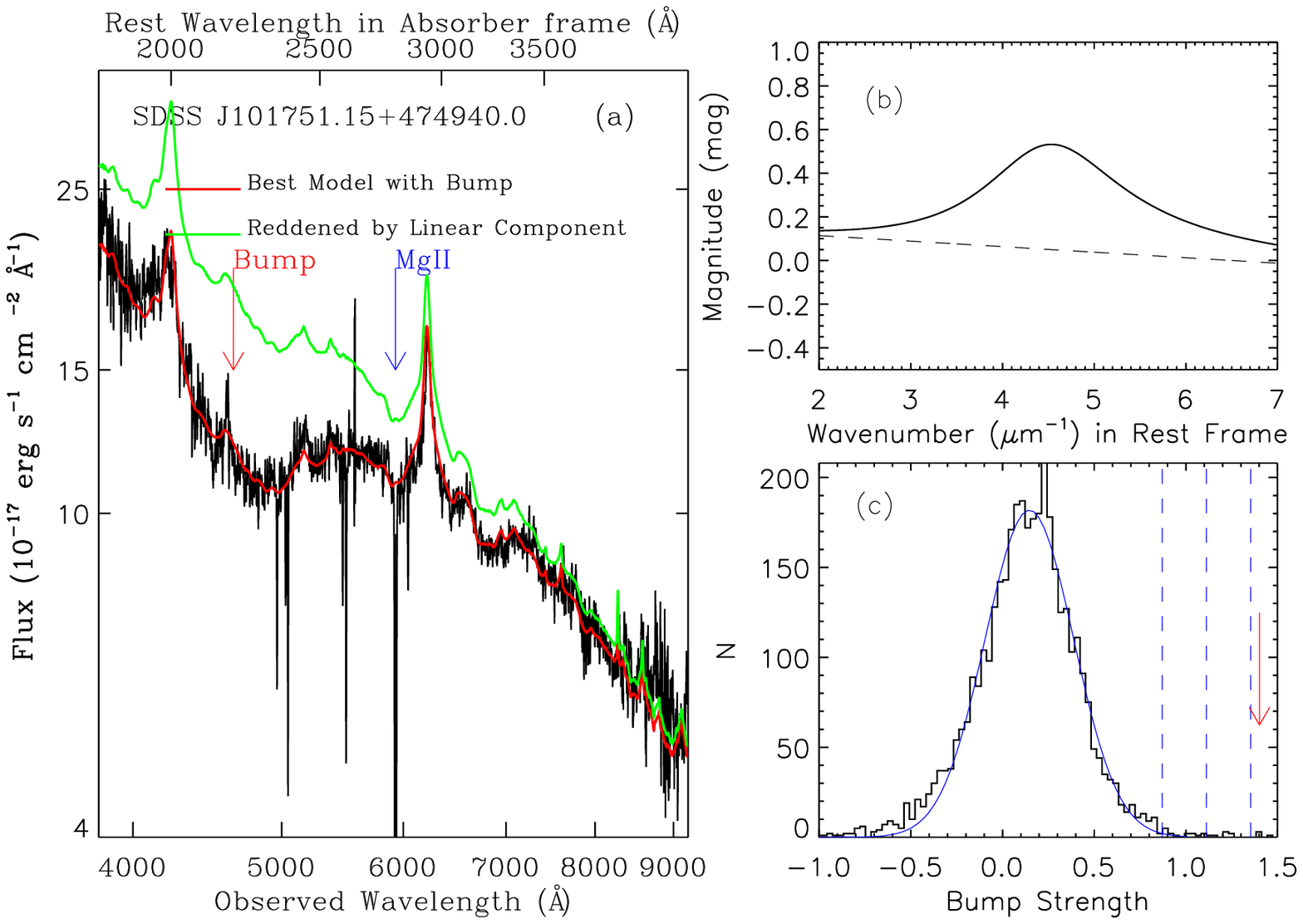}
\caption{(ONLINE ONLY)
The best fitted extinction model for J1017+4749. (a). Red solid line is the best fitted model.
Green solid line is reddened composite quasar spectrum
by using the linear component of best model only. (b). The best fitted
extinction curve. (c). Histogram of fitted bump strength of the control sample for J1017+4749.
\label{fig12}}
\end{figure}
\clearpage

\clearpage
\begin{figure}\epsscale{1.0}
\plotone{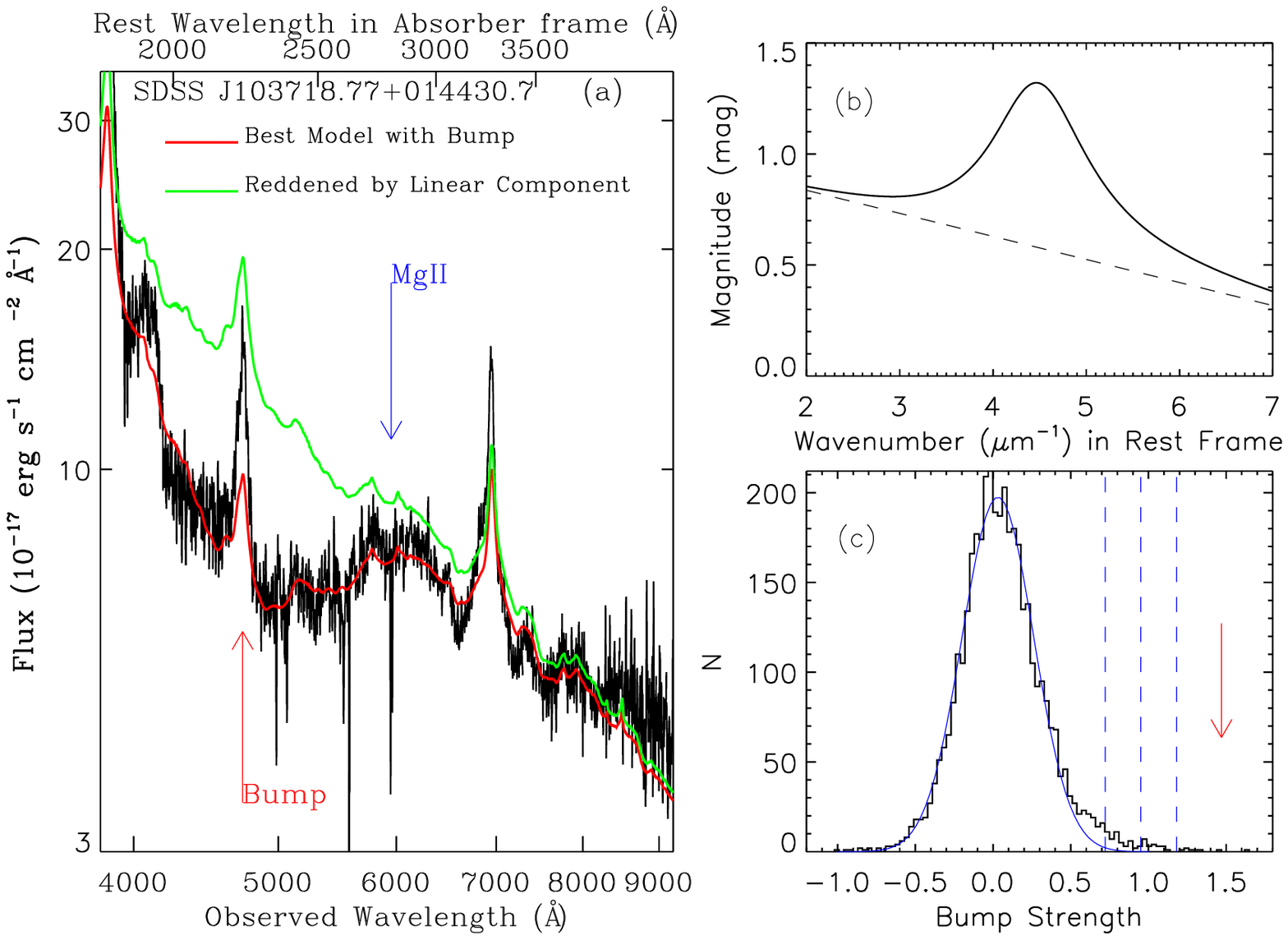}
\caption{(ONLINE ONLY)
The best fitted extinction model for J1037+0144. (a). Red solid line is the best fitted model.
Green solid line is reddened composite quasar spectrum
by using the linear component of best model only. (b). The best fitted
extinction curve. (c). Histogram of fitted bump strength of the control sample for J1037+0144.
\label{fig13}}
\end{figure}
\clearpage

\clearpage
\begin{figure}\epsscale{1.0}
\plotone{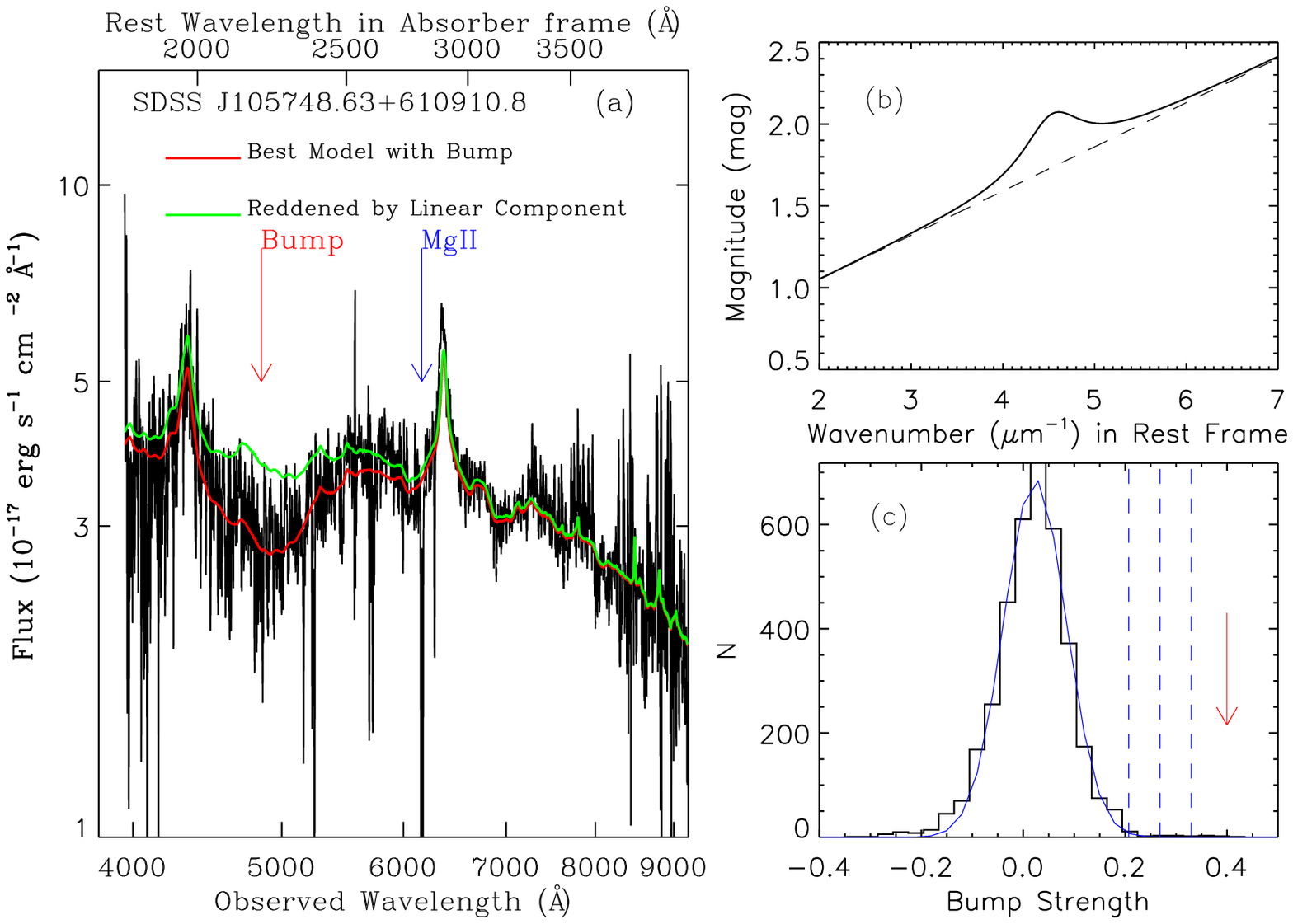}
\caption{(ONLINE ONLY)
The best fitted extinction model for J1057+6109. (a). Red solid line is the best fitted model.
Green solid line is reddened composite quasar spectrum
by using the linear component of best model only. (b). The best fitted
extinction curve. (c). Histogram of fitted bump strength of the control sample for J1057+6109.
\label{fig14}}
\end{figure}
\clearpage

\clearpage
\begin{figure}\epsscale{1.0}
\plotone{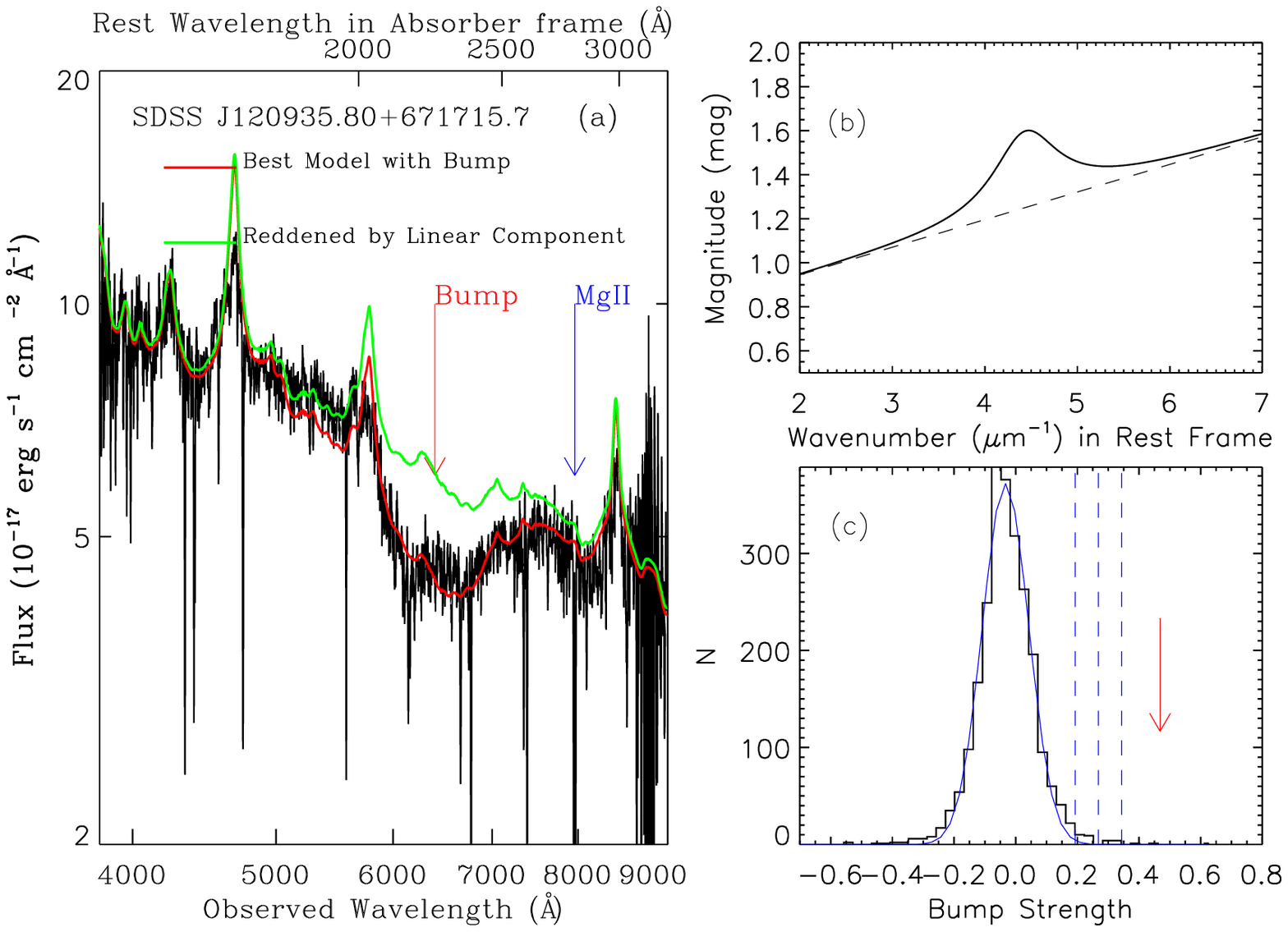}
\caption{(ONLINE ONLY)
The best fitted extinction model for J1209+6717. (a). Red solid line is the best fitted model.
Green solid line is reddened composite quasar spectrum
by using the linear component of best model only. (b). The best fitted
extinction curve. (c). Histogram of fitted bump strength of the control sample for J1209+6717.
\label{fig15}}
\end{figure}
\clearpage

\clearpage
\begin{figure}\epsscale{1.0}
\plotone{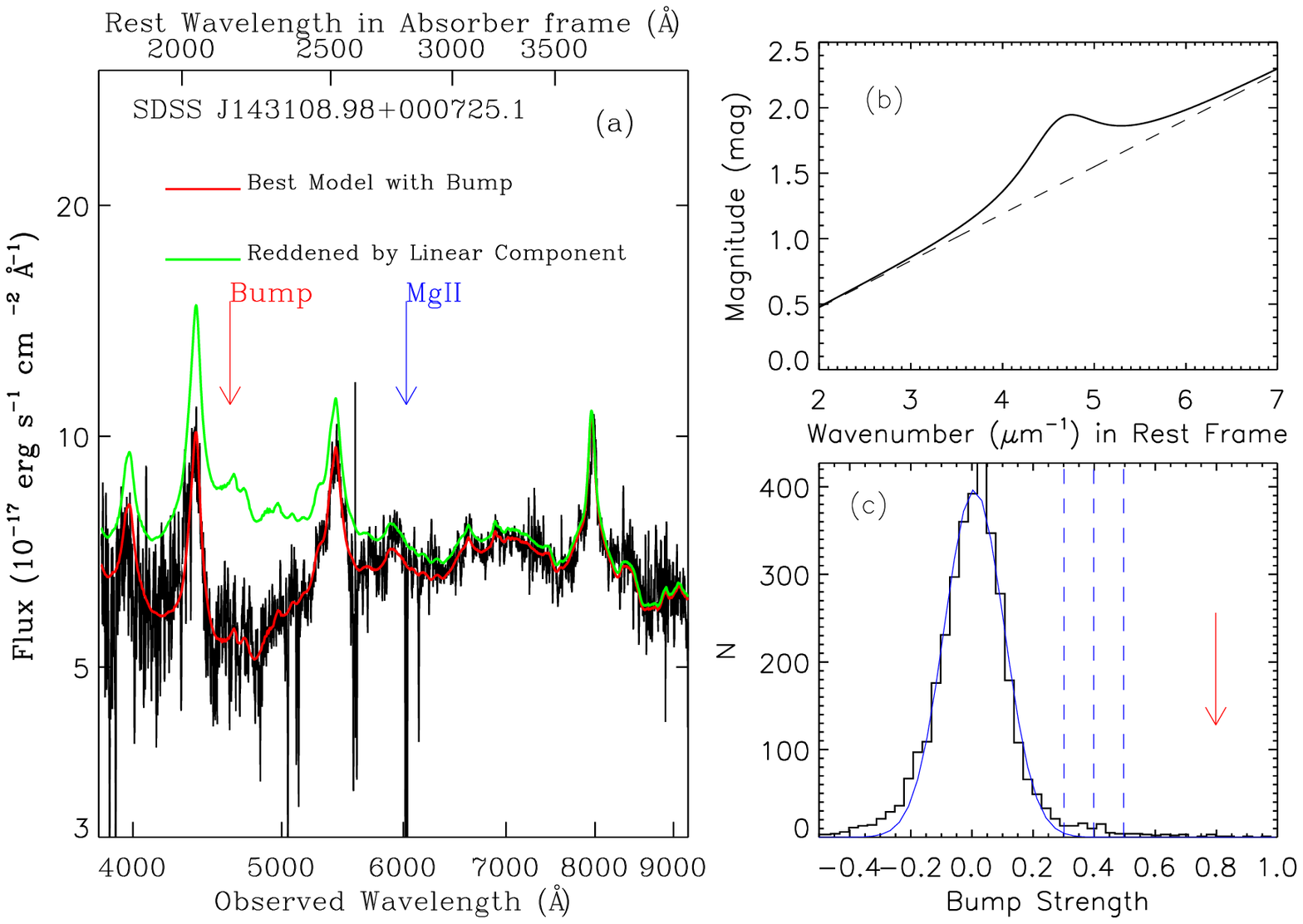}
\caption{(ONLINE ONLY)
The best fitted extinction model for J1431+0007. (a). Red solid line is the best fitted model.
Green solid line is reddened composite quasar spectrum
by using the linear component of best model only. (b). The best fitted
extinction curve. (c). Histogram of fitted bump strength of the control sample for J1431+0007.
\label{fig16}}
\end{figure}
\clearpage

\clearpage
\begin{figure}\epsscale{1.0}
\plotone{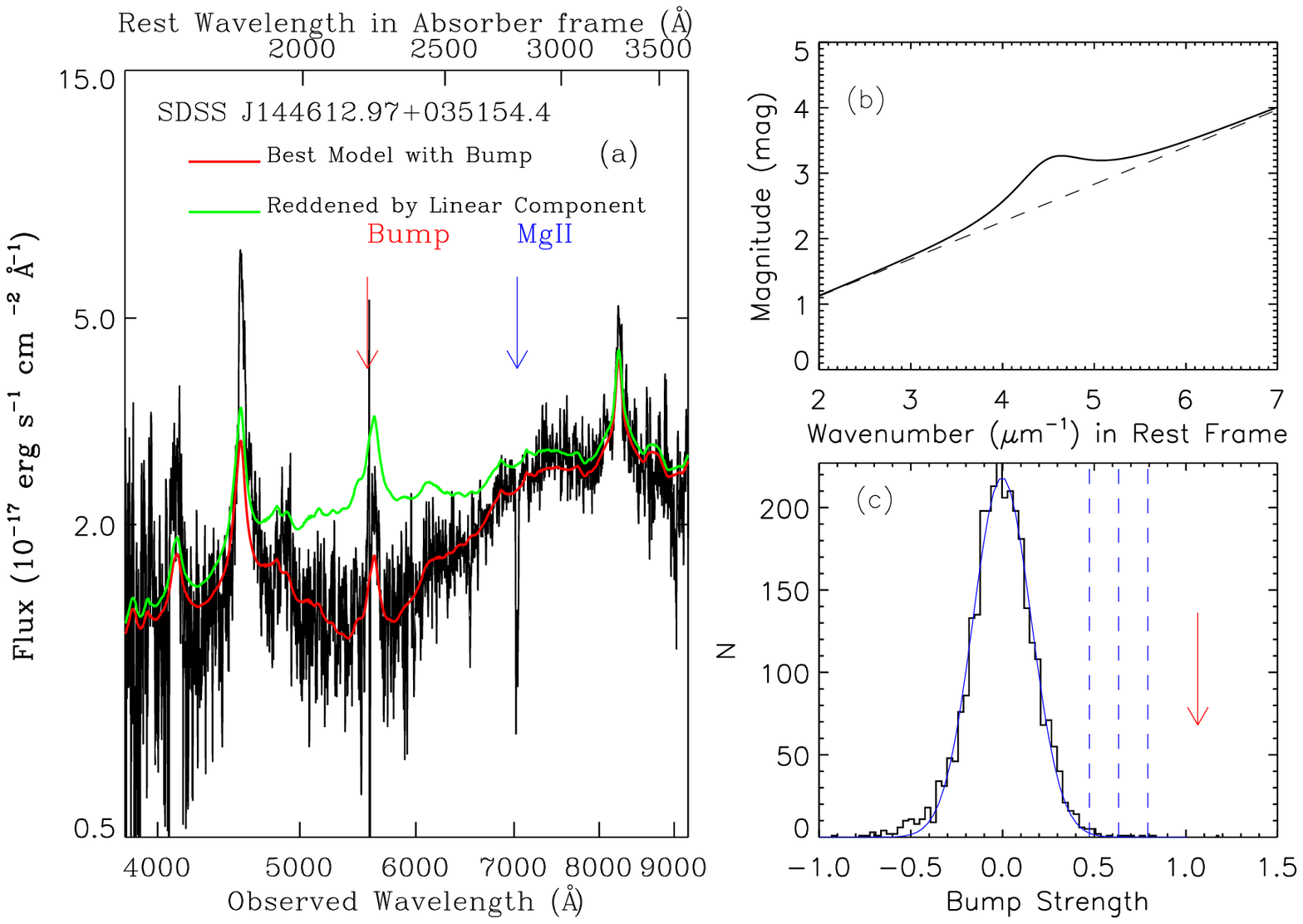}
\caption{(ONLINE ONLY)
The best fitted extinction model for J1446+0351. (a). Red solid line is the best fitted model.
Green solid line is reddened composite quasar spectrum
by using the linear component of best model only. (b). The best fitted
extinction curve. (c). Histogram of fitted bump strength of the control sample for J1446+0351.
\label{fig17}}
\end{figure}
\clearpage

\clearpage
\begin{figure}\epsscale{1.0}
\plotone{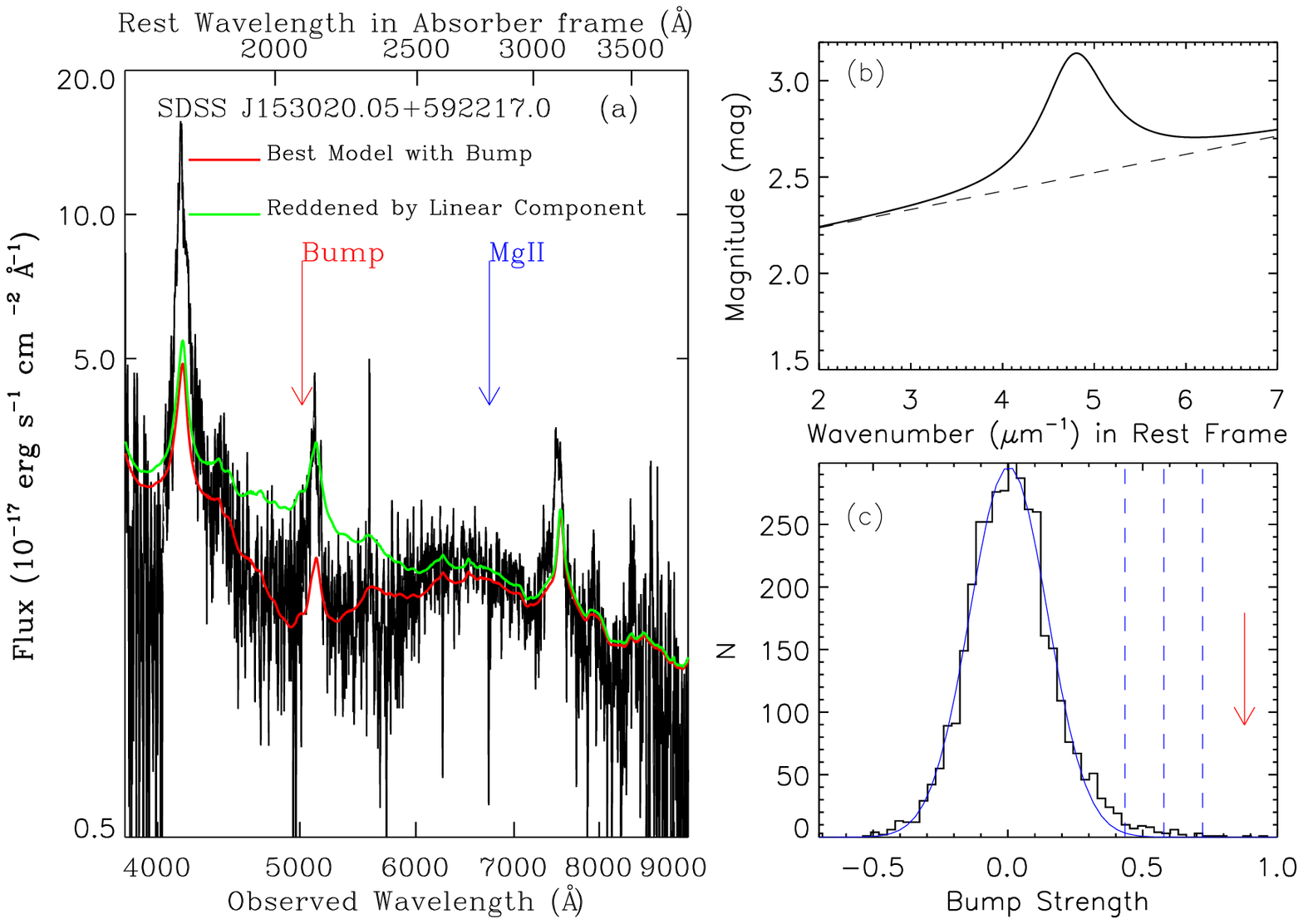}
\caption{(ONLINE ONLY)
The best fitted extinction model for J1530+5922. (a). Red solid line is the best fitted model.
Green solid line is reddened composite quasar spectrum
by using the linear component of best model only. (b). The best fitted
extinction curve. (c). Histogram of fitted bump strength of the control sample for J1530+5922.
\label{fig18}}
\end{figure}
\clearpage

\clearpage
\begin{figure}\epsscale{1.0}
\plotone{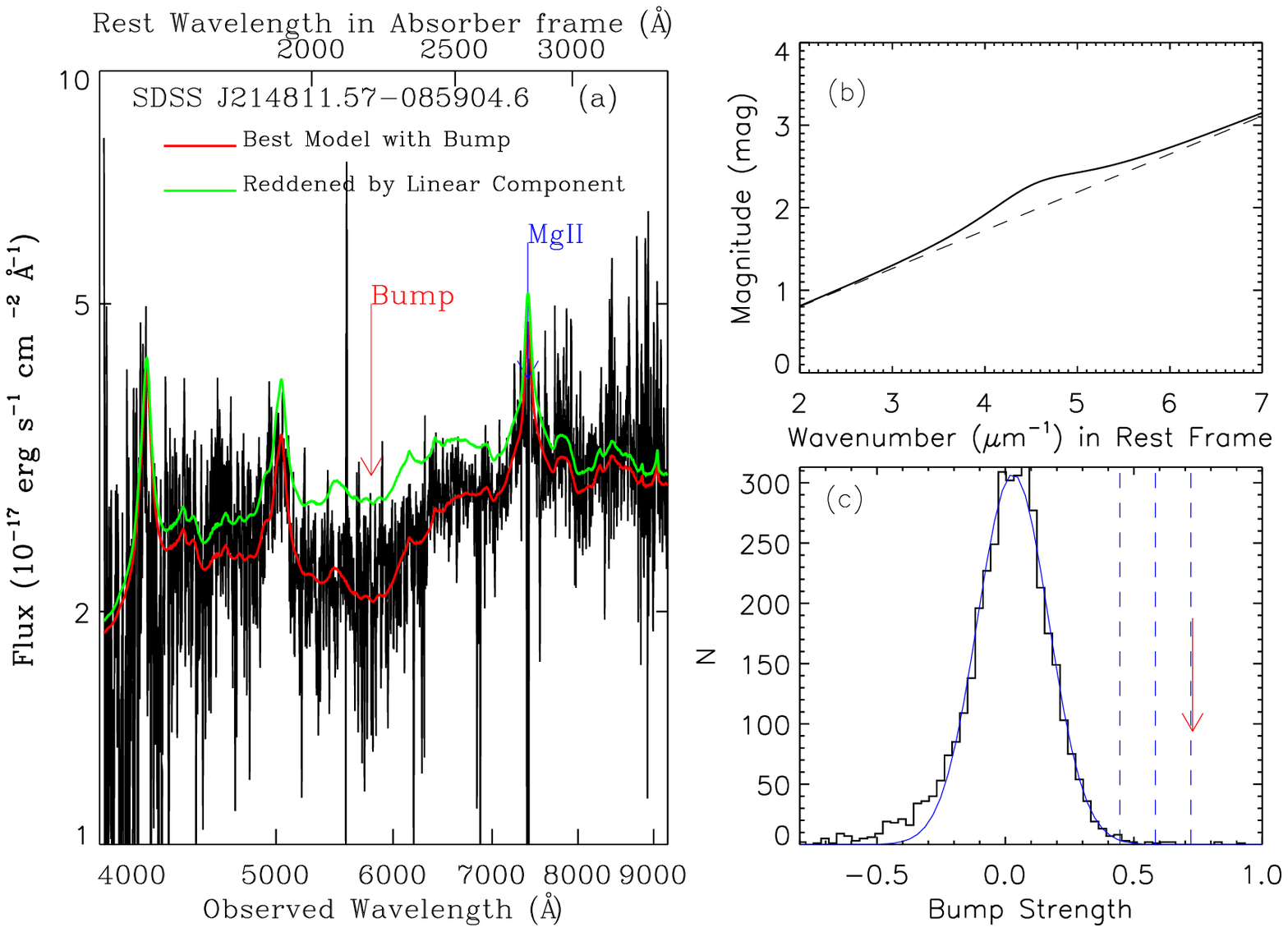}
\caption{(ONLINE ONLY)
The best fitted extinction model for J2148$-$0859. (a). Red solid line is the best fitted model.
Green solid line is reddened composite quasar spectrum
by using the linear component of best model only. (b). The best fitted
extinction curve. (c). Histogram of fitted bump strength of the control sample for J2148$-$0859.
\label{fig19}}
\end{figure}
\clearpage

\clearpage
\begin{figure}\epsscale{1.0}
\plotone{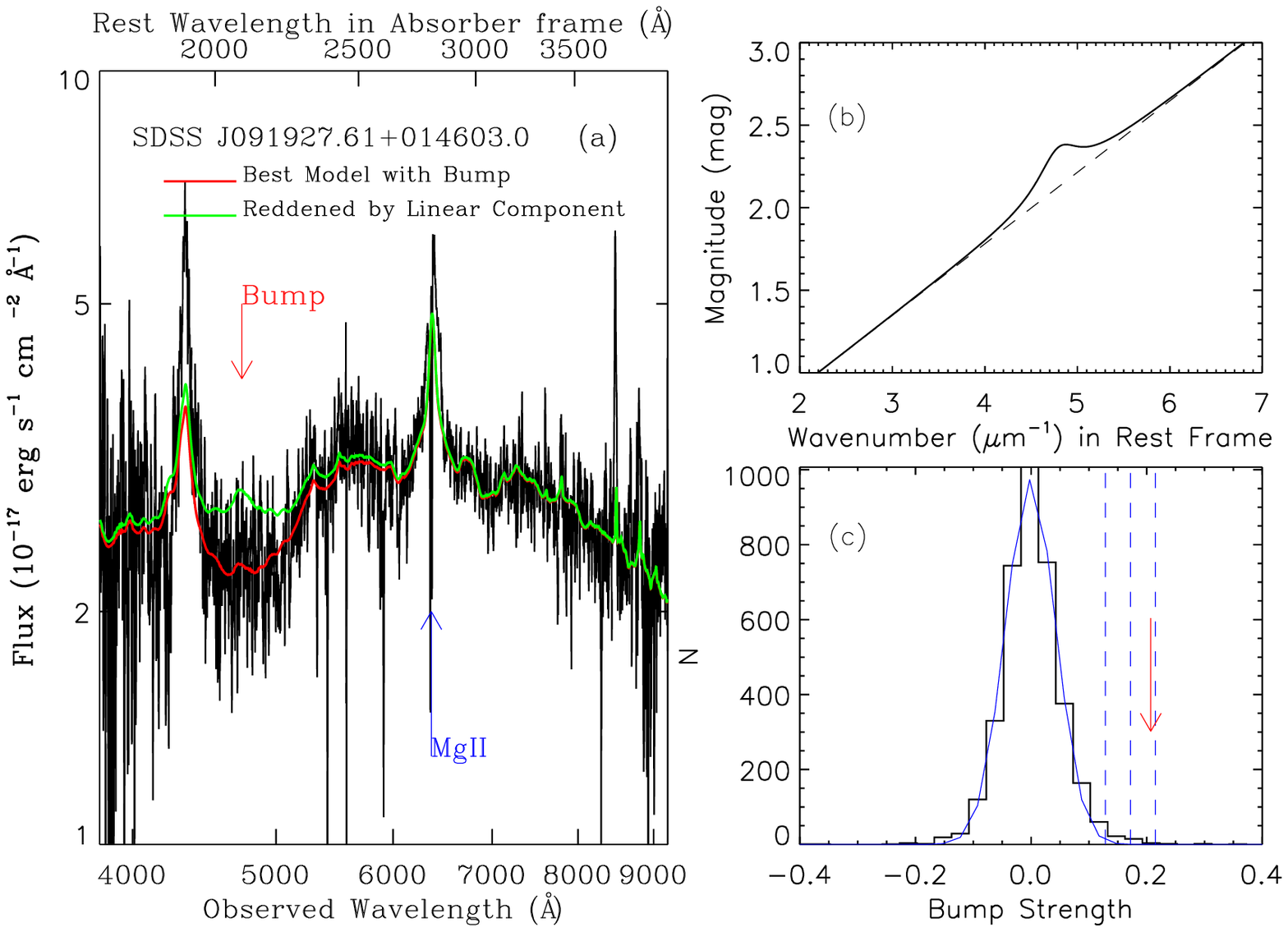}
\caption{(ONLINE ONLY)
The best fitted extinction model for J0919+0146. (a). Red solid line is the best fitted model.
Green solid line is reddened composite quasar spectrum
by using the linear component of best model only. (b). The best fitted
extinction curve. (c). Histogram of fitted bump strength of the control sample for J0919+0146.
\label{fig20}}
\end{figure}
\clearpage

\clearpage
\begin{figure}\epsscale{1.0}
\plotone{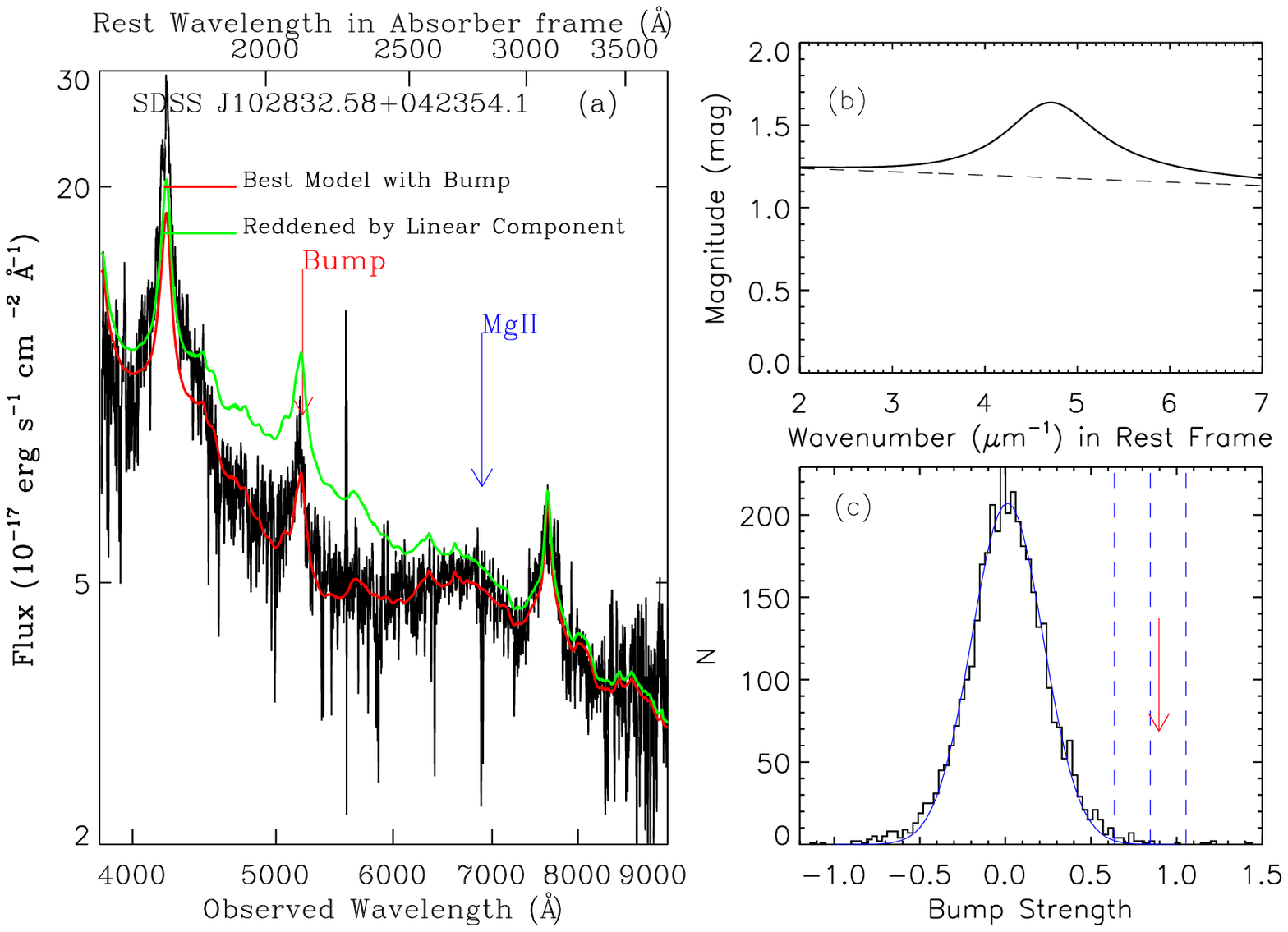}
\caption{(ONLINE ONLY)
The best fitted extinction model for J1028+0423. (a). Red solid line is the best fitted model.
Green solid line is reddened composite quasar spectrum
by using the linear component of best model only. (b). The best fitted
extinction curve. (c). Histogram of fitted bump strength of the control sample for J1028+0423.
\label{fig21}}
\end{figure}
\clearpage

\clearpage
\begin{figure}\epsscale{1.0}
\plotone{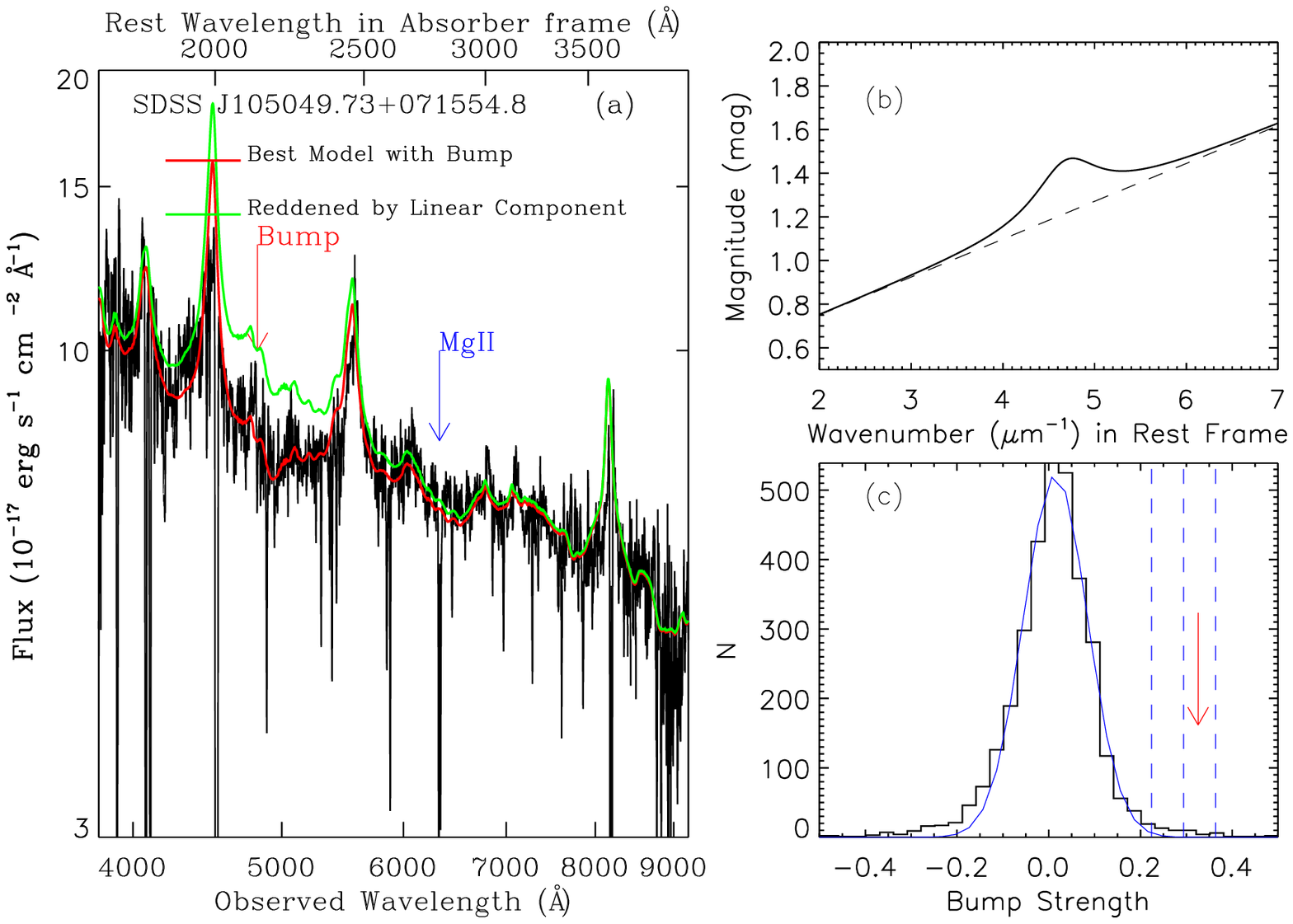}
\caption{(ONLINE ONLY)
The best fitted extinction model for J1050+0715. (a). Red solid line is the best fitted model.
Green solid line is reddened composite quasar spectrum
by using the linear component of best model only. (b). The best fitted
extinction curve. (c). Histogram of fitted bump strength of the control sample for J1050+0715.
\label{fig22}}
\end{figure}
\clearpage

\clearpage
\begin{figure}\epsscale{1.0}
\plotone{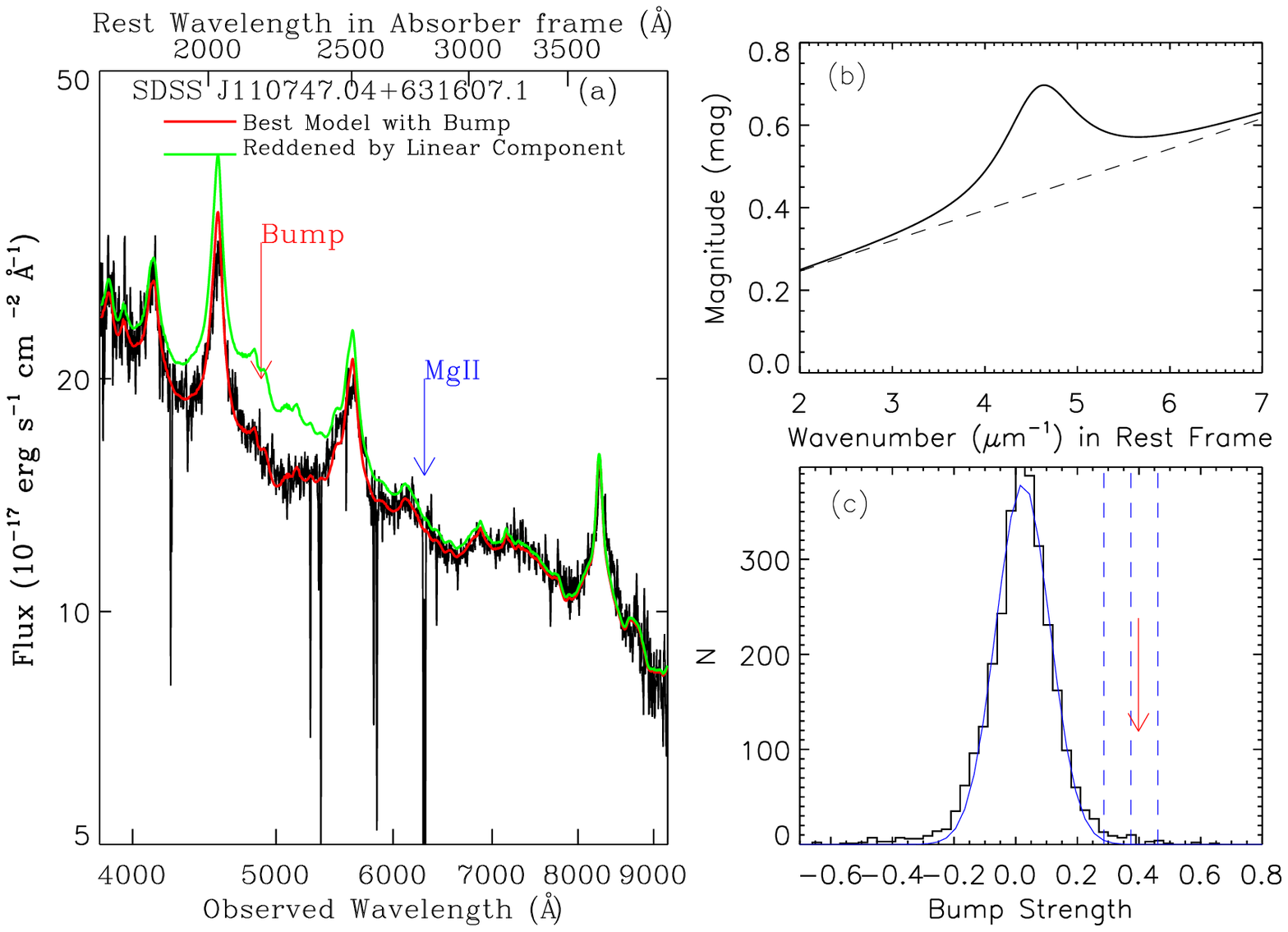}
\caption{(ONLINE ONLY)
The best fitted extinction model for J1107+6316. (a). Red solid line is the best fitted model.
Green solid line is reddened composite quasar spectrum
by using the linear component of best model only. (b). The best fitted
extinction curve. (c). Histogram of fitted bump strength of the control sample for J1107+6316.
\label{fig23}}
\end{figure}
\clearpage

\clearpage
\begin{figure}\epsscale{1.0}
\plotone{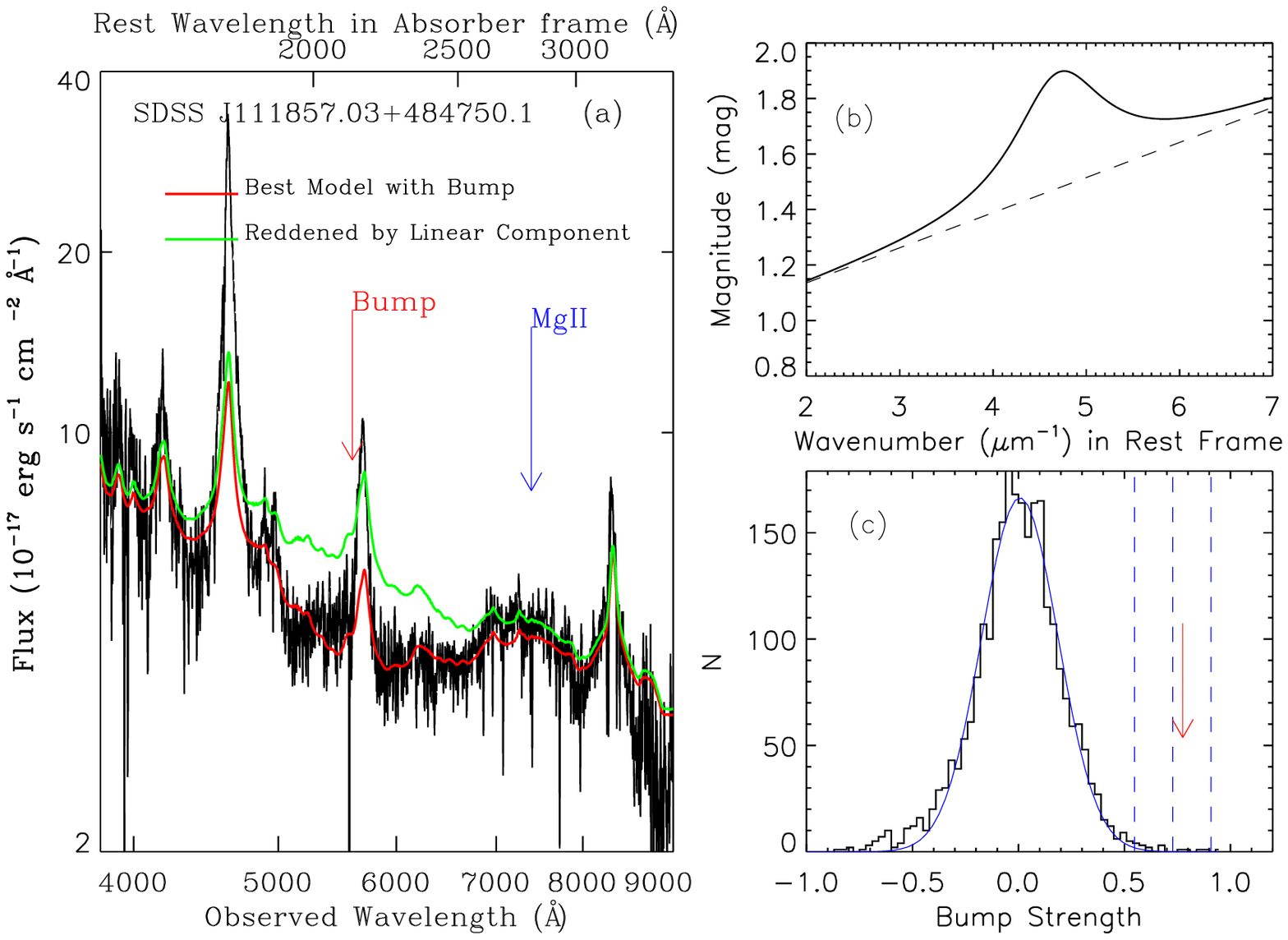}
\caption{(ONLINE ONLY)
The best fitted extinction model for J1118+4847. (a). Red solid line is the best fitted model.
Green solid line is reddened composite quasar spectrum
by using the linear component of best model only. (b). The best fitted
extinction curve. (c). Histogram of fitted bump strength of the control sample for J1118+4847.
\label{fig24}}
\end{figure}
\clearpage

\clearpage
\begin{figure}\epsscale{1.0}
\plotone{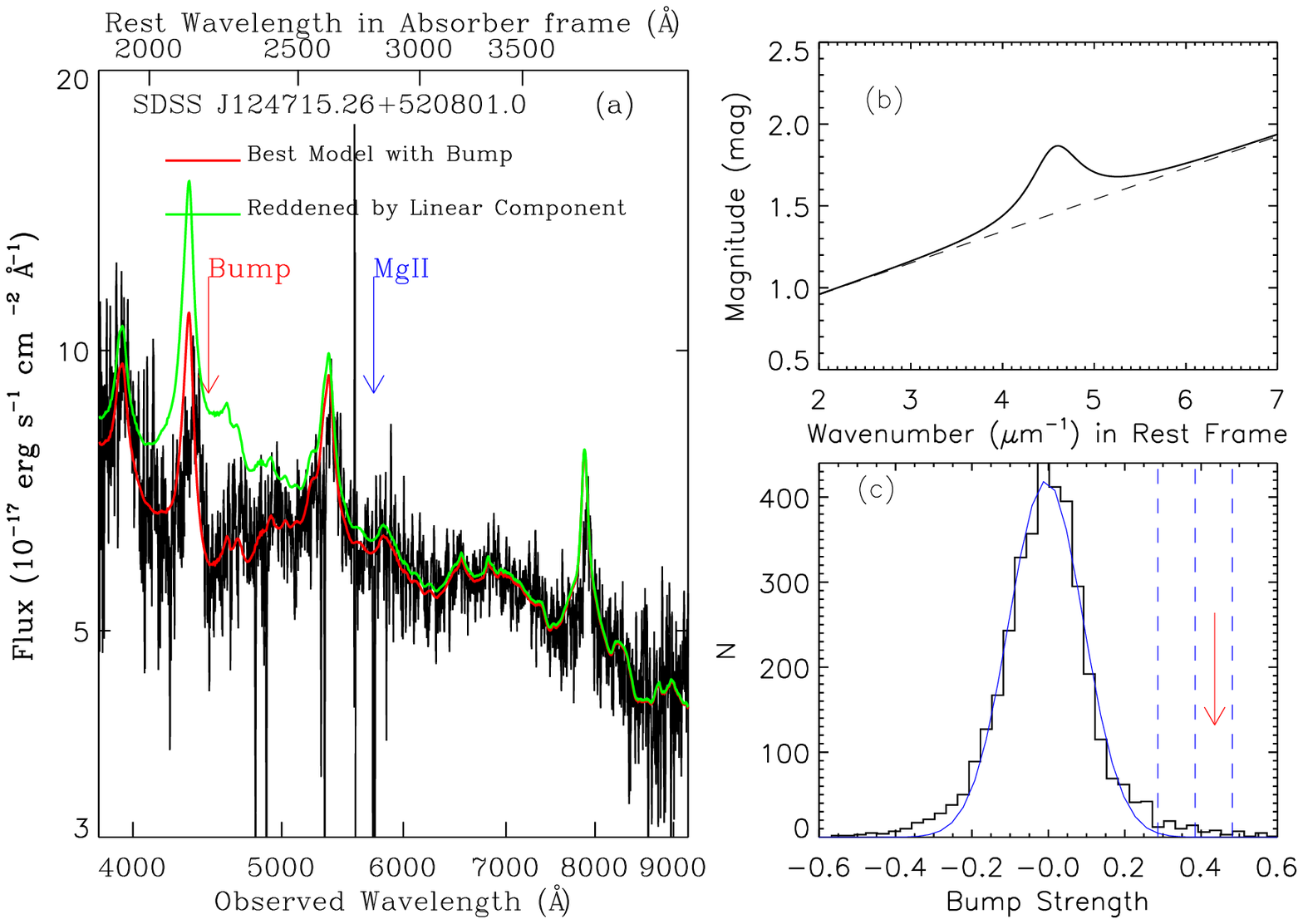}
\caption{(ONLINE ONLY)
The best fitted extinction model for J1247+5208. (a). Red solid line is the best fitted model.
Green solid line is reddened composite quasar spectrum
by using the linear component of best model only. (b). The best fitted
extinction curve. (c). Histogram of fitted bump strength of the control sample for J1247+5208.
\label{fig25}}
\end{figure}
\clearpage

\clearpage
\begin{figure}\epsscale{1.0}
\plotone{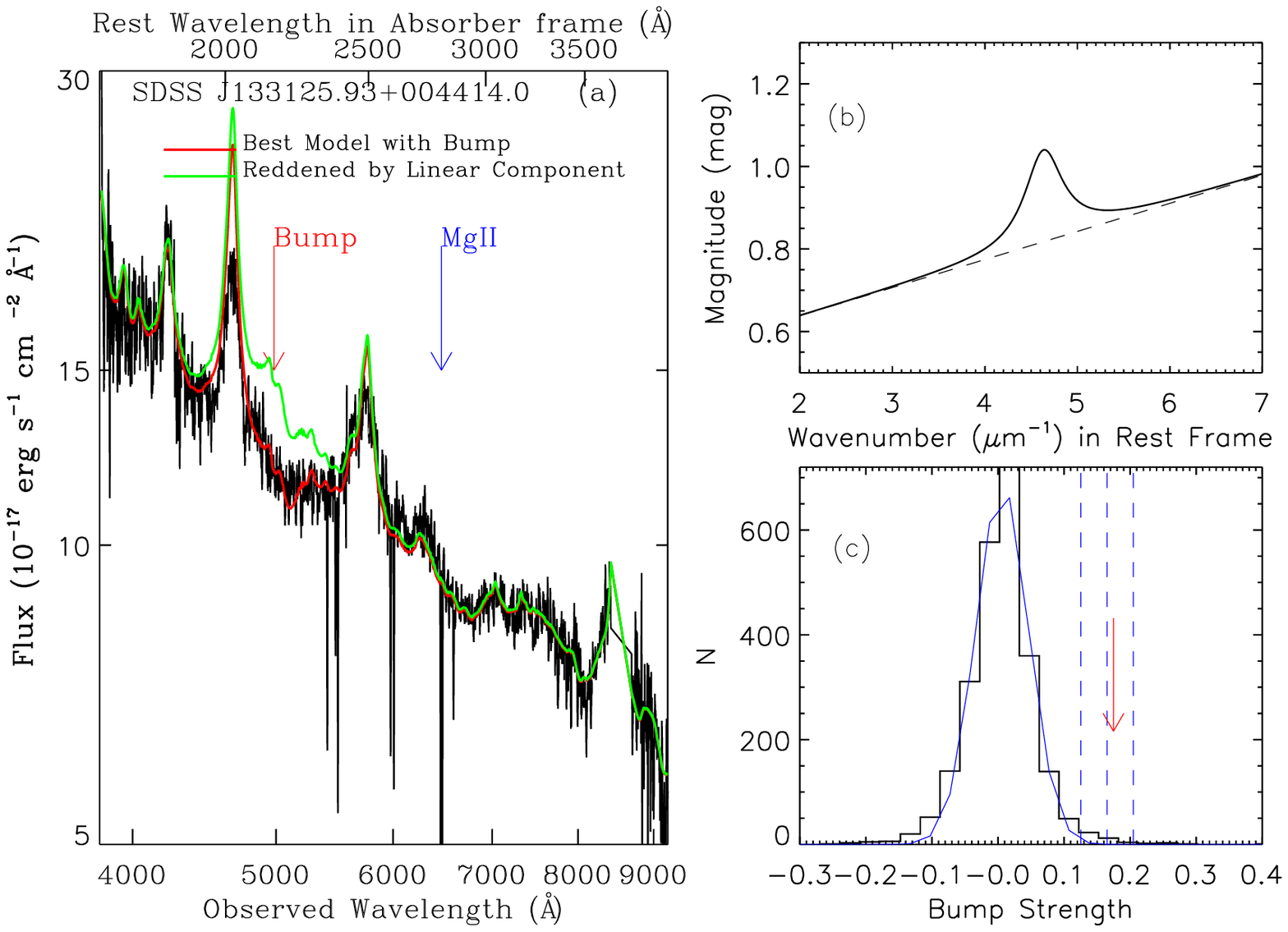}
\caption{(ONLINE ONLY)
The best fitted extinction model for J1331+0044. (a). Red solid line is the best fitted model.
Green solid line is reddened composite quasar spectrum
by using the linear component of best model only. (b). The best fitted
extinction curve. (c). Histogram of fitted bump strength of the control sample for J1331+0044.
\label{fig26}}
\end{figure}
\clearpage

\begin{figure}\epsscale{1.0}
\plotone{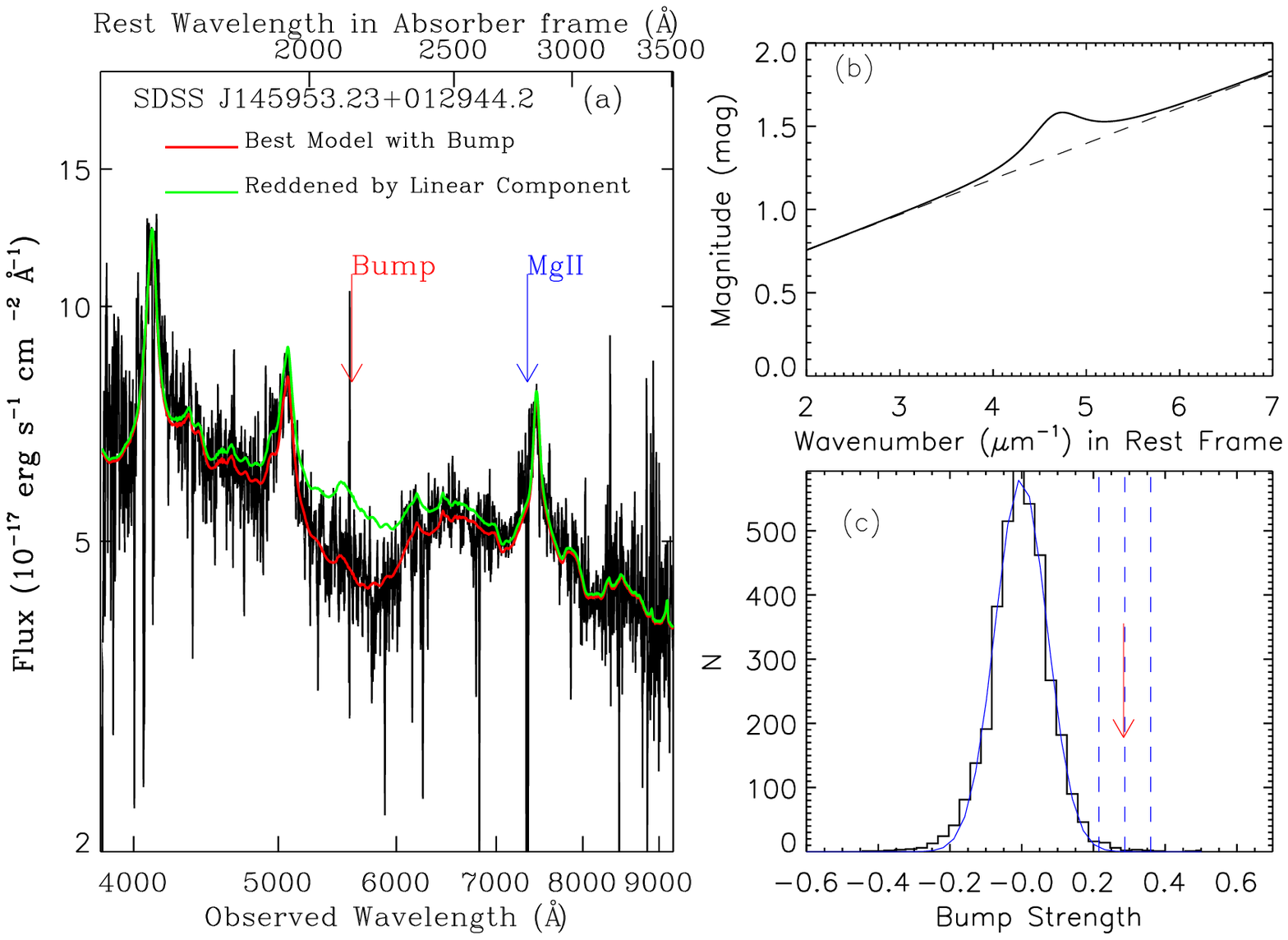}
\caption{(ONLINE ONLY)
The best fitted extinction model for J1459+0129. (a). Red solid line is the best fitted model.
Green solid line is reddened composite quasar spectrum
by using the linear component of best model only. (b). The best fitted
extinction curve. (c). Histogram of fitted bump strength of the control sample for J1459+0129.
\label{fig27}}
\end{figure}
\clearpage

\begin{figure}\epsscale{1.0}
\plotone{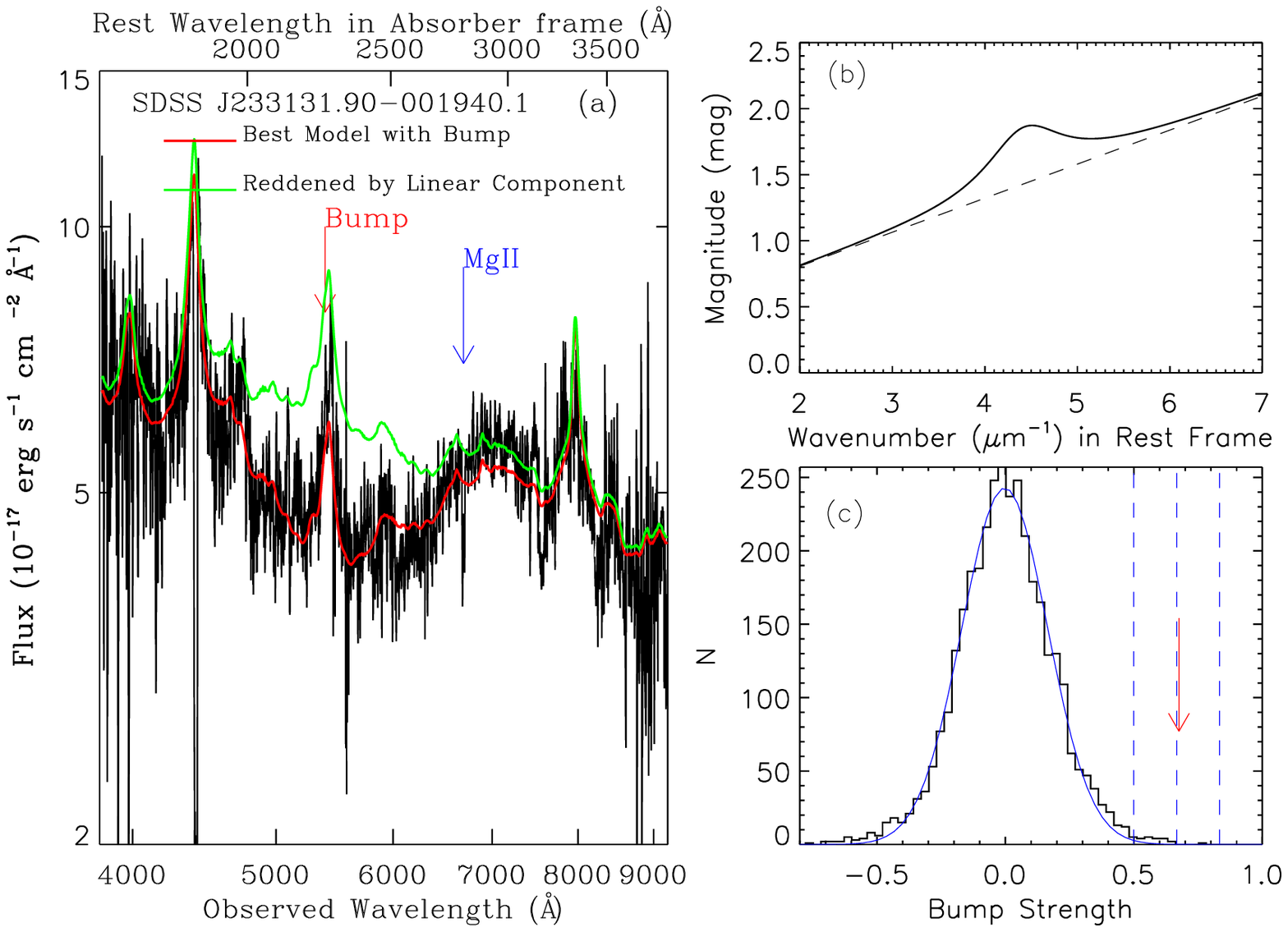}
\caption{(ONLINE ONLY)
The best fitted extinction model for J2331$-$0019. (a). Red solid line is the best fitted model.
Green solid line is reddened composite quasar spectrum
by using the linear component of best model only. (b). The best fitted
extinction curve. (c). Histogram of fitted bump strength of the control sample for J2331$-$0019.
\label{fig28}}
\end{figure}
\clearpage
\clearpage
\begin{figure}\epsscale{1.0}
\plotone{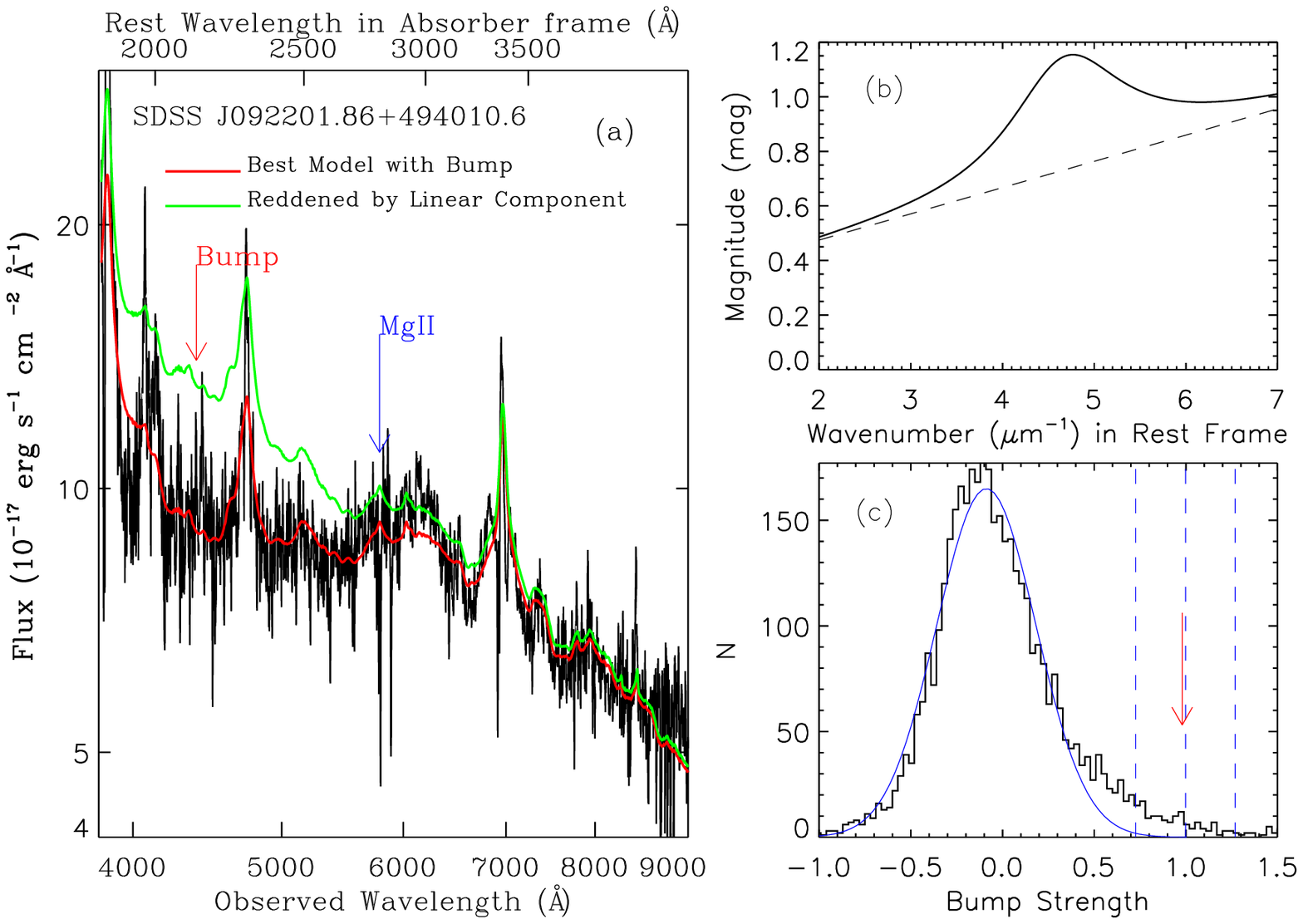}
\caption{(ONLINE ONLY)
The best fitted extinction model for J0922+4940. (a). Red solid line is the best fitted model.
Green solid line is reddened composite quasar spectrum
by using the linear component of best model only. (b). The best fitted
extinction curve. (c). Histogram of fitted bump strength of the control sample for J0922+4940.
\label{fig29}}
\end{figure}
\clearpage

\clearpage
\begin{figure}\epsscale{1.0}
\plotone{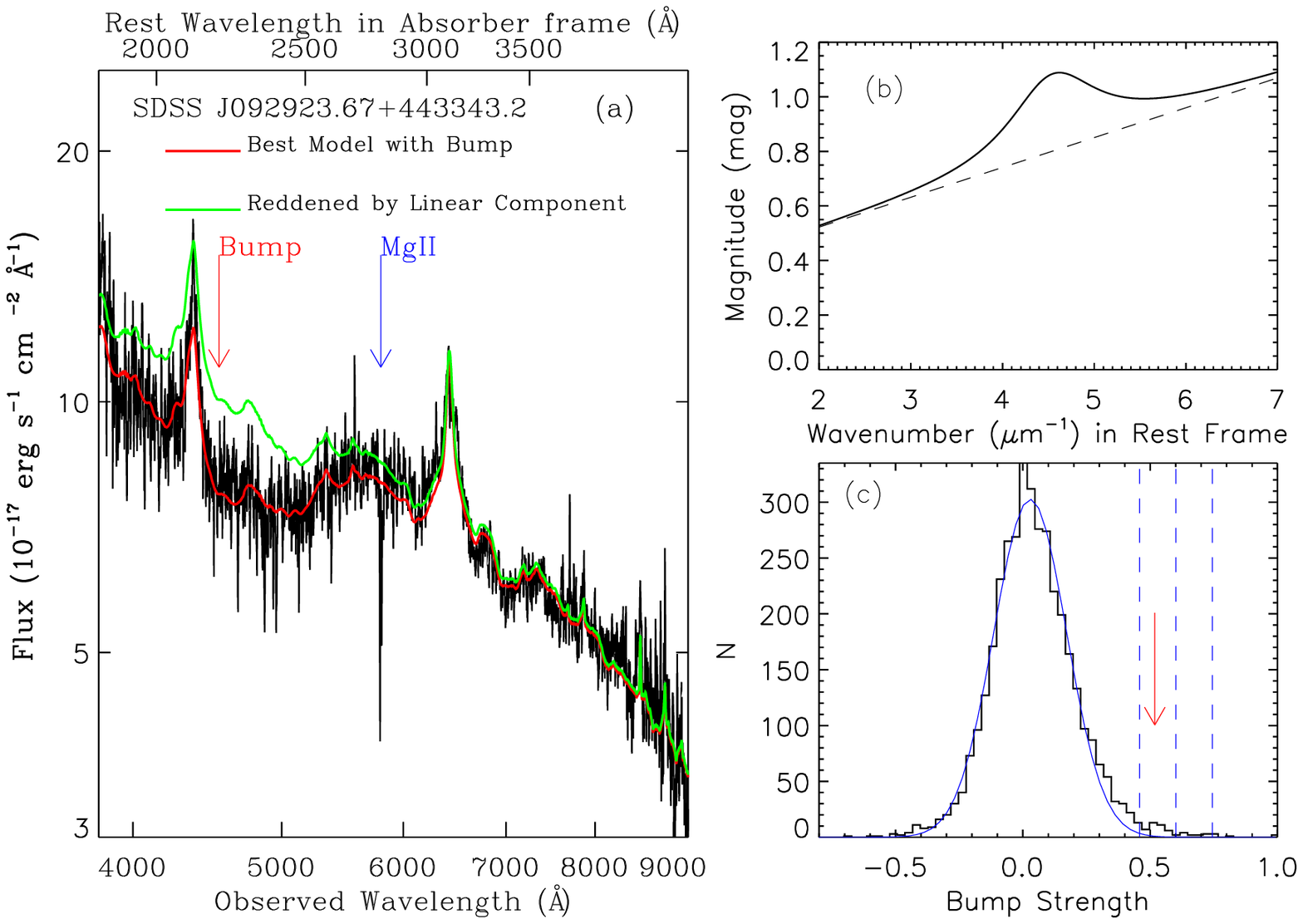}
\caption{(ONLINE ONLY)
The best fitted extinction model for J0929+4433. (a). Red solid line is the best fitted model.
Green solid line is reddened composite quasar spectrum
by using the linear component of best model only. (b). The best fitted
extinction curve. (c). Histogram of fitted bump strength of the control sample for J0929+4433.
\label{fig30}}
\end{figure}
\clearpage

\clearpage
\begin{figure}\epsscale{1.0}
\plotone{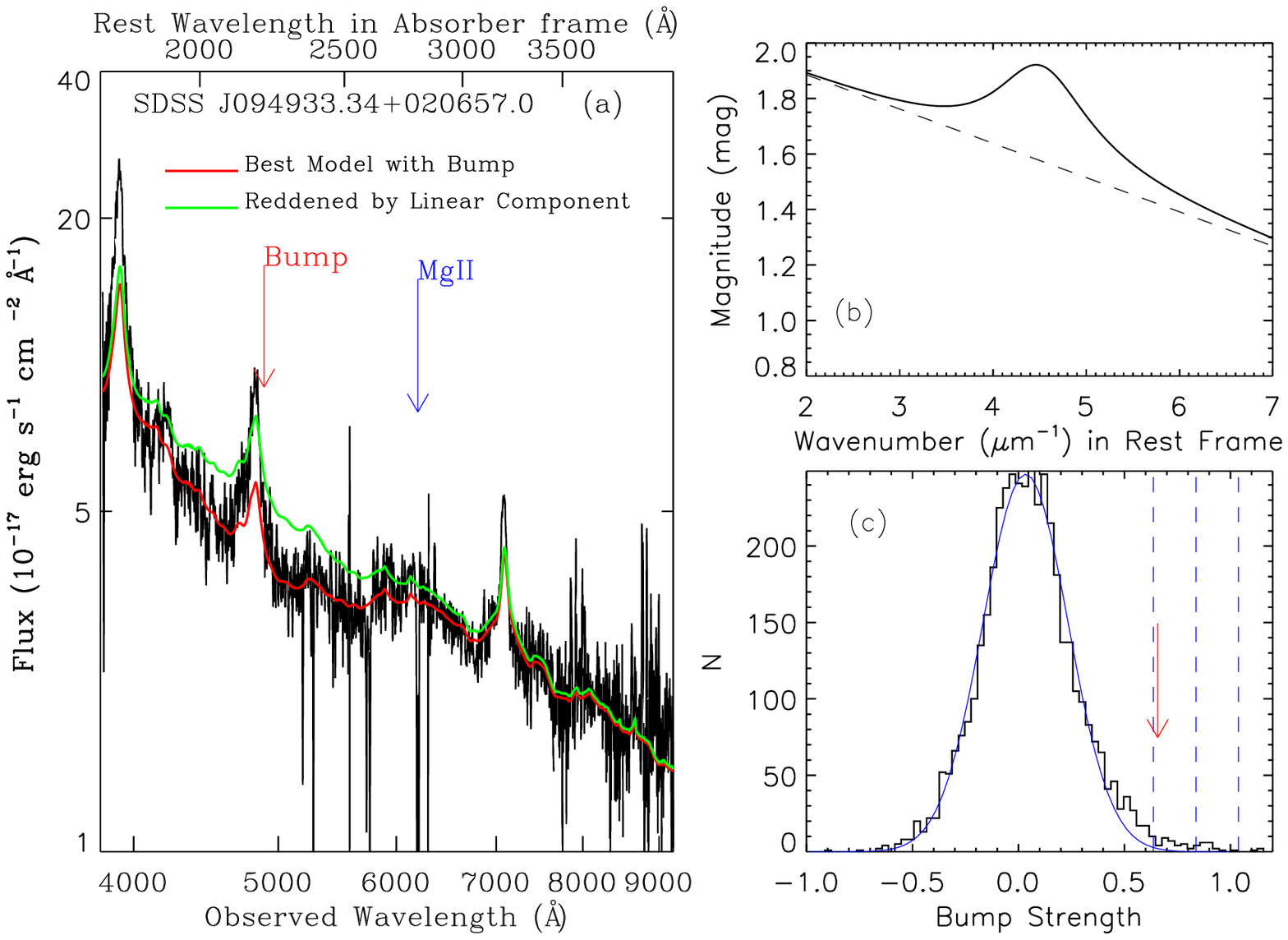}
\caption{(ONLINE ONLY)
The best fitted extinction model for J0949+0206. (a). Red solid line is the best fitted model.
Green solid line is reddened composite quasar spectrum
by using the linear component of best model only. (b). The best fitted
extinction curve. (c). Histogram of fitted bump strength of the control sample for J0949+0206.
\label{fig31}}
\end{figure}
\clearpage

\clearpage
\begin{figure}\epsscale{1.0}
\plotone{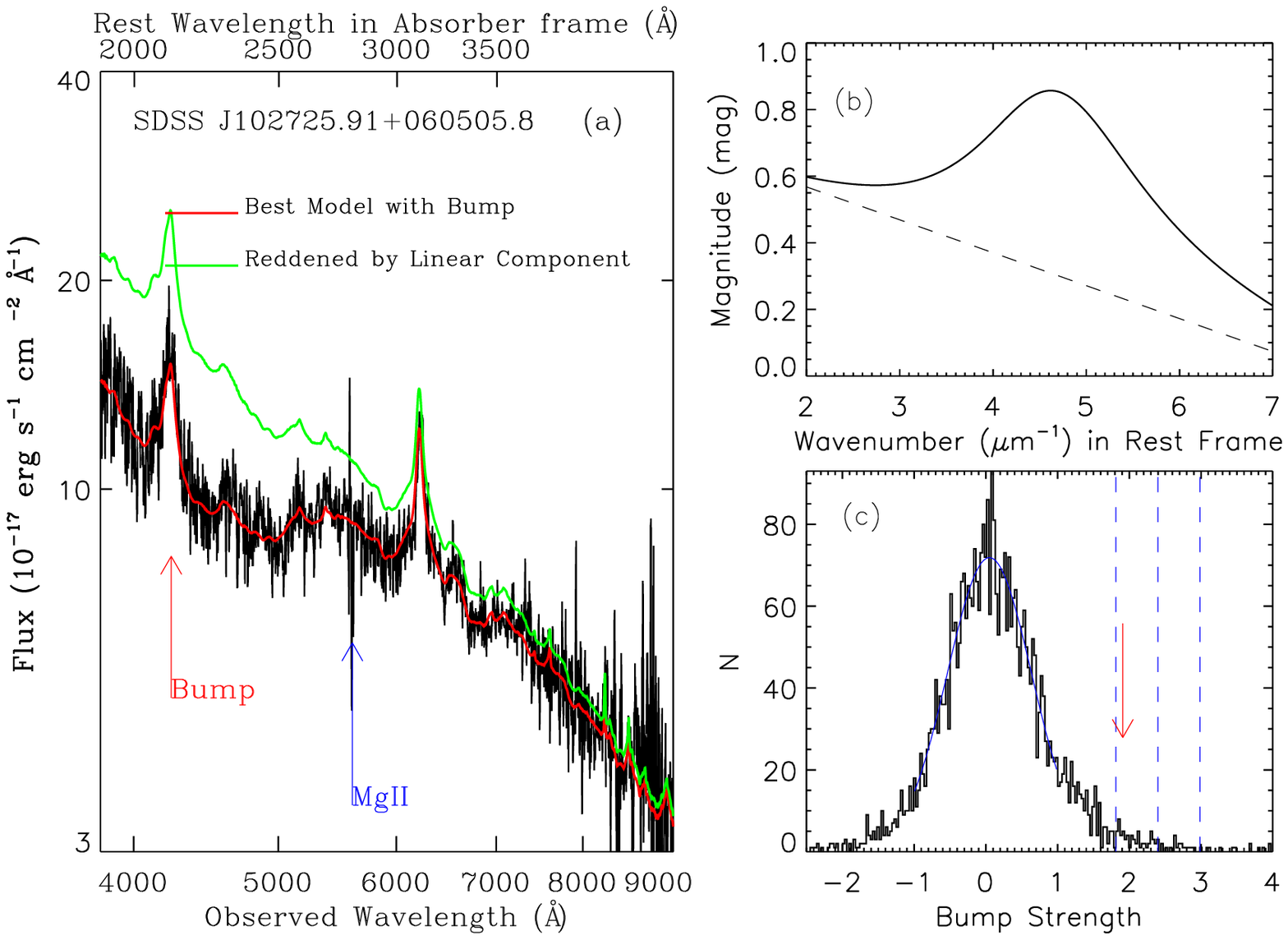}
\caption{(ONLINE ONLY)
The best fitted extinction model for J1027+0605. (a). Red solid line is the best fitted model.
Green solid line is reddened composite quasar spectrum
by using the linear component of best model only. (b). The best fitted
extinction curve. (c). Histogram of fitted bump strength of the control sample for J1027+0605.
\label{fig32}}
\end{figure}
\clearpage

\clearpage
\begin{figure}\epsscale{1.0}
\plotone{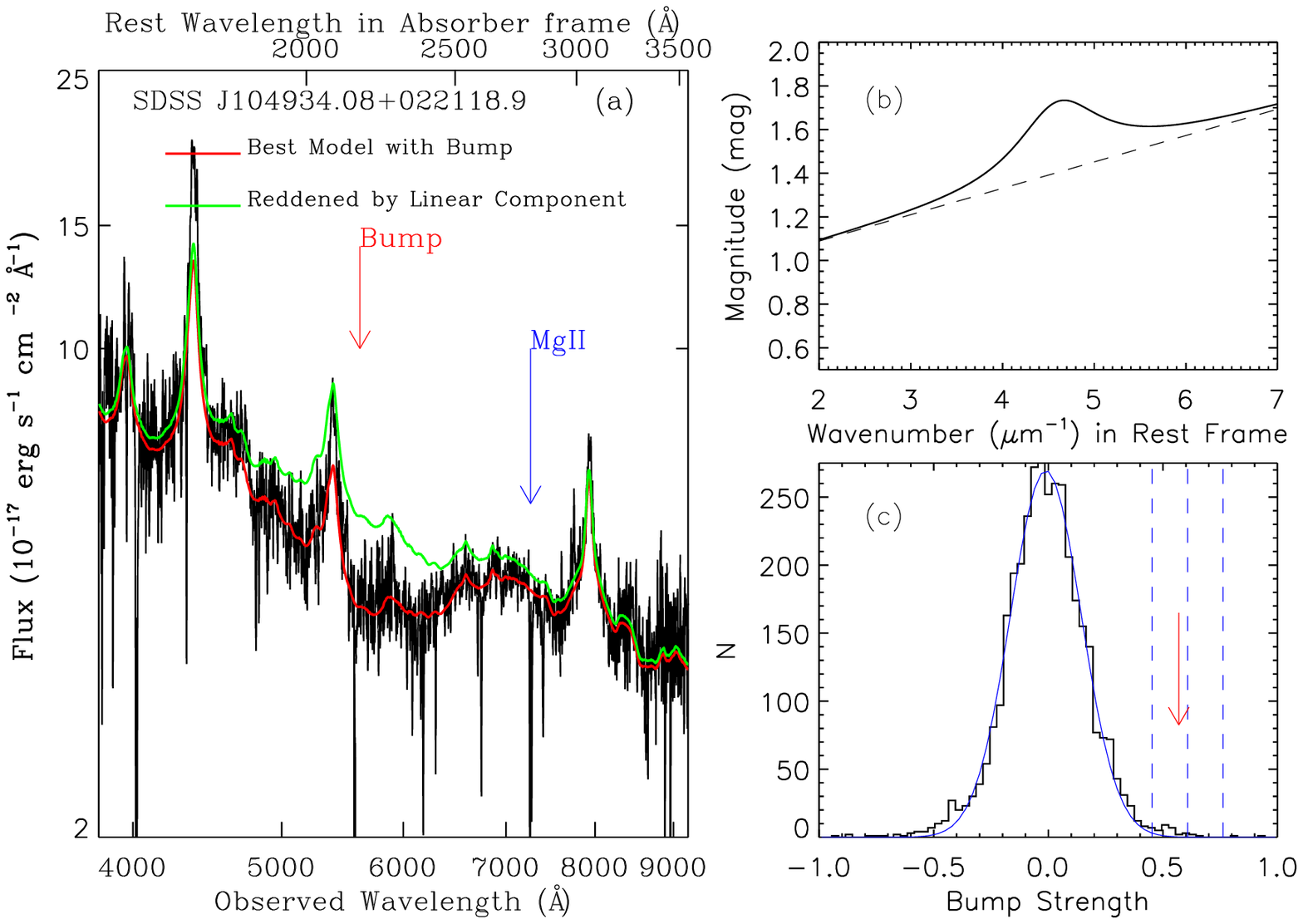}
\caption{(ONLINE ONLY)
The best fitted extinction model for J1049+0221. (a). Red solid line is the best fitted model.
Green solid line is reddened composite quasar spectrum
by using the linear component of best model only. (b). The best fitted
extinction curve. (c). Histogram of fitted bump strength of the control sample for J1049+0221.
\label{fig33}}
\end{figure}
\clearpage

\clearpage
\begin{figure}\epsscale{1.0}
\plotone{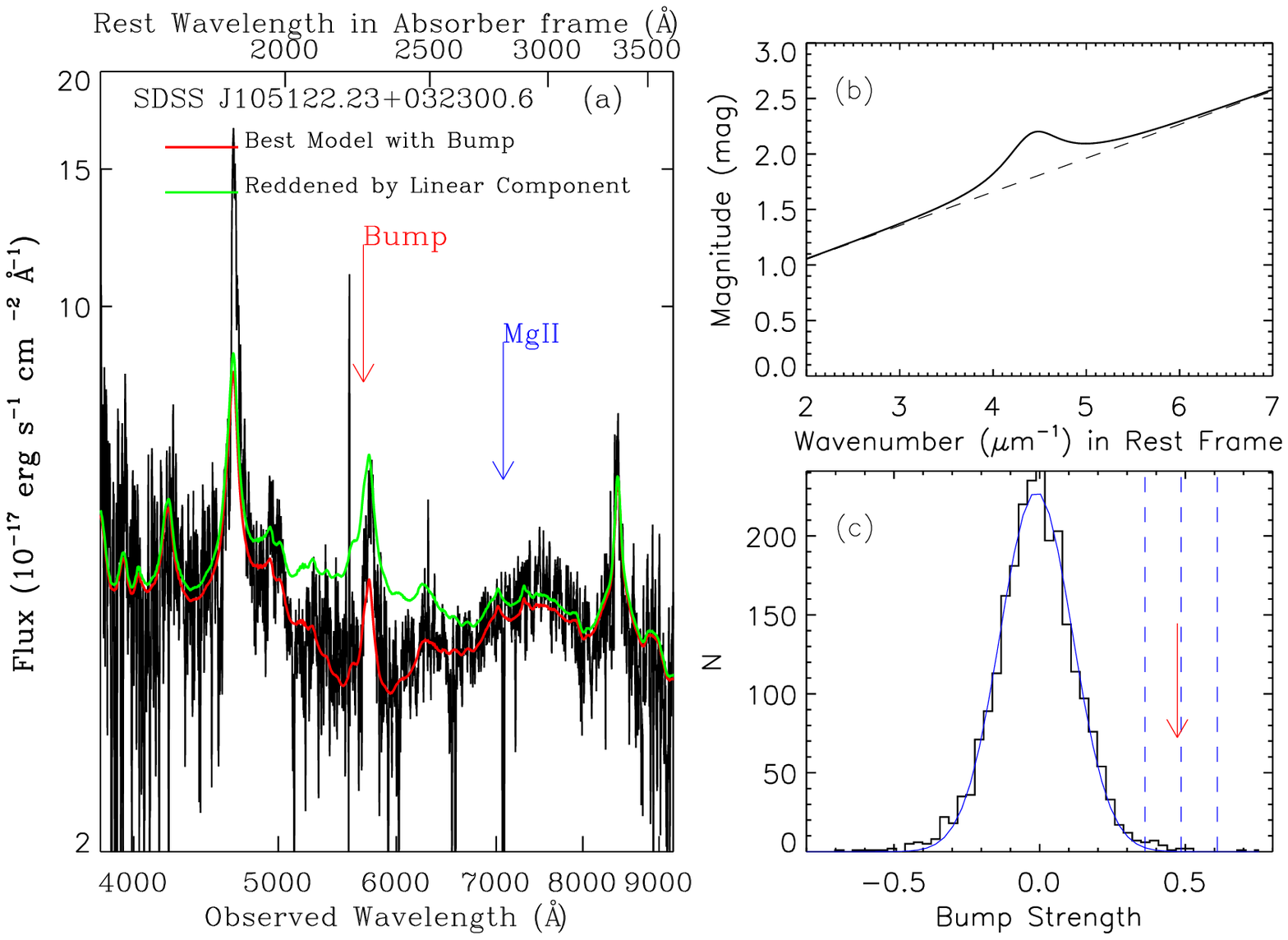}
\caption{(ONLINE ONLY)
The best fitted extinction model for J1051+0323. (a). Red solid line is the best fitted model.
Green solid line is reddened composite quasar spectrum
by using the linear component of best model only. (b). The best fitted
extinction curve. (c). Histogram of fitted bump strength of the control sample for J1051+0323.
\label{fig34}}
\end{figure}
\clearpage

\clearpage
\begin{figure}\epsscale{1.0}
\plotone{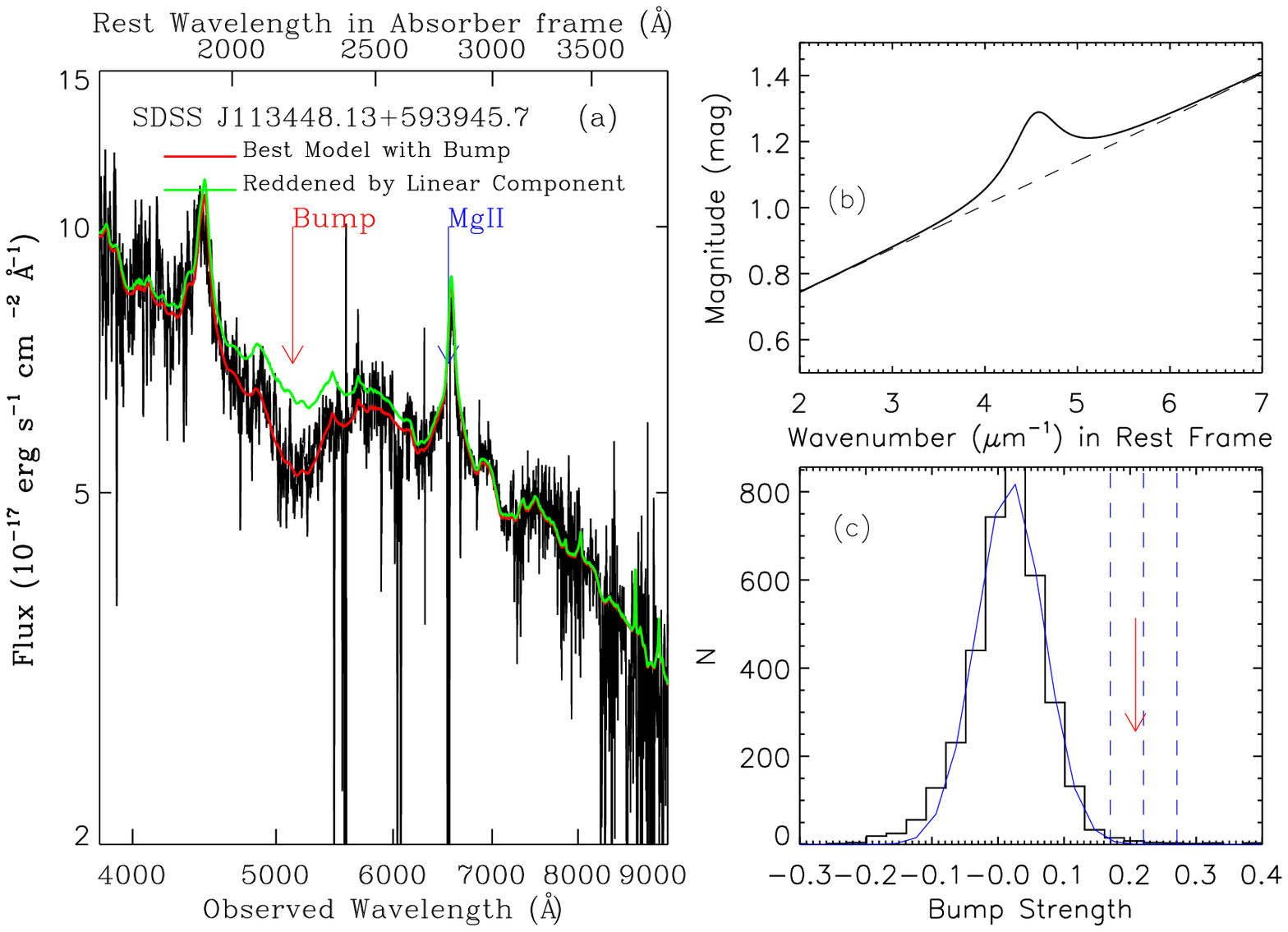}
\caption{(ONLINE ONLY)
The best fitted extinction model for J1134+5939. (a). Red solid line is the best fitted model.
Green solid line is reddened composite quasar spectrum
by using the linear component of best model only. (b). The best fitted
extinction curve. (c). Histogram of fitted bump strength of the control sample for J1134+5939.
\label{fig35}}
\end{figure}
\clearpage

\clearpage
\begin{figure}\epsscale{1.0}
\plotone{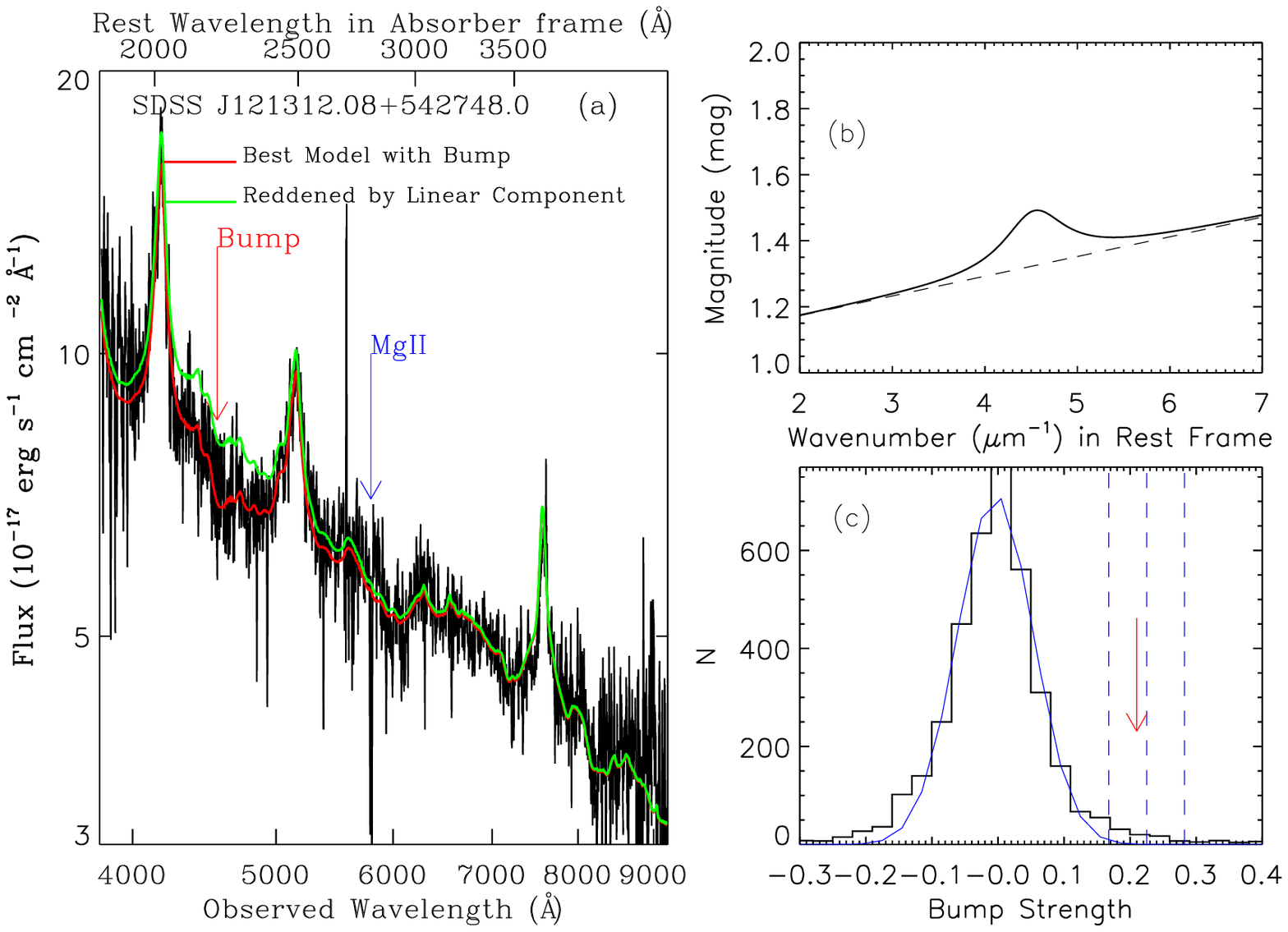}
\caption{(ONLINE ONLY)
The best fitted extinction model for J1213+5427. (a). Red solid line is the best fitted model.
Green solid line is reddened composite quasar spectrum
by using the linear component of best model only. (b). The best fitted
extinction curve. (c). Histogram of fitted bump strength of the control sample for J1213+5427.
\label{fig36}}
\end{figure}
\clearpage

\clearpage
\begin{figure}\epsscale{1.0}
\plotone{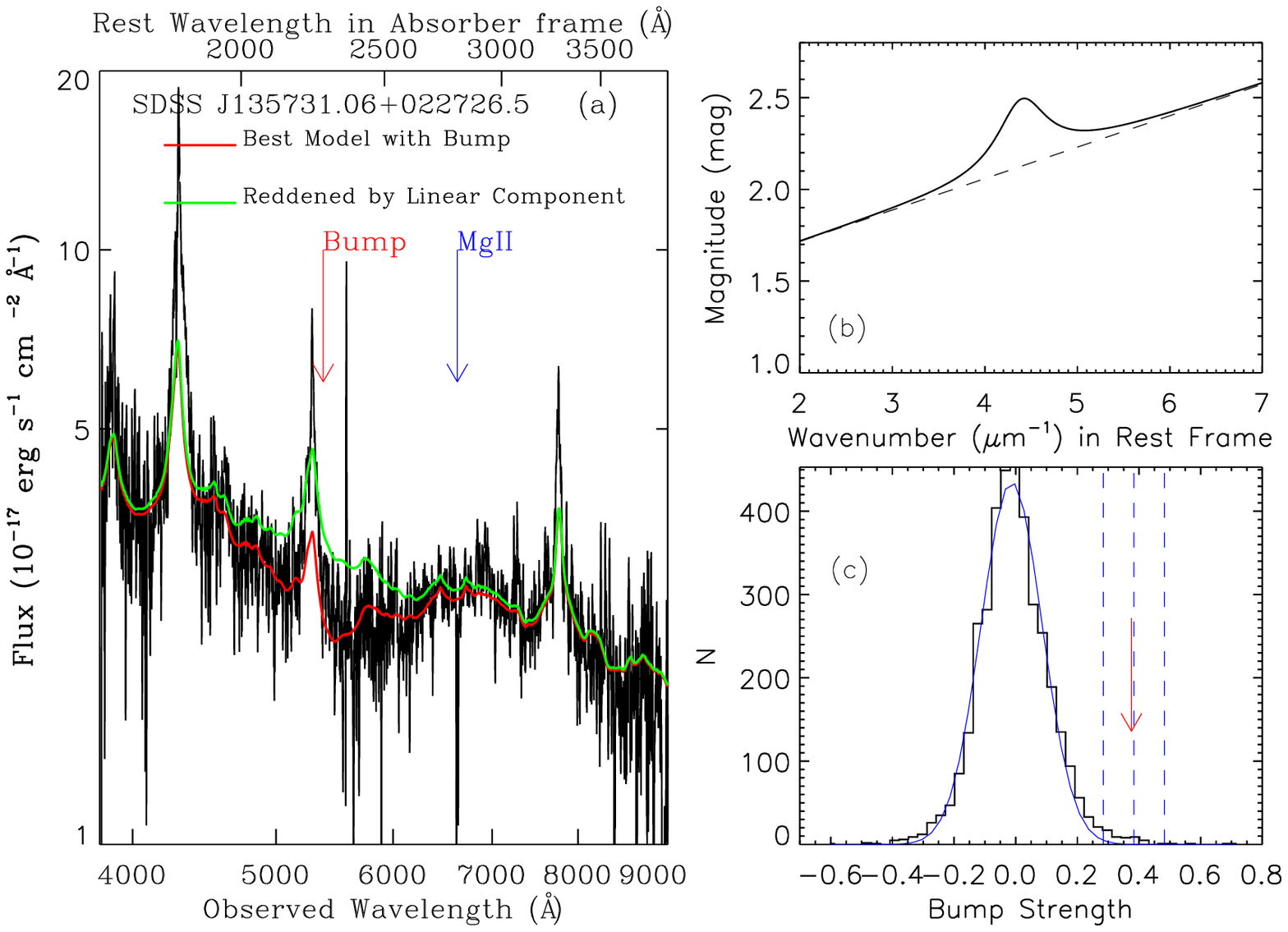}
\caption{(ONLINE ONLY) 
The best fitted extinction model for J1357+0227. (a). Red solid line is the best fitted model.
Green solid line is reddened composite quasar spectrum
by using the linear component of best model only. (b). The best fitted
extinction curve. (c). Histogram of fitted bump strength of the control sample for J1357+0227.
\label{fig37}}
\end{figure}
\clearpage

\clearpage
\begin{figure}\epsscale{1.0}
\plotone{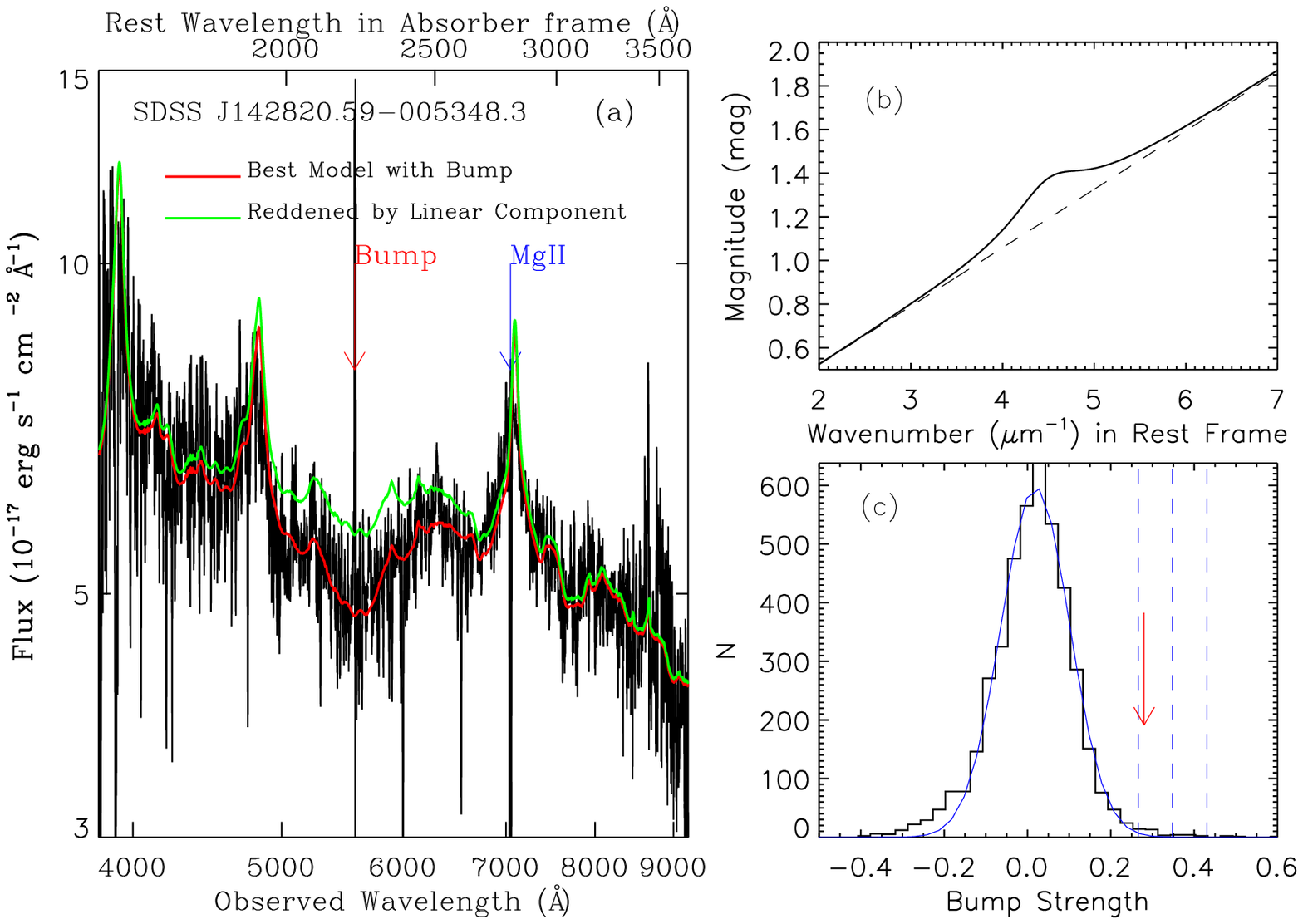}
\caption{(ONLINE ONLY)
The best fitted extinction model for J1428$-$0053. (a). Red solid line is the best fitted model.
Green solid line is reddened composite quasar spectrum
by using the linear component of best model only. (b). The best fitted
extinction curve. (c). Histogram of fitted bump strength of the control sample for J1428$-$0053.
\label{fig38}}
\end{figure}
\clearpage

\clearpage
\begin{figure}\epsscale{1.0}
\plotone{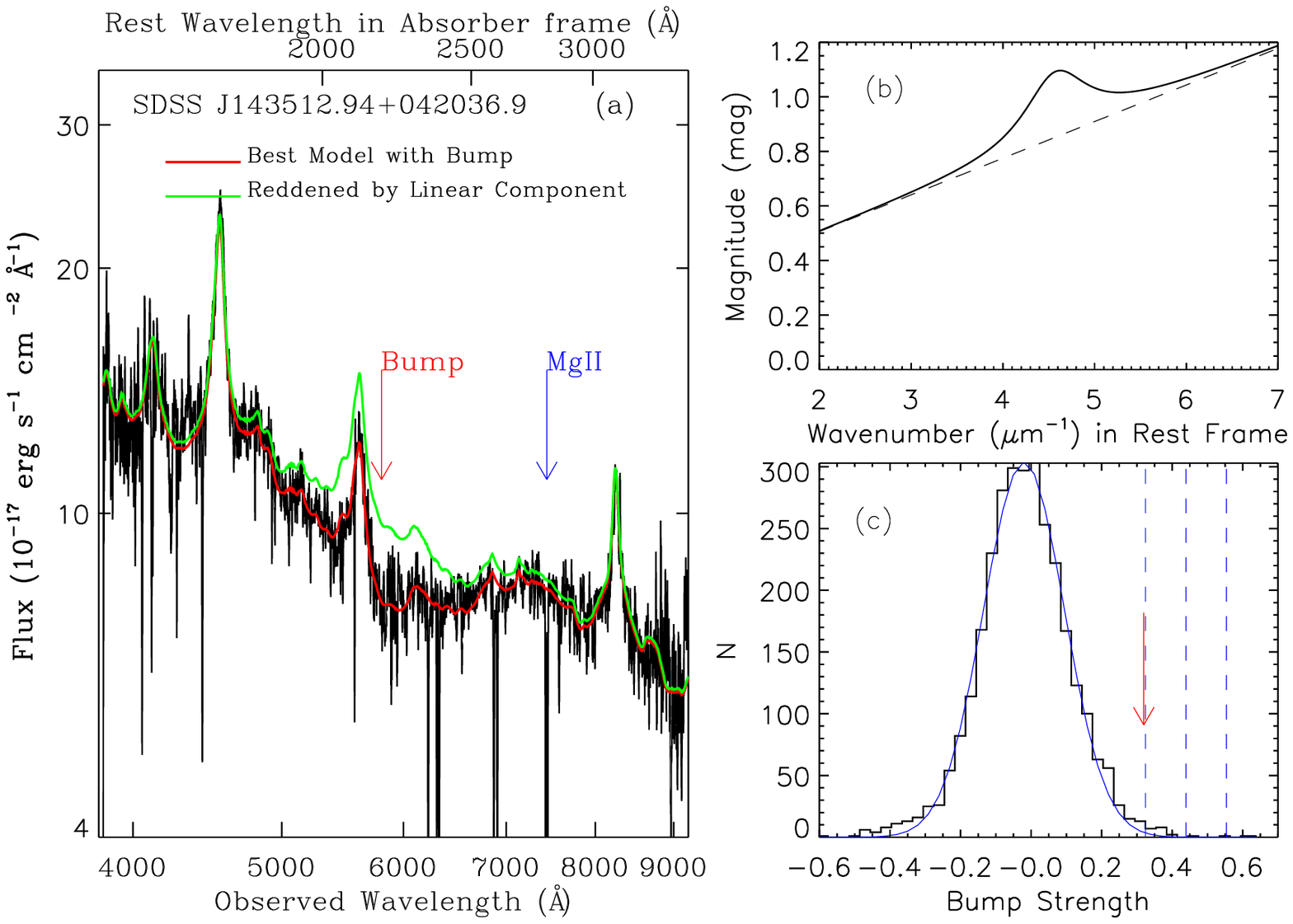}
\caption{(ONLINE ONLY)
The best fitted extinction model for J1435+0420. (a). Red solid line is the best fitted model.
Green solid line is reddened composite quasar spectrum
by using the linear component of best model only. (b). The best fitted
extinction curve. (c). Histogram of fitted bump strength of the control sample for J1435+0420.
\label{fig39}}
\end{figure}
\clearpage

\clearpage
\begin{figure}\epsscale{1.0}
\plotone{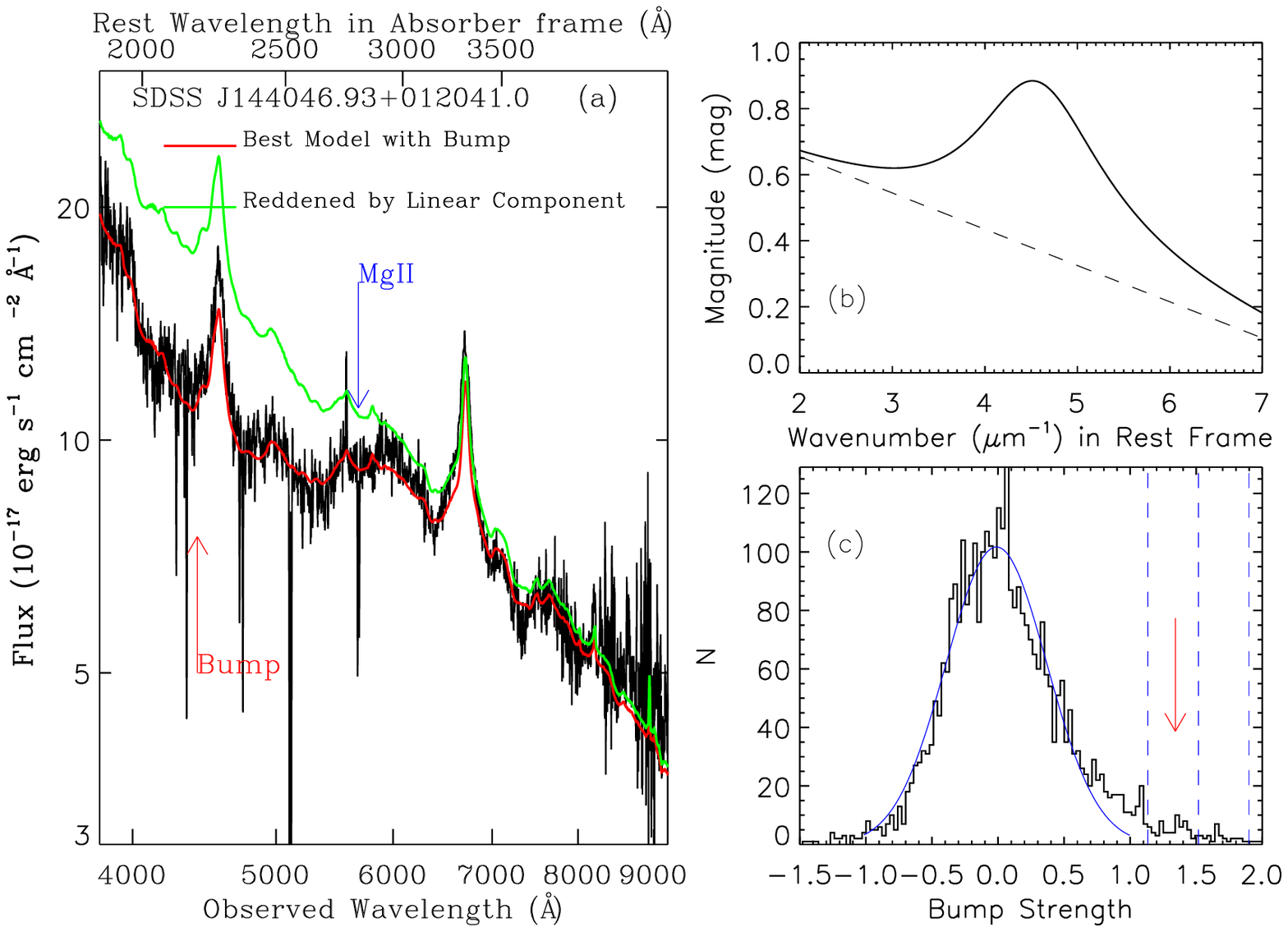}
\caption{(ONLINE ONLY)
The best fitted extinction model for J1440+0120. (a). Red solid line is the best fitted model.
Green solid line is reddened composite quasar spectrum
by using the linear component of best model only. (b). The best fitted
extinction curve. (c). Histogram of fitted bump strength of the control sample for J1440+0120.
\label{fig40}}
\end{figure}
\clearpage

\begin{figure}\epsscale{1.0}
\plotone{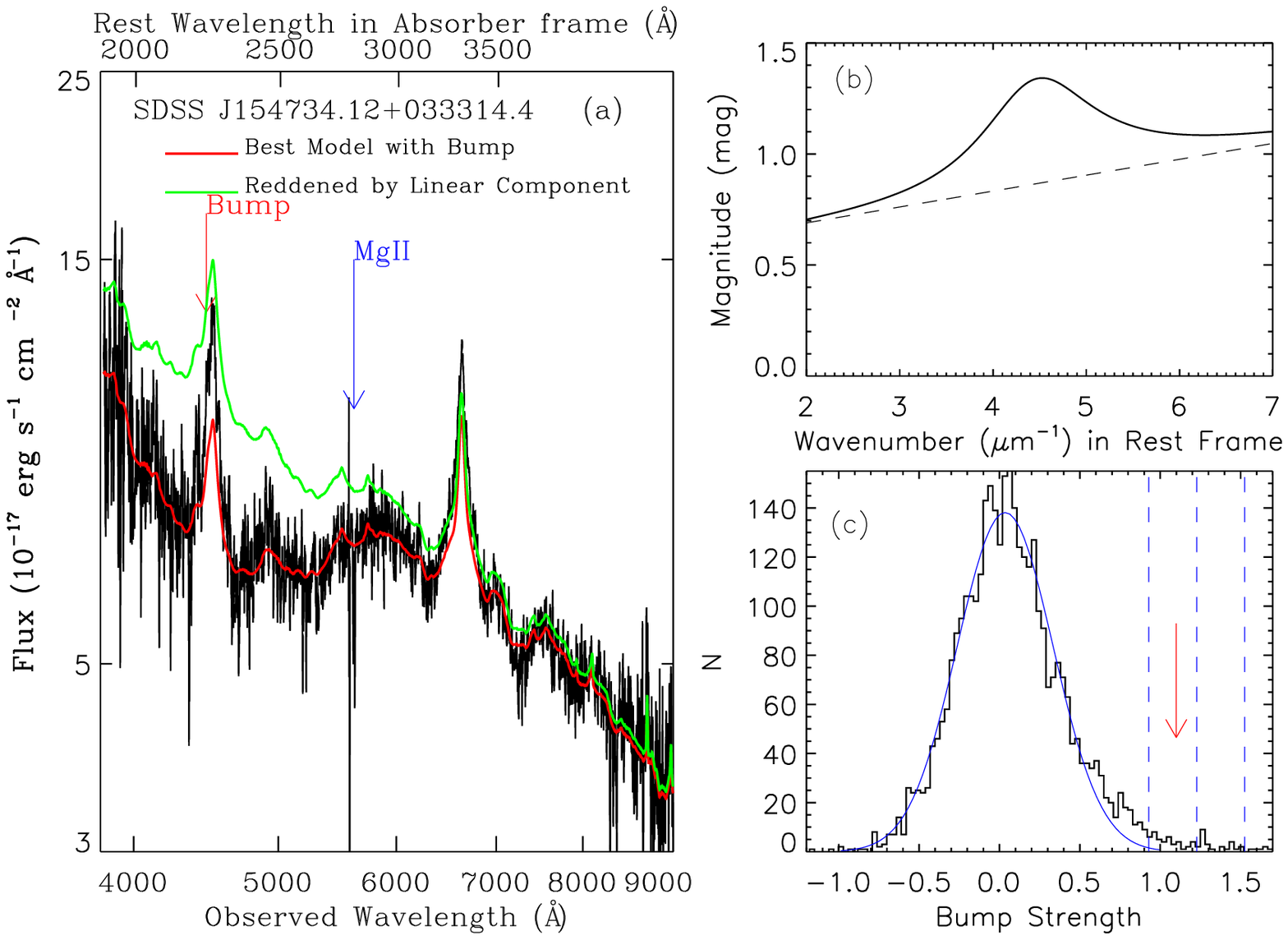}
\caption{(ONLINE ONLY)
The best fitted extinction model for J1547+0333. (a). Red solid line is the best fitted model.
Green solid line is reddened composite quasar spectrum
by using the linear component of best model only. (b). The best fitted
extinction curve. (c). Histogram of fitted bump strength of the control sample for J1547+0333.
\label{fig41}}
\end{figure}
\clearpage

\begin{figure}\epsscale{1.0}
\plotone{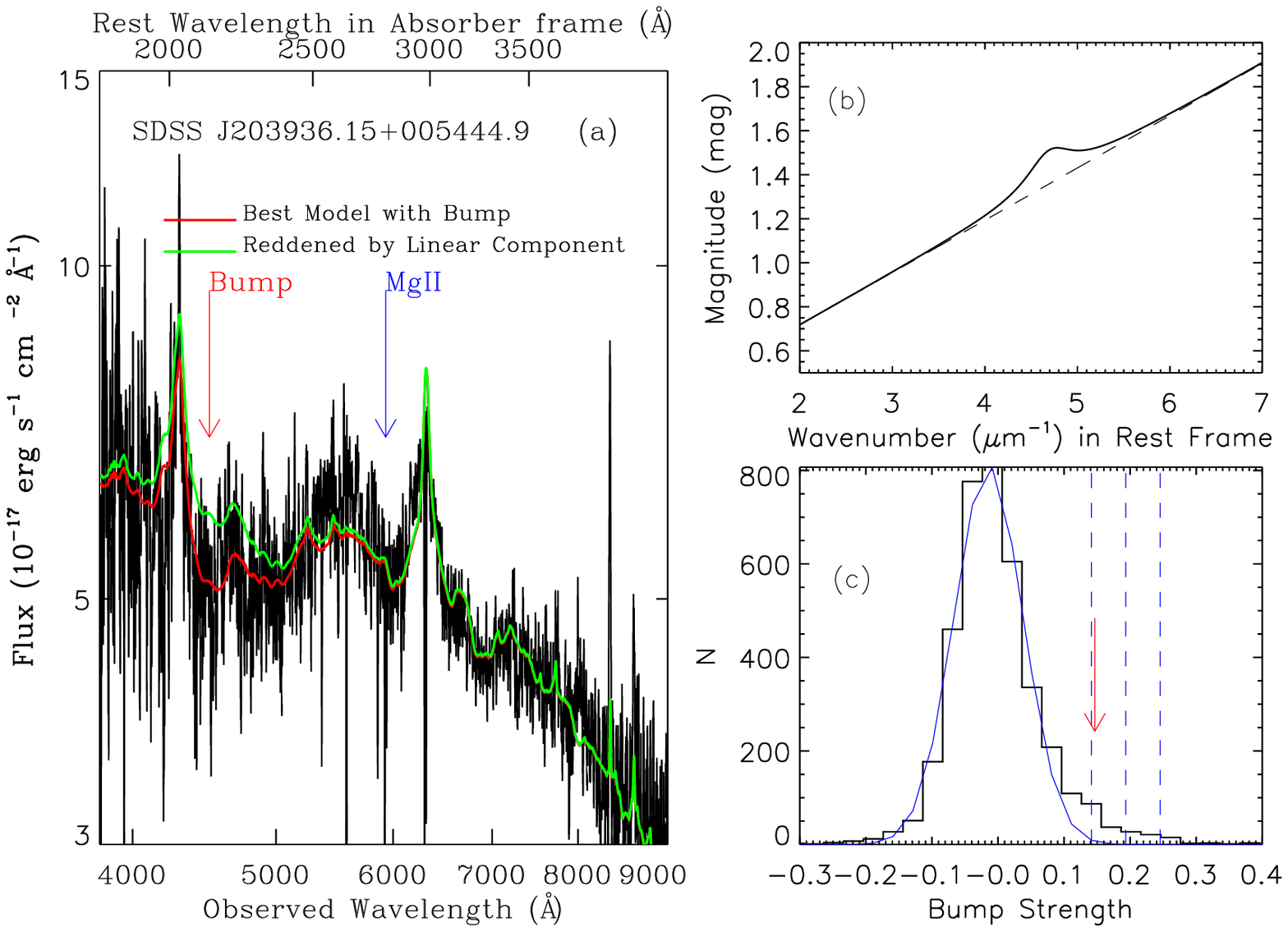}
\caption{(ONLINE ONLY)
The best fitted extinction model for J2039+0054. (a). Red solid line is the best fitted model.
Green solid line is reddened composite quasar spectrum
by using the linear component of best model only. (b). The best fitted
extinction curve. (c). Histogram of fitted bump strength of the control sample for J2039+0054.
\label{fig42}}
\end{figure}
\clearpage

\begin{figure}\epsscale{1.0}
\plotone{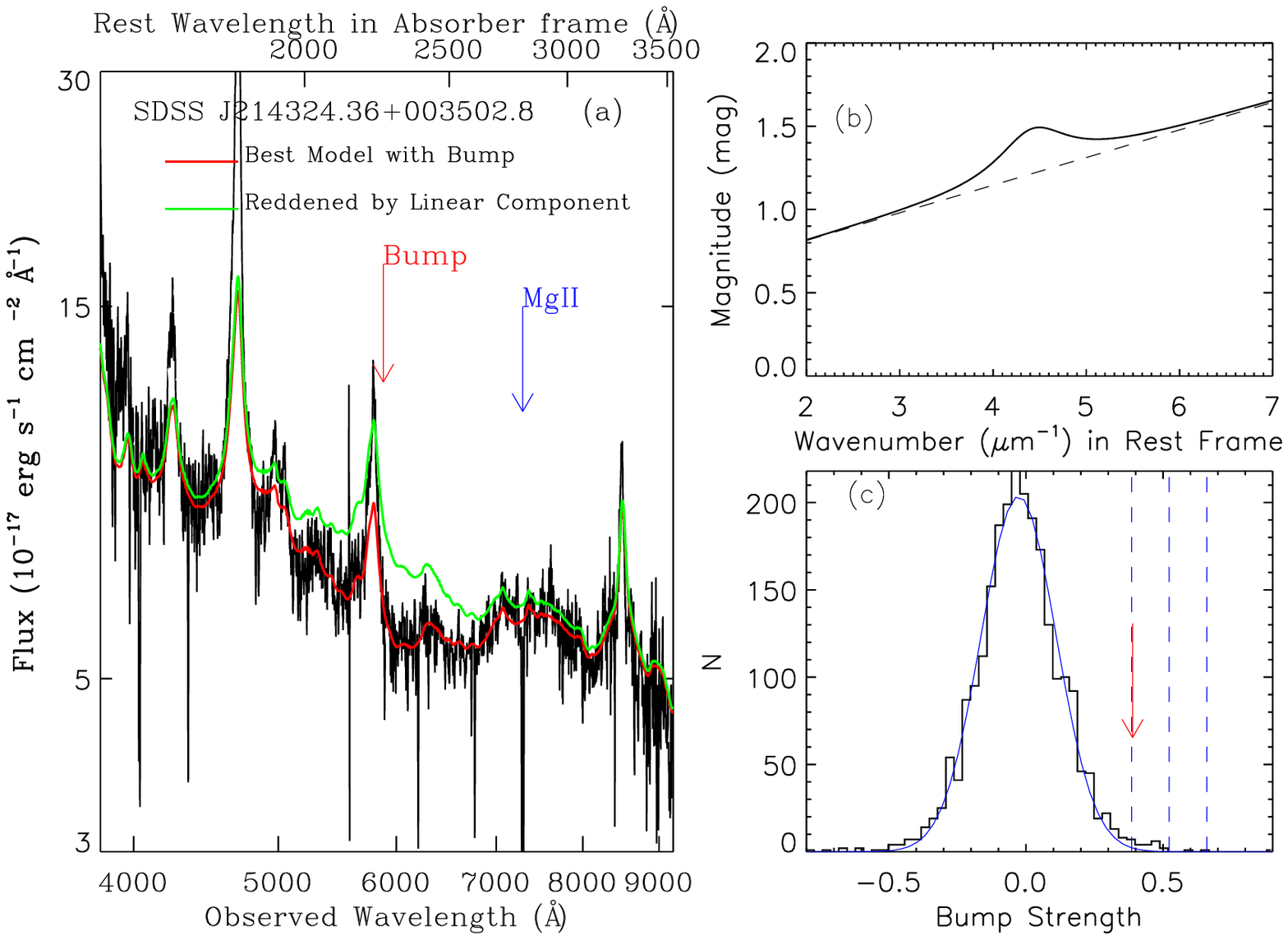}
\caption{(ONLINE ONLY)
The best fitted extinction model for J2143+0035. (a). Red solid line is the best fitted model.
Green solid line is reddened composite quasar spectrum
by using the linear component of best model only. (b). The best fitted
extinction curve. (c). Histogram of fitted bump strength of the control sample for J2143+0035.
\label{fig43}}
\end{figure}
\clearpage

\begin{figure}\epsscale{1.0}
\plotone{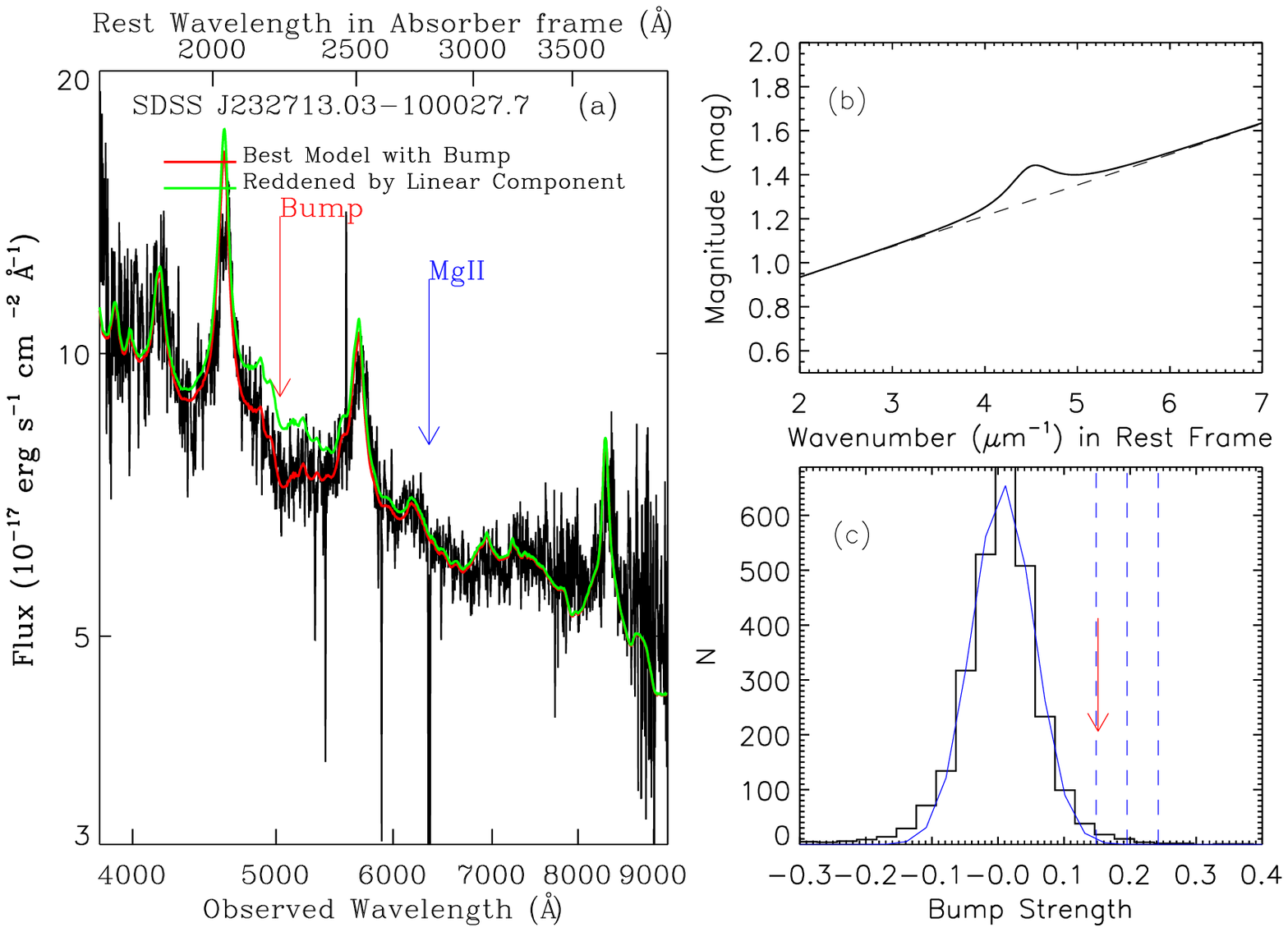}
\caption{(ONLINE ONLY)
The best fitted extinction model for J2327$-$1000. (a). Red solid line is the best fitted model.
Green solid line is reddened composite quasar spectrum
by using the linear component of best model only. (b). The best fitted
extinction curve. (c). Histogram of fitted bump strength of the control sample for J2327$-$1000.
\label{fig44}}
\end{figure}
\clearpage

\clearpage
\begin{figure}\epsscale{1.0}
\plotone{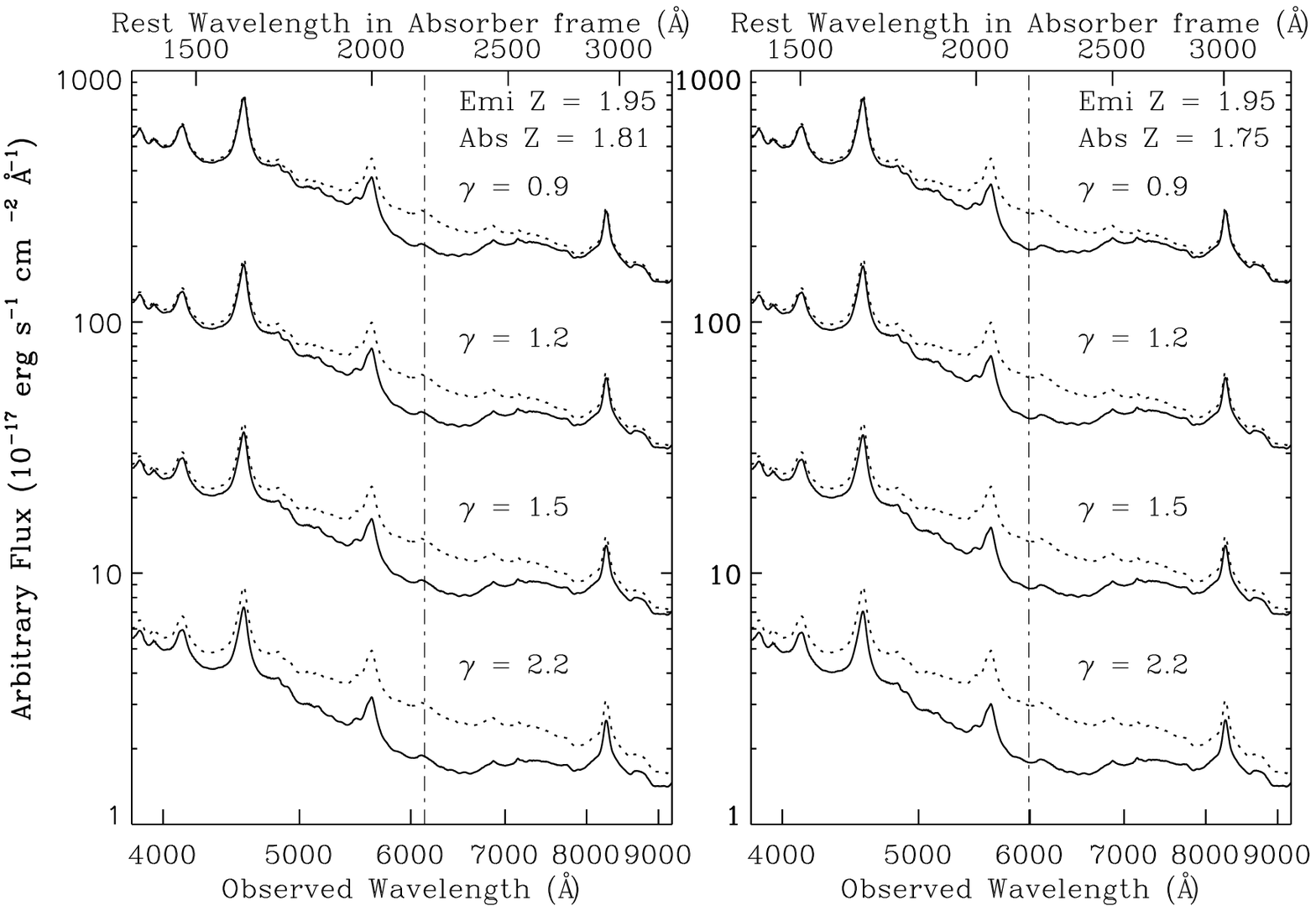}
\caption{(ONLINE ONLY) The dotted spectra are SDSS
quasar composite spectra. The solid ones are the spectra reddened by bumps.
The spectra for different x$_0^{qso}$s are organized in separated panels; the
spectra in the same panel are for increasing $\gamma$ from top to bottom.
The dot-dashed lines indicate the center positions of the bump.
\label{fig45}}
\end{figure}
\clearpage

\begin{figure}\epsscale{1.0}
\plotone{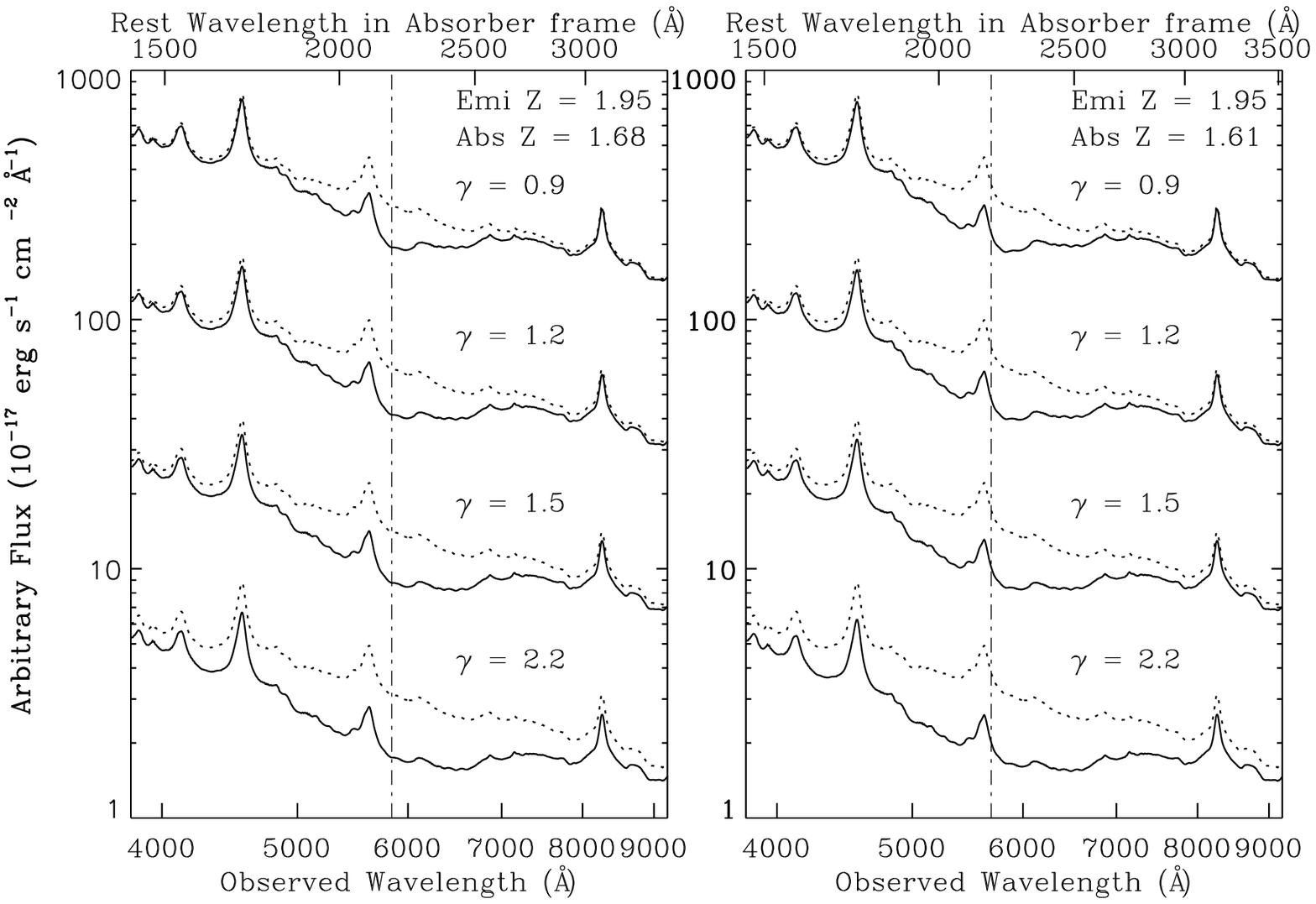}
\caption{(ONLINE ONLY) The dotted spectra are SDSS
quasar composite spectra. The solid ones are the spectra reddened by bumps.
The spectra for different x$_0^{qso}$s are organized in separated panels; the
spectra in the same panel are for increasing $\gamma$ from top to bottom.
The dot-dashed lines indicate the center positions of the bump.
\label{fig46}}
\end{figure}
\clearpage

\begin{figure}\epsscale{1.0}
\plotone{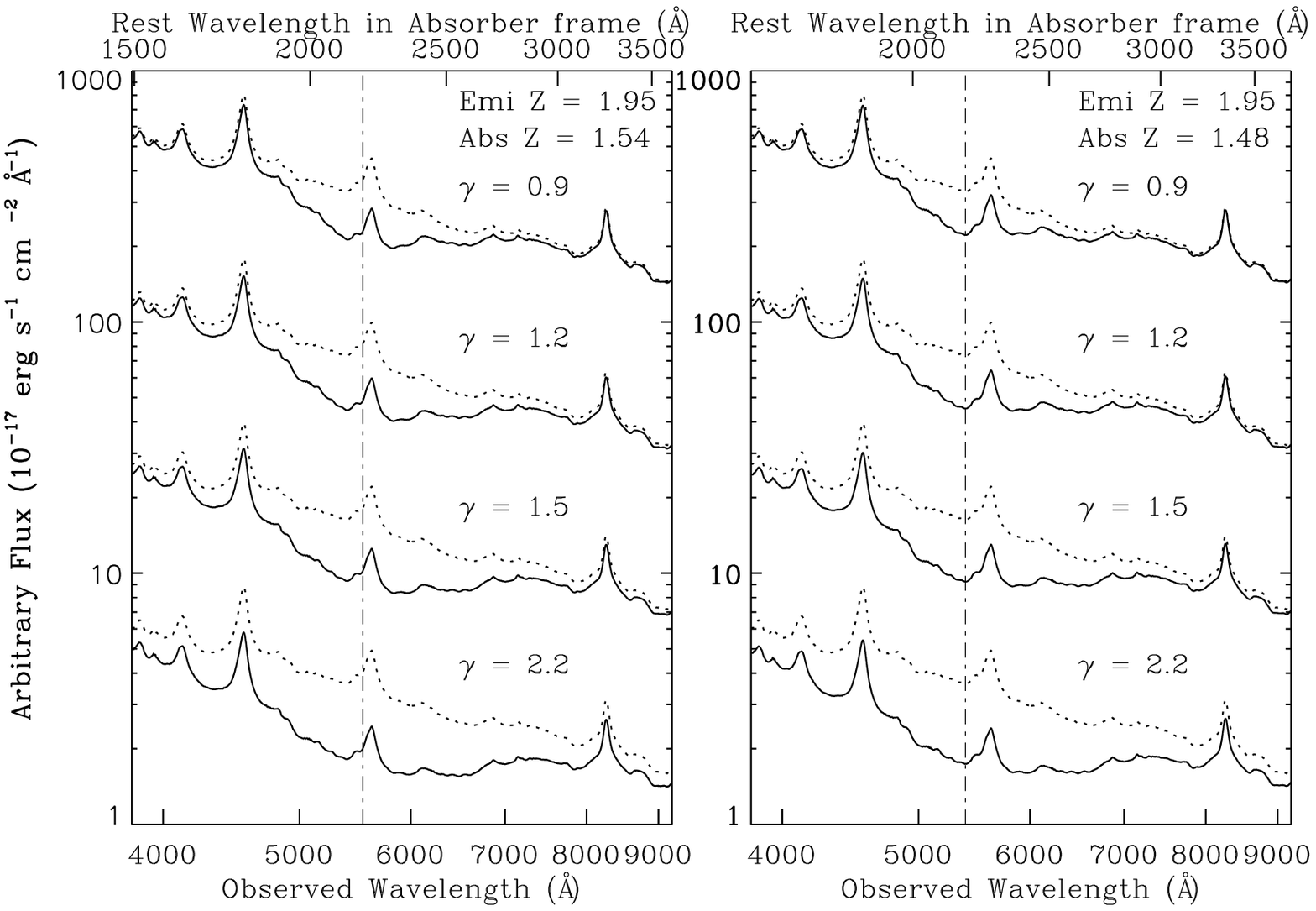}
\caption{(ONLINE ONLY) The dotted spectra are SDSS
quasar composite spectra. The solid ones are the spectra reddened by bumps.
The spectra for different x$_0^{qso}$s are organized in separated panels; the
spectra in the same panel are for increasing $\gamma$ from top to bottom.
The dot-dashed lines indicate the center positions of the bump.
\label{fig47}}
\end{figure}
\clearpage

\begin{figure}\epsscale{1.0}
\plotone{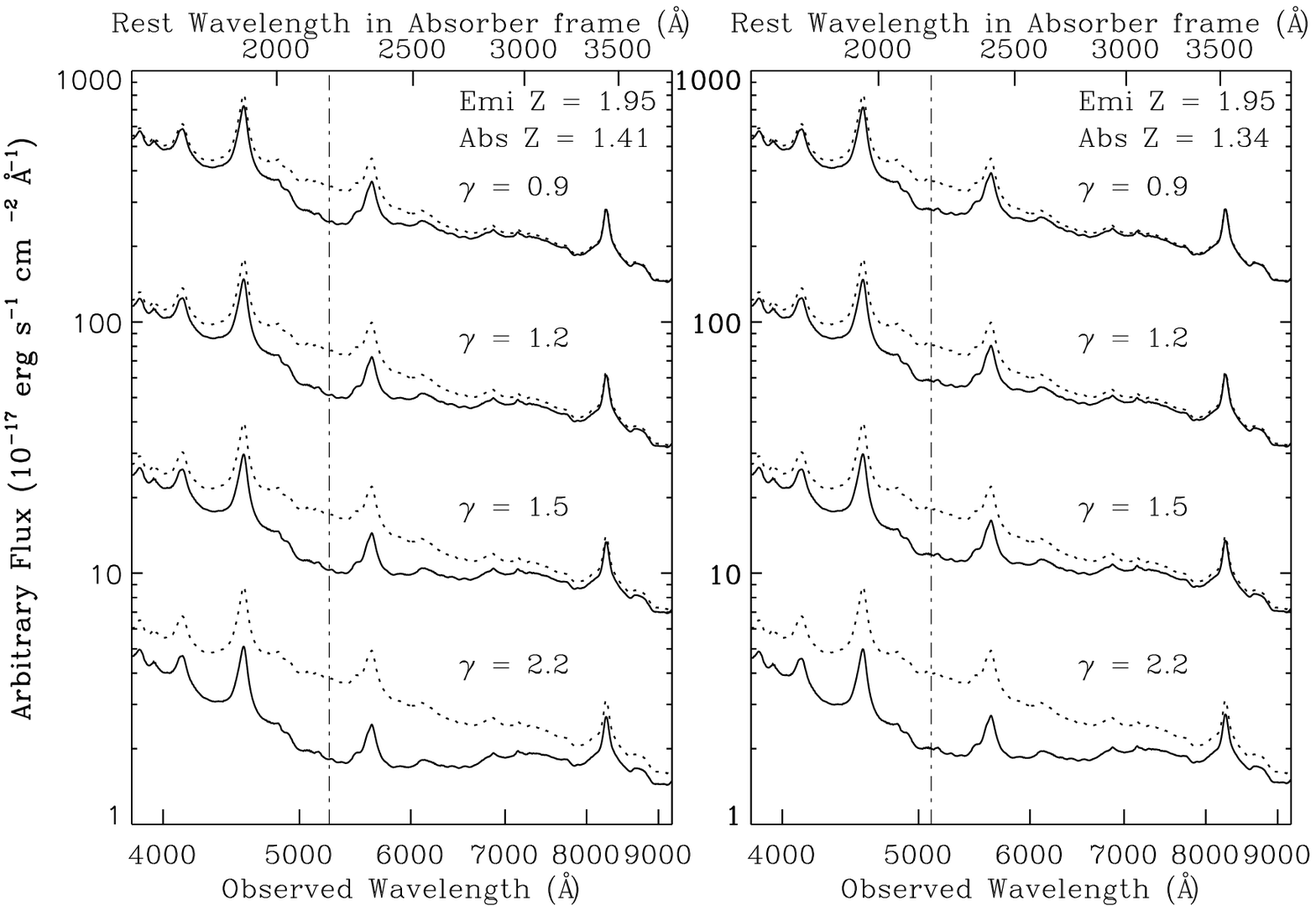}
\caption{(ONLINE ONLY) The dotted spectra are SDSS
quasar composite spectra. The solid ones are the spectra reddened by bumps.
The spectra for different x$_0^{qso}$s are organized in separated panels; the
spectra in the same panel are for increasing $\gamma$ from top to bottom.
The dot-dashed lines indicate the center positions of the bump.
\label{fig48}}
\end{figure}
\clearpage

\begin{figure}\epsscale{1.0}
\plotone{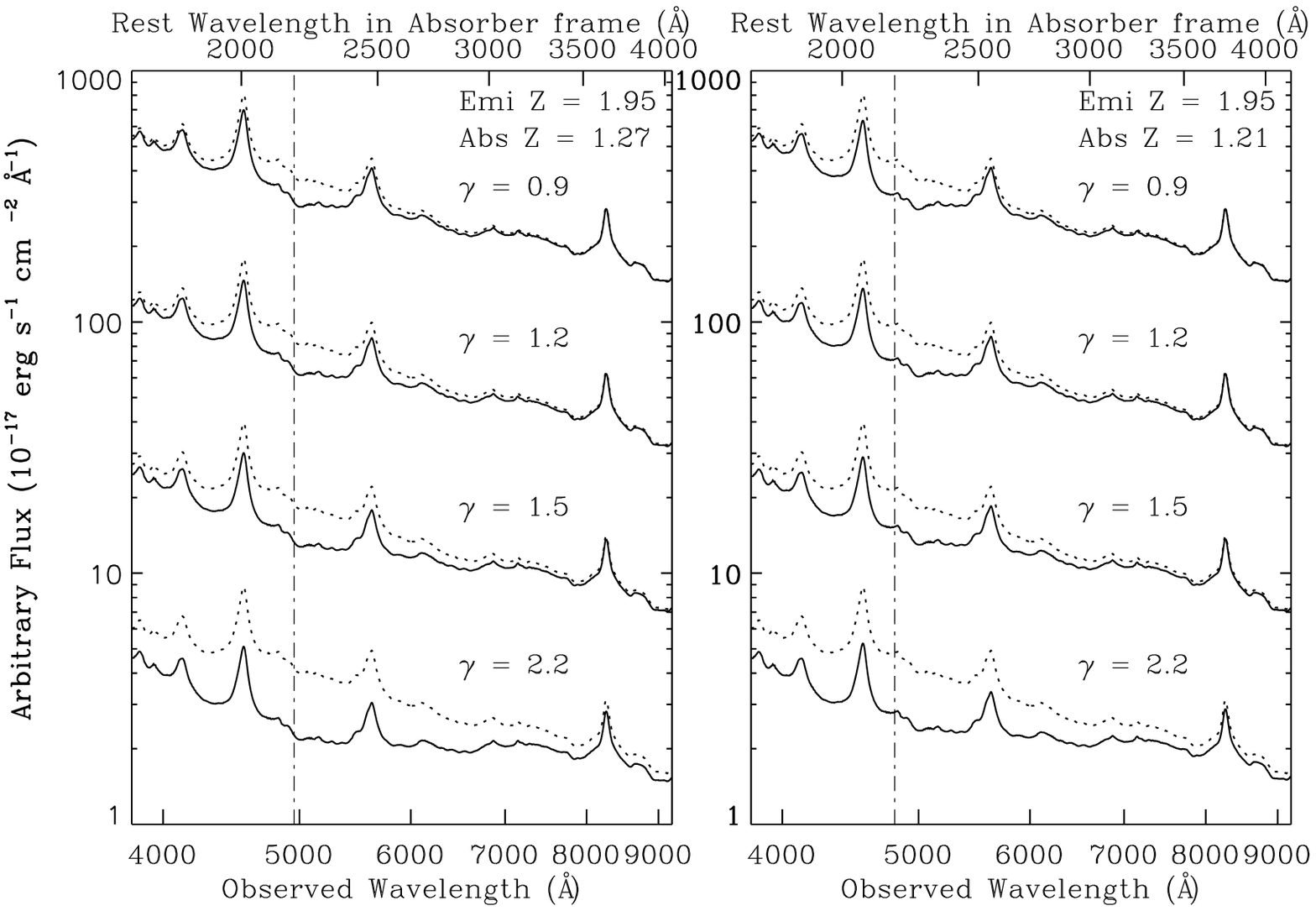}
\caption{(ONLINE ONLY) The dotted spectra are SDSS
quasar composite spectra. The solid ones are the spectra reddened by bumps.
The spectra for different x$_0^{qso}$s are organized in separated panels; the
spectra in the same panel are for increasing $\gamma$ from top to bottom.
The dot-dashed lines indicate the center positions of the bump.
\label{fig49}}
\end{figure}
\clearpage

\begin{figure}\epsscale{1.0}
\plotone{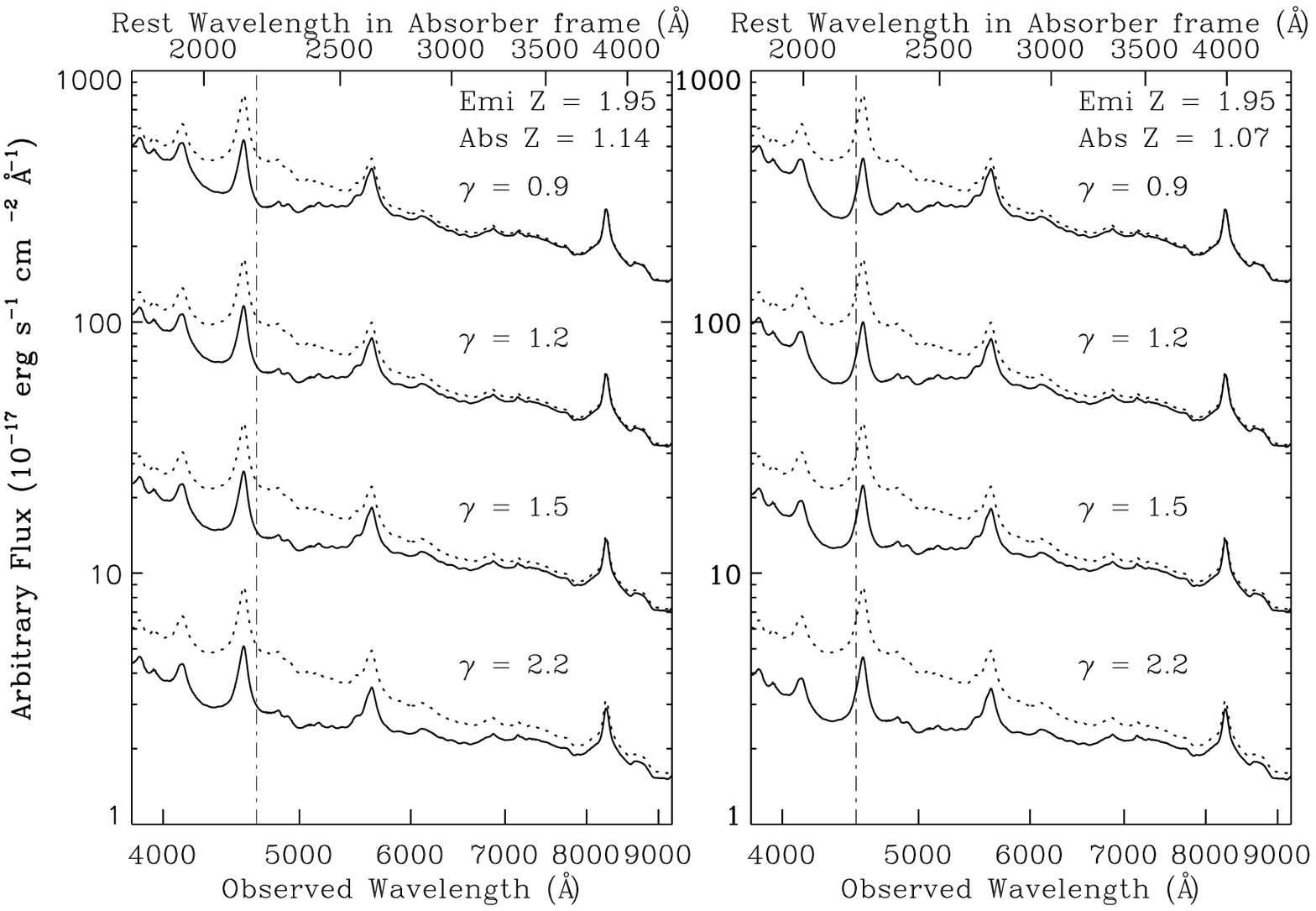}
\caption{(ONLINE ONLY) The dotted spectra are SDSS
quasar composite spectra. The solid ones are the spectra reddened by bumps.
The spectra for different x$_0^{qso}$s are organized in separated panels; the
spectra in the same panel are for increasing $\gamma$ from top to bottom.
The dot-dashed lines indicate the center positions of the bump.
\label{fig50}}
\end{figure}
\end{document}